\documentclass[a4paper,fleqn]{cas-sc}

\usepackage[numbers,sort&compress]{natbib}
\usepackage{float}
\usepackage{tabularx}
\usepackage{comment}

\usepackage{etoolbox}

\csletcs{savedAffiliationAlph}{int_to_alph:n}

\AddToHook{cmd/maketitle/before}{%
  \csletcs{int_to_alph:n}{int_to_arabic:n}%
}

\AddToHook{cmd/maketitle/after}{%
  \csletcs{int_to_alph:n}{savedAffiliationAlph}%
}

\def\tsc#1{\csdef{#1}{\textsc{\lowercase{#1}}\xspace}}
\tsc{WGM}
\tsc{QE}

\makeatletter
\let\ps@pprintTitle\ps@empty
\makeatother

\usepackage{comment}

\def\tsc#1{\csdef{#1}{\textsc{\lowercase{#1}}\xspace}}
\tsc{WGM}
\tsc{QE}

\ExplSyntaxOn
\cs_gset:Npn \__first_footerline:
{
  \group_begin:
  \small
  \sffamily
  \group_end:
}
\ExplSyntaxOff

\begin{document}
\let\WriteBookmarks\relax
\def\floatpagepagefraction{1}
\def\textpagefraction{.001}

\shorttitle{}    

\shortauthors{}  

\title [mode = title]{Physics-Informed Deep Learning in Oncology}

\author[1]{Scindhiya Laxmi}


\ead{scindhiya@iitism.ac.in}


\credit{}

\affiliation[1]{organization={Department of Mathematics and Computing},
         addressline={Indian Institute of Technology (Indian School of Mines)},
          city={Dhanbad},
         postcode={826004},
           state={Jharkhand},
            country={India}}

\author[2]{Vivek S Yadav}
\ead{vyadav06@mail.ubc.ca}

\author[3]{Ameya D. Jagtap}
\ead{ajagtap@wpi.edu}
\cormark[1]

\credit{}

\affiliation[2]{organization={Department of Mechanical Engineering},
            addressline={University of British Columbia}, 
            city={Vancouver },
            postcode={V6T 1Z4}, 
            state={BC},
            country={Canada}}
            
\affiliation[3]{organization={Aerospace Engineering Department},
            addressline={Worcester Polytechnic Institute}, 
            city={Worcester},
            postcode={01609}, 
            state={MA},
            country={USA}}
\cortext[1]{Corresponding author}



\begin{abstract}
Deep learning has enabled substantial advances in oncology, including cancer detection and diagnosis, medical image analysis, tumor characterization, treatment planning, response assessment, and outcome prediction. However, conventional data-driven models often require large annotated datasets, exhibit limited generalization across heterogeneous clinical settings, and lack explicit mechanisms for incorporating the physical, physiological, and biological principles underlying cancer development and treatment. Physics-informed deep learning (PIDL) integrates governing physical laws, mechanistic models, physiological constraints, and domain knowledge into deep learning frameworks, providing a principled approach for learning from sparse, noisy, and heterogeneous oncological data. This review presents a task-oriented synthesis of PIDL methods across oncology, encompassing neural networks, Kolmogorov–Arnold networks, convolutional and vision architectures, neural operators, generative models, and hybrid physics–data-driven frameworks. Their applications are examined across cancer detection and diagnosis, image reconstruction and analysis, tumor characterization, quantitative imaging, tumor progression, treatment response, radiation therapy, treatment planning, and multimodal clinical modeling. Based on the synthesized methodological and application-level evidence, we develop task-optimized PIDL workflows that align the clinical task with appropriate data representations, physical constraints, model architectures, and validation strategies. We further analyze key challenges, including physical-model fidelity, computational scalability, uncertainty quantification, domain generalization, standardized benchmarking, and clinical validation. Emerging directions involving multimodal foundation models, patient-specific digital twins, uncertainty-aware learning, and agentic PIDL are also discussed. This review provides a unified framework for understanding and designing task-specific PIDL approaches for oncology and identifies methodological requirements for their reliable clinical translation.
\end{abstract}

\begin{NoHyper}
\maketitle
\end{NoHyper}

\section{Introduction}

Cancer remains one of the leading causes of morbidity and mortality worldwide, demanding accurate, robust, and interpretable imaging technologies for early diagnosis, treatment planning, therapeutic monitoring, and prognosis \cite{bray2024global}. Modern oncologic imaging modalities, including magnetic resonance imaging (MRI), computed tomography (CT), positron emission tomography (PET), ultrasound, and hybrid imaging systems, provide rich anatomical, functional, and molecular information essential for precision oncology \cite{min2021decade}. However, these imaging modalities fundamentally operate as inverse problems, where clinically meaningful images or quantitative biomarkers must be inferred from incomplete, noisy, and often indirect physical measurements governed by complex acquisition physics \cite{baguer2020computed}. Therefore, computational oncology has become increasingly important in contemporary cancer care, transforming heterogeneous biomedical data into diagnostically and clinically actionable information through mathematical modeling, optimisation, and machine learning. Recent advances in deep learning have further revolutionised the analysis of oncological data, enabling improved disease characterisation, tumour localisation, progression modeling, and clinical decision-making \cite{izhar2025medical}. Convolutional neural networks (CNNs), U-Nets, residual architectures, transformer-based models, and hybrid attention frameworks have achieved remarkable success across tumor segmentation, lesion detection, cancer classification, radiomic characterisation, image reconstruction, motion correction, and radiotherapy planning \cite{ronneberger2015u, tran2021deep,shamshad2023transformers,azad2024medical}. These approaches benefit from hierarchical representation learning and end-to-end optimisation, enabling substantial improvements over traditional handcrafted machine learning pipelines. In particular, deep architectures have demonstrated strong performance in delineating complex tumor boundaries, accelerating reconstruction from undersampled acquisitions, and automating clinically intensive tasks \cite{fotopoulos2026explainable, adegboro2026artificial}. 

\begin{figure}[pos=htbp]
    \centering
    \includegraphics[scale=0.7,clip=true]{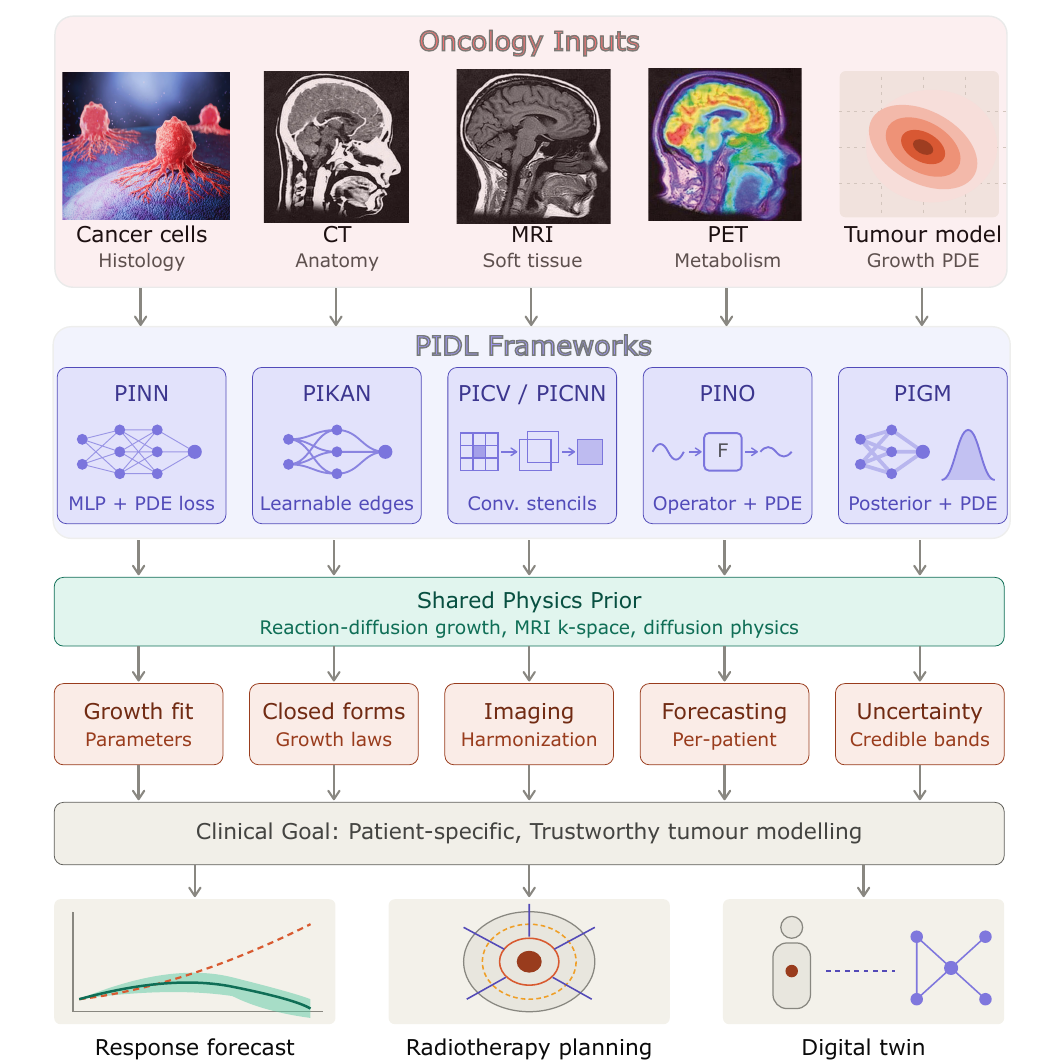}
    \caption{PIDL pipeline for oncology: multimodal inputs and tumour growth models are processed by PIDL-family frameworks under shared physics priors to enable parameter fitting, growth-law discovery, imaging harmonization, patient-specific forecasting, and uncertainty quantification, supporting response forecasting, radiotherapy planning, and digital twins.}
    \label{fig:onco1}
\end{figure}
Despite these advances, purely data-driven deep learning models face fundamental limitations that restrict reliable clinical deployment. Their performance is heavily dependent on large and high-quality annotated datasets, which are scarce in oncology due to annotation costs, privacy constraints, institutional fragmentation, and disease heterogeneity \cite{yao2025deep}. Models trained on limited or institution-specific datasets often exhibit poor generalisation across scanners, protocols, and patient populations. Moreover, most deep learning models operate as black-box function approximators, offering limited interpretability and weak uncertainty quantification \cite{ennab2024enhancing}. At the fundamental level, deep learning lacks explicit mechanisms to enforce the physical principles that govern image formation, signal generation, and tissue dynamics. Consequently, visually convincing predictions frequently violate established physical and physiological constraints, resulting in hallucinated anatomical structures, unstable image reconstructions, and unreliable quantitative biomarkers \cite{arridge2019solving}. These limitations have motivated the emergence of physics-informed deep learning (PIDL) \cite{raissi2019physics}, a paradigm that integrates governing physical knowledge directly into the learning process; see Figure~\ref{fig:onco1}. Rather than relying solely on empirical data, PIDL embeds scientific priors, including governing partial differential equations (PDEs), imaging forward operators, conservation laws, constitutive relationships, biomechanical models, pharmacokinetic constraints, and acquisition physics, within neural network architectures or optimisation objectives \cite{banerjee2025physics}. By constraining the solution space with established physical principles, PIDL leads to predictions that are both data-driven and physically consistent, thus improving generalisation under limited supervision, improving interpretability, and reducing dependence on large datasets \cite{cuomo2022scientific}.

Oncological modeling and data analysis are especially well suited to PIDL, as tumor initiation, progression, and therapeutic response are governed by well-established biological, physiological, and physical mechanisms \cite{das2026physics}. Many computational oncology problems, including image reconstruction, quantitative parameter estimation, kinetic modeling, radiation dosimetry, and tumor progression analysis, are formulated as forward or inverse problems constrained by known physical laws but are often challenged by sparse, noisy, or indirect measurements \cite{ahmadi2026physics}. Incorporating these mechanistic models into deep learning enables physically admissible inference while preserving the expressive power of neural networks. Building on the foundational concept of physics-informed neural networks (PINNs) \cite{raissi2019physics,jagtap2020adaptive,jagtap2020locally,abbasi2025challenges,abbasi2025history,jagtap2022physics,jagtap2022deepD,jagtap2022deep,menon2026intelligent,mhaskar2026approximation}, the field has rapidly evolved to include eXtended PINNs (XPINNs) \cite{jagtap2020extended}, conservative PINNs (cPINNs) \cite{cpinn_intro}, scientific foundational models \cite{menon2026scientific}, neural operators \cite{lu2019deeponet,zhang2026bubbleokan,goswami2024learning}, physics-informed convolutional neural networks (PICNNs) \cite{zhao2023physics}, physics-informed Kolmogorov-Arnold Networks \cite{menon2026fekan}, hybrid mechanistic learning frameworks \cite{poovesh2025bio}, transformer-based physics-guided architectures \cite{zhao2023pinnsformer} and Bayesian formulations for uncertainty quantification \cite{gou2023bayesian}. Collectively, these developments have substantially expanded the applicability of PIDL across diverse oncologic imaging tasks \cite{ahmadi2026physics,banerjee2025physics}.

Despite the growing body of review literature on PIDL and its applications in biomedical research \cite{hammernik2023physics, nemirovskyrotman2024explicit,yogita2025advances,ahmadi2025physics,banerjee2025physics,ahmadi2026physics}, a dedicated synthesis of PIDL for computational oncology remains lacking. Existing reviews lack a unified perspective on the end-to-end computational cancer research workflow and do not systematically examine how imaging physics and modality-specific acquisition mechanisms shape the design and application of PIDL across different modalities. Furthermore, the methodological foundations, complementary capabilities, and limitations of PINNs, PICNNs, PIKANs, neural operator-based methods, and uncertainty-aware mechanistic models have not been systematically compared within the context of oncologic data analysis. Current review literature also provides limited discussion of the challenges associated with clinical translation, including uncertainty quantification, domain generalisation, standardised benchmarking, and prospective clinical validation. To address these gaps, this review presents a comprehensive and unified perspective on PIDL in oncology. It systematically synthesizes methodological advances, task-oriented and modality-specific applications, benchmark datasets, evaluation protocols, translational challenges and future research opportunities. The principal contributions are as follows:

\begin{enumerate}
\item \textbf{Systematic evolution of PIDL}: We trace the methodological progression from conventional PINNs to contemporary architectures, including PICNNs, PIKANs, physics-informed neural operators (PINO), physics-informed generative models, transformer-based PIDL, Bayesian multi-fidelity frameworks and hybrid mechanistic-deep learning approaches.
\item \textbf{Task-oriented taxonomy of physics-informed learning in oncology}: We organize PIDL applications according to clinically relevant oncology tasks, namely reconstruction, denoising, registration, biomechanical motion correction, super-resolution, tumor segmentation, detection, metabolic quantification, treatment planning and therapy monitoring.
\item \textbf{Modality-oriented  taxonomy of physics-informed learning in oncology}: We systematically examine PIDL across CT, MRI, PET, ultrasound, and thermal imaging, extending the analysis to hybrid imaging and multiphysics models. We highlight how modality-specific acquisition mechanisms,  and coupled physical processes influence model design, computational challenges, and reliability in oncological applications.
\item \textbf{Comparison with purely data-driven deep learning}: We contrast conventional CNN- and transformer-based oncological task pipelines with physics-informed paradigms, highlighting their respective limitations in data dependence, interpretability, robustness and physical consistency.
\item \textbf{Translational challenges}: We critically examine the key methodological, computational, and clinical barriers to translating PIDL into cancer research and clinical practice.
\item \textbf{Design principles and future research directions}: We identify suitable PIDL strategies for diverse oncological tasks and discuss emerging research directions, including multimodal foundation models, digital twins, Bayesian multi-fidelity learning, and agentic and trustworthy PIDL, to advance precision oncology.
\end{enumerate}
Together, these contributions establish a unified framework that connects methodological advances with clinical translation, providing a foundation for the next generation of reliable, interpretable and clinically deployable physics-informed AI in oncologic imaging.

\section{How to make neural network models physics-informed}
Conventional deep-learning models learn a mapping from inputs to outputs by minimizing a data-driven loss. PINNs instead embed governing laws directly into the learning problem, so that the network is trained to satisfy the governing equations, initial conditions and boundary conditions as well as the observations \citep{raissi2019physics,jagtap2020extended,karniadakis2021physics,jagtap2026limits}. 

\begin{figure}[pos=htbp]
    \centering
    \includegraphics[trim= 0.2cm 0.3cm 0.2cm 0cm, scale=0.9,clip=true]{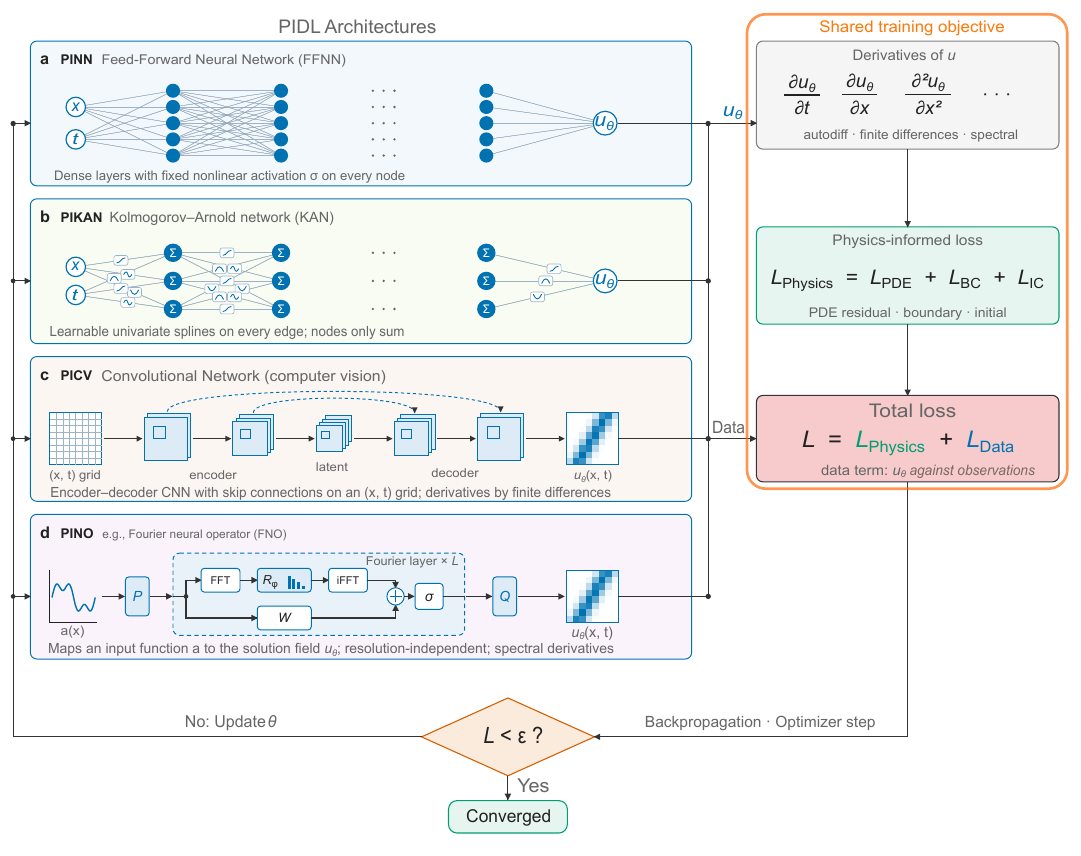}
    \caption{\textbf{The PIDL family}:
  Four architectures share one training loop.
  \textbf{a}, Physics-informed neural network (PINN): a feed-forward network maps $(x,t)$ to $u_{\theta}$.
  \textbf{b}, Physics-informed Kolmogorov-Arnold network (PIKAN): learnable splines on edges.
  \textbf{c}, Physics-informed computer-vision network (PICV): a convolutional encoder-decoder with skip connections.
  \textbf{d}, Physics-informed neural operator (PINO): Fourier layers map an input function $a$ to the solution field.
  In each case, derivatives of $u_{\theta}$ give the physics loss
($\mathcal{L}_{\mathrm{PDE}}+\mathcal{L}_{\mathrm{BC}}+\mathcal{L}_{\mathrm{IC}}$),
  which is added to a data loss; the parameters $\theta$ are updated by backpropagation until $\mathcal{L}<\varepsilon$.}
    \label{fig:pinn_architecture}
\end{figure}

Consider a physical system governed by a time-dependent differential equation on a spatial domain $\Omega\subset\mathbb{R}^{d}$ over a time interval $t\in[0,T]$,
\begin{equation}
\mathcal{F}\left(\mathbf{x},t,u,\nabla u,\nabla^2u,\ldots\right)=f(\mathbf{x},u),
\qquad (\mathbf{x},t)\in\Omega\times(0,T],
\label{eq:pde}
\end{equation}
where $u(\mathbf{x},t)$ is the unknown field and $\mathcal{F}$ is the governing differential operator. The problem is closed by an initial condition, $u(\mathbf{x},0)=u_0(\mathbf{x}),\mathbf{x}\in \Omega \cup \partial \Omega$, and a boundary condition, $\mathcal{B}(\mathbf{x},t,u,\nabla u,\ldots)=g(\mathbf{x},t)$ on $\partial\Omega\times(0,T]$. A neural network $u_\theta(\mathbf{x},t)$ with trainable parameters $\theta$ approximates the solution. Substituting this approximation into equation~(\ref{eq:pde}) defines the physics residual
\begin{equation}
r_\theta(\mathbf{x},t)=\mathcal{F}\left(\mathbf{x},t,u_\theta,\nabla u_\theta,\nabla^2u_\theta,\ldots\right)-f(\mathbf{x},u_\theta),
\label{eq:residual}
\end{equation}
whose derivatives are evaluated exactly by automatic differentiation \citep{baydin2018automatic}. Enforcing the residual at $N_f$ collocation points, the initial condition at $N_0$ points and the boundary condition at $N_b$ points gives
\begin{align}
\mathcal{L}_{\mathrm{phys}}&=\frac{1}{N_f}\sum_{i=1}^{N_f}\left|r_\theta(\mathbf{x}_f^{(i)},t_f^{(i)})\right|^2,\\
\mathcal{L}_{\mathrm{IC}}&=\frac{1}{N_0}\sum_{i=1}^{N_0}\left|u_\theta(\mathbf{x}_0^{(i)},0)-u_0(\mathbf{x}_0^{(i)})\right|^2,\\
\mathcal{L}_{\mathrm{BC}}&=\frac{1}{N_b}\sum_{i=1}^{N_b}\left|\mathcal{B}\left(\mathbf{x}_b^{(i)},t_b^{(i)},u_\theta,\nabla u_\theta,\ldots\right)-g(\mathbf{x}_b^{(i)},t_b^{(i)})\right|^2.
\end{align}
When measurements or high-fidelity simulations $u_d^{(i)}$ are available at $N_d$ points, a data-consistency term
$\mathcal{L}_{\mathrm{data}}=N_d^{-1}\sum_{i}|u_\theta(\mathbf{x}_d^{(i)},t_d^{(i)})-u_d^{(i)}|^2$
is added. The physics-informed objective is the weighted sum
\begin{equation}
{
\mathcal{L}_{\mathrm{total}}=
\lambda_{\mathrm{phys}}\mathcal{L}_{\mathrm{phys}}+
\lambda_{\mathrm{IC}}\mathcal{L}_{\mathrm{IC}}+
\lambda_{\mathrm{BC}}\mathcal{L}_{\mathrm{BC}}+
\lambda_{\mathrm{data}}\mathcal{L}_{\mathrm{data}}},
\qquad
\theta^{*}=\underset{\theta}{\arg\min}\;\mathcal{L}_{\mathrm{total}}(\theta),
\label{eq:total_loss}
\end{equation}
in which the coefficients $\lambda$ set the relative influence of each constraint (Fig.~\ref{fig:pinn_architecture}). The balance is not trivial. Competing loss terms can produce imbalanced gradients and stiff optimization, and the choice of weights, sampling and architecture strongly affects whether training converges \citep{wang2021understanding,krishnapriyan2021characterizing}. 

The central distinction from conventional deep learning is that a physics-informed model does not rely on observational examples alone. Known physical laws regularize the learning problem, so that limited, noisy or expensive data can be complemented by prior knowledge and the network is steered towards physically consistent solutions. The same principle has since been realized in several architectural families, each of which replaces or augments the multilayer perceptron (MLP) ansatz in a different way: Kolmogorov-Arnold networks, CNNs, neural operators and generative models; see Fig. \ref{fig:pinn_architecture}. 

\subsection{Physics-informed Kolmogorov-Arnold networks}
Kolmogorov-Arnold networks (KANs) \citep{liu2025kan,menon2026fekan} offer an alternative to the MLP for representing nonlinear functions. Motivated by the Kolmogorov-Arnold representation theorem, a KAN decomposes a multivariate function into compositions of learnable univariate functions. Whereas an MLP places fixed activation functions on nodes, a KAN places trainable univariate functions on edges, commonly parameterized with B-spline bases so that the nonlinearity is learned directly from data \citep{patra2025physicsinformed,rigas2024adaptive}. The $l$-th layer maps $n_l$ inputs to $n_{l+1}$ outputs as
\begin{equation}
x^{(l+1)}_j=\sum_{i=1}^{n_l}\phi^{(l)}_{j,i}\left(x^{(l)}_i\right),\qquad j=1,\ldots,n_{l+1},
\end{equation}
with edge functions $\phi^{(l)}_{j,i}(x)=\sum_m c^{(l)}_{j,i,m}B_m(x)$, where $B_m$ are spline basis functions and $c^{(l)}_{j,i,m}$ are trainable coefficients \citep{rigas2024adaptive}.

In a physics-informed KAN (PIKAN), this representation replaces the MLP ansatz of a PINN, $u_\theta(\mathbf{x},t)=\mathcal{N}^{\mathrm{KAN}}_\theta(\mathbf{x},t)$, with $\theta$ now denoting the spline coefficients. The residual $r_\theta$ is formed exactly as in equation~(\ref{eq:residual}), and training minimizes
\begin{equation}
\mathcal{L}_{\mathrm{PIKAN}}=\lambda_{\mathrm{PDE}}\mathcal{L}_{\mathrm{PDE}}+\lambda_{\mathrm{IC}}\mathcal{L}_{\mathrm{IC}}+\lambda_{\mathrm{BC}}\mathcal{L}_{\mathrm{BC}},\qquad
\mathcal{L}_{\mathrm{PDE}}=\frac{1}{N_r}\sum_{i=1}^{N_r}\left|r_\theta(\mathbf{x}_i,t_i)\right|^2 .
\end{equation}
PIKANs therefore retain the physics-constrained formulation of PINNs while exchanging the solution representation. Recent work addresses their computational and optimization challenges through efficient-KAN implementations, wavelet-based bases, adaptive grid refinement and augmented-Lagrangian optimization \citep{patra2025physicsinformed,rigas2024adaptive,zhang2025physicsinformed}. Systematic comparisons of KAN and MLP representations for differential equations suggest that the benefits are problem dependent, which calls for careful benchmarking \citep{shukla2024comprehensive,hao2024pinnacle}. 

\subsection{Physics-informed computer vision}
CNNs are the natural architecture for grid-based scientific and imaging data. Physics-informed CNNs (PICNNs) \citep{zhao2023physics} exploit this by approximating differential operators directly on discrete image or feature maps with convolution kernels or finite-difference stencils \citep{ren2022phycrnet}, rather than by automatic differentiation of a coordinate-based network \citep{baydin2018automatic}. This makes physical constraints compatible with standard convolutional pipelines. An early hybrid framework of this kind, introduced by Fang \citep{fang2021high}, replaced automatic differentiation with approximations of differential operators for PDE-constrained learning.

Physics-informed computer vision (PICV) generalizes the idea to vision-based learning, requiring predictions to be consistent with the imaging physics or governing equations. Let $I$ be an observed image and $u_\theta=f_\theta(I)$ the predicted physical field. The objective combines the image-analysis task with a physics penalty,
\begin{equation}
\mathcal{L}=\mathcal{L}_{\mathrm{task}}+\lambda_{\mathrm{phys}}\mathcal{L}_{\mathrm{phys}},\qquad
\mathcal{L}_{\mathrm{phys}}=\frac{1}{N}\sum_{i=1}^{N}\left|r_\theta(\mathbf{x}_i,t_i)\right|^2,
\end{equation}
where $r_\theta=\mathcal{F}(u_\theta,\nabla u_\theta,\nabla^2u_\theta,\partial u_\theta/\partial t)$. On a regular grid with spacings $\Delta x$, $\Delta y$ and $\Delta t$, central differences give
\begin{equation}
\frac{\partial u}{\partial x}\bigg|_{i,j}\approx\frac{u_{i+1,j}-u_{i-1,j}}{2\Delta x},\qquad
\frac{\partial^2u}{\partial x^2}\bigg|_{i,j}\approx\frac{u_{i+1,j}-2u_{i,j}+u_{i-1,j}}{\Delta x^2},
\end{equation}
with analogous expressions in $y$. These are equivalent to fixed convolution kernels, for example
\begin{equation}
K_x=\frac{1}{2\Delta x}\begin{bmatrix}-1&0&1\end{bmatrix},\qquad
K_{xx}=\frac{1}{\Delta x^2}\begin{bmatrix}1&-2&1\end{bmatrix},
\end{equation}
and their transposes $K_y$ and $K_{yy}$ in the second direction. For dynamic data, temporal derivatives follow from consecutive frames, using either a forward difference, $(u^{n+1}_{i,j}-u^{n}_{i,j})/\Delta t$, or a higher-order central difference, $(u^{n+1}_{i,j}-u^{n-1}_{i,j})/2\Delta t$, which correspond to the one-dimensional kernels $K_t^{\mathrm{forward}}=\Delta t^{-1}[-1,\,1]$ and $K_t^{\mathrm{central}}=(2\Delta t)^{-1}[-1,\,0,\,1]$. For the diffusion equation $\partial_t u-D\nabla^2u=0$, the discrete convolutional residual is therefore
\begin{equation}
r_\theta^n=\frac{u_\theta^{n+1}-u_\theta^{n}}{\Delta t}-D\left(K_{xx}*u_\theta^{n}+K_{yy}*u_\theta^{n}\right),
\end{equation}
where $*$ denotes convolution operation. This residual enters $\mathcal{L}_{\mathrm{phys}}$ and is optimized jointly with the task objective, so that differential operators are evaluated inside the CNN pipeline without altering its computational structure.

Zhao et al.\ \citep{zhao2023physics} used this approach for label-free thermal surrogate modeling, showing that physics-guided learning can be effective without explicit supervision. Ren et al.\ \citep{ren2022phycrnet} extended it to convolutional-recurrent encoder-decoder architectures for spatiotemporal dynamical systems. These developments are particularly relevant to oncologic imaging, where acquisition fidelity and biological consistency are essential for tumour characterization. Representative applications include physics-informed CT harmonization for stabilizing quantitative biomarkers across scanners \citep{zarei2024physics}, physics-guided diffusion MRI denoising for prostate tumour imaging \citep{lin2025novel}, and CNN-based reconstruction with MRI $k$-space consistency for accelerated acquisition \citep{ahmadi2026physics}. More broadly, embedding acquisition physics in convolutional architectures supports robust reconstruction and analysis when training data are limited. 
\subsection{Physics-informed neural operators}
Operator-learning methods \citep{kovachki2023neural,hao2023gnot} learn mappings between function spaces rather than individual solutions, and can therefore generalize across physical parameters, acquisition settings or patient-specific conditions. DeepONet \citep{lu2019deeponet} introduced this paradigm through a branch network that encodes the input function and a trunk network that represents the evaluation coordinates. Neural operators extend the concept by learning global solution operators through parameterized integral or spectral transformations, as in Fourier neural operators (FNOs) \citep{kovachki2021universal,li2023fourier}. Physics-informed DeepONets show that governing equations can supplement or even replace paired training data in this setting \citep{wang2021learning}. 

The physics-informed neural operator (PINO) combines operator learning with PDE constraints. For a dynamical system
\begin{equation}
\frac{\partial u}{\partial t}=\mathcal{R}(u),\qquad u(\mathbf{x},0)=a(\mathbf{x}),\qquad u|_{\partial\mathcal{D}}=g,
\end{equation}
where $a$ is the initial condition and $\mathcal{R}$ a spatial differential operator, the goal is to learn $\mathcal{G}^{\dagger}:a\mapsto u(\mathbf{x},t)$ for $t\in[0,T]$ with a neural operator $\mathcal{G}_\theta$, typically an FNO. In contrast to a PINN, which fits one PDE instance, PINO represents a whole family of solutions. The PDE, boundary and initial-condition constraints form
\begin{equation}
\mathcal{L}_{\mathrm{PDE}}=\int_0^T\!\!\int_{\mathcal{D}}\left|\partial_t\mathcal{G}_\theta(a)-\mathcal{R}(\mathcal{G}_\theta(a))\right|^2 d\mathbf{x}\,dt
+\alpha\int_0^T\!\!\int_{\partial\mathcal{D}}\left|\mathcal{G}_\theta(a)-g\right|^2 d\mathbf{x}\,dt
+\beta\int_{\mathcal{D}}\left|\mathcal{G}_\theta(a)(\mathbf{x},0)-a\right|^2 d\mathbf{x}.
\end{equation}
Given paired data $\{(a_j,u_j)\}_{j=1}^{N}$, PINO adds the supervised loss $\mathcal{J}_{\mathrm{data}}=N^{-1}\sum_j\|u_j-\mathcal{G}_\theta(a_j)\|_U^2$ to the expected PDE loss $\mathcal{J}_{\mathrm{PDE}}=\mathbb{E}_{a\sim\mu}[\mathcal{L}_{\mathrm{PDE}}(a,\mathcal{G}_\theta(a))]$:
\begin{equation}
{\mathcal{J}_{\mathrm{PINO}}=\mathcal{J}_{\mathrm{data}}+\lambda_{\mathrm{PDE}}\mathcal{J}_{\mathrm{PDE}}},
\end{equation}
and either term may be used alone when data or observations are unavailable. A distinctive feature is multi-resolution enforcement of physics: the operator can be trained on coarse solution data while the residual is evaluated on a much finer grid.

PINO also supports instance-wise fine-tuning. A pretrained operator $\mathcal{G}_{\theta_0}$ provides the initial ansatz $\mathcal{G}_\theta(a)$ for a specific initial condition, which is then refined using the PDE residual. An anchor loss $\mathcal{L}_{\mathrm{op}}=\|\mathcal{G}_\theta(a)-\mathcal{G}_{\theta_0}(a)\|_U^2$ prevents excessive drift from the learned operator, giving
\begin{equation}
{\mathcal{J}_{\mathrm{FT}}=\mathcal{L}_{\mathrm{PDE}}+\alpha_{\mathrm{op}}\mathcal{L}_{\mathrm{op}}},
\end{equation}
and progressive grid refinement further improves accuracy at high resolution. The resulting two-stage strategy, which pairs coarse data and a high-resolution PDE constraint for operator learning with a PDE loss and anchor loss for fine-tuning, is well suited to personalized modeling.

In oncology, this separation maps naturally onto patient-specific dynamics such as tumour growth, diffusion, invasion and treatment response. The initial tumour state and biological parameters are the operator input, the predicted longitudinal tumour state is the output, and a reaction--diffusion or other mechanistic growth model supplies the PDE residual. Operator learning thus captures a family of tumour dynamics, while fine-tuning adapts it to an individual patient. Physics-informed operator methods further offer opportunities for generalized inverse problems, including rapid reconstruction, parameter estimation and adaptation across imaging protocols \citep{tripura2023wavelet,zeng2023neural}.

\subsection{Physics-informed generative models}
Generative models learn a distribution over solution fields rather than a single deterministic prediction, which makes them natural tools for ill-posed problems, sparse or noisy observations and uncertainty quantification. Physics-informed generative models (PIGMs) impose the governing equations on the generated samples, so that every draw from the learned distribution is both statistically plausible and physically consistent \citep{yang2020physics,jacobsen2025cocogen,bastek2025physics,mondal2026genvoid}. Let $c$ denote the conditioning information (for example sparse measurements, an initial condition or an image) and $u\sim p_\theta(u\,|\,c)$ the generated field. The residual of equation~(\ref{eq:residual}) is evaluated on each sample, and its expectation defines the physics loss
\begin{equation}
\mathcal{L}_{\mathrm{phys}}=\mathbb{E}_{c}\,\mathbb{E}_{u\sim p_\theta(u|c)}\!\left[\frac{1}{N_f}\sum_{i=1}^{N_f}\left|r(\mathbf{x}_f^{(i)},t_f^{(i)};u)\right|^2\right],
\end{equation}
which is added to the generative objective of the chosen model class. The families differ in where the distribution is represented and how it is sampled.

Bayesian PINNs (B-PINNs) \citep{yang2021b} are the most direct probabilistic extension of the PINN and can be regarded as generative in the classical statistical sense. Network parameters and physical variables are treated as random variables with a prior $p(\theta)$, and the likelihood combines the observations, the physics residual and the initial and boundary conditions,
\begin{equation}
p(\theta\,|\,\mathcal{D})\;\propto\;p(\theta)\,p(\mathcal{D}_{\mathrm{data}}\,|\,\theta)\,p(\mathcal{D}_{\mathrm{phys}}\,|\,\theta)\,p(\mathcal{D}_{\mathrm{IC/BC}}\,|\,\theta).
\end{equation}
Samples $\theta\sim p(\theta\,|\,\mathcal{D})$, obtained by Hamiltonian Monte Carlo, variational inference or deep ensembles, induce a distribution over solution fields $u_\theta$ and propagate data and model uncertainty through the governing physics \citep{linka2022bayesian,molnar2022flow}. This gives robustness to noise, sparsity and model mismatch. B-PINNs perform instance-specific inference in weight space, which is principled but costly to sample and difficult to scale to high-dimensional parameter spaces. Deep generative models take the complementary approach of learning an explicit, amortized sampler over fields.

In physics-informed generative adversarial networks (PIGANs), a generator $G_\theta$ maps a latent variable $z\sim p(z)$ and the condition $c$ to a field $\hat{u}=G_\theta(z,c)$ and is trained against a discriminator $D_\phi$ that distinguishes generated fields from reference data. The generator is additionally penalized for violating the physics,
\begin{equation}
\min_{\theta}\max_{\phi}\;\mathcal{L}_{\mathrm{GAN}}(\theta,\phi)+\lambda_{\mathrm{phys}}\mathcal{L}_{\mathrm{phys}}(\theta)+\lambda_{\mathrm{IC}}\mathcal{L}_{\mathrm{IC}}(\theta)+\lambda_{\mathrm{BC}}\mathcal{L}_{\mathrm{BC}}(\theta),
\end{equation}
where $\mathcal{L}_{\mathrm{GAN}}=\mathbb{E}_{u}[\log D_\phi(u)]+\mathbb{E}_{z}[\log(1-D_\phi(G_\theta(z,c)))]$. The residual is computed by automatic differentiation of the generator output, and sampling different latent variables yields an ensemble of physically consistent solutions from which predictive uncertainty can be estimated \citep{yang2020physics,yang2021b}. Adversarial training is, however, prone to instability and mode collapse, and is sensitive to the balance between the adversarial and physics terms.

Diffusion and score-based models currently offer the most flexible route. A forward noising process $u_\tau=\alpha_\tau u_0+\sigma_\tau\epsilon$, with $\epsilon\sim\mathcal{N}(0,\mathbf{I})$ and diffusion time $\tau\in[0,1]$, is inverted by a network $\epsilon_\theta(u_\tau,\tau,c)$, equivalently a score $s_\theta=\nabla_{u_\tau}\log p_\tau(u_\tau|c)$, trained by denoising score matching,
\begin{equation}
\mathcal{L}_{\mathrm{DSM}}=\mathbb{E}_{u_0,\tau,\epsilon}\left\|\epsilon_\theta(u_\tau,\tau,c)-\epsilon\right\|^2 .
\end{equation}
Samples are generated by integrating the reverse-time stochastic or probability-flow differential equation from noise to data. Physics enters in two complementary ways. During training, the one-step denoised estimate $\hat{u}_0=(u_\tau-\sigma_\tau\epsilon_\theta)/\alpha_\tau$ is passed through the residual,
\begin{equation}
\mathcal{J}_{\mathrm{PIDM}}=\mathcal{L}_{\mathrm{DSM}}+\lambda_{\mathrm{phys}}\,\mathbb{E}_{u_0,\tau,\epsilon}\left\|r(\hat{u}_0)\right\|^2,
\end{equation}
so that the learned distribution concentrates on solutions of the PDE \citep{bastek2025physics}. During sampling, the physics acts as guidance. For observations $y=\mathcal{A}(u)+\eta$, with $\mathcal{A}$ an observation operator, the conditional score is approximated by
\begin{equation}
\nabla_{u_\tau}\log p(u_\tau|y)\approx s_\theta(u_\tau,\tau)-\zeta_{\mathrm{obs}}\nabla_{u_\tau}\left\|\mathcal{A}(\hat{u}_0)-y\right\|^2-\zeta_{\mathrm{PDE}}\nabla_{u_\tau}\left\|r(\hat{u}_0)\right\|^2,
\end{equation}
where $\zeta_{\mathrm{obs}}$ and $\zeta_{\mathrm{PDE}}$ set the guidance strengths. A single pretrained prior can then serve both forward and inverse problems, such as reconstruction from sparse sensors or recovery of unknown coefficients, without retraining \citep{jacobsen2025cocogen,huang2024diffusionpde}. Flow-matching formulations follow the same pattern with a learned velocity field in place of the score and allow faster sampling.

These formulations are complementary to the architectures discussed above: the network, generator or denoiser can itself be an MLP, a KAN, a CNN or a neural operator, with the probabilistic layer supplying a description of uncertainty. Compared with B-PINNs, deep generative models scale more readily to high-dimensional, image-like outputs and sample quickly once trained, at the price of requiring training data and, for diffusion models, iterative sampling. Open challenges include the cost of sampling, the need to differentiate the residual through the sampler or denoiser, the sensitivity of physical consistency to the guidance weights, and the fact that soft constraints guarantee consistency only approximately. Hard-constraint projection and consistency-based sampling are active responses to these issues. Recent advances in scalability and reliability are making these approaches increasingly attractive for high-stakes settings such as medical imaging, disease modeling and treatment planning \citep{casali2025physics,de2025quantification,shih2026uncertainty}.

In oncology, PIGMs offer a route to uncertainty-aware, patient-specific inference. Conditioned on sparse longitudinal imaging, a Bayesian or generative physics-informed model can sample an ensemble of tumour states and reaction-diffusion parameters that are consistent with both the scans and the mechanistic growth law, thereby quantifying uncertainty in predicted progression or treatment response. The same mechanism supports physically constrained image reconstruction, synthesis and super-resolution when acquisition is limited, and complements operator learning by providing calibrated distributions rather than point forecasts. By reporting predictive confidence alongside physical consistency, physics-informed generative learning is an important step towards trustworthy scientific machine learning and clinically deployable PIDL systems.

\begin{figure}[pos=htbp]
    \centering
    \includegraphics[width=1.0\textwidth]{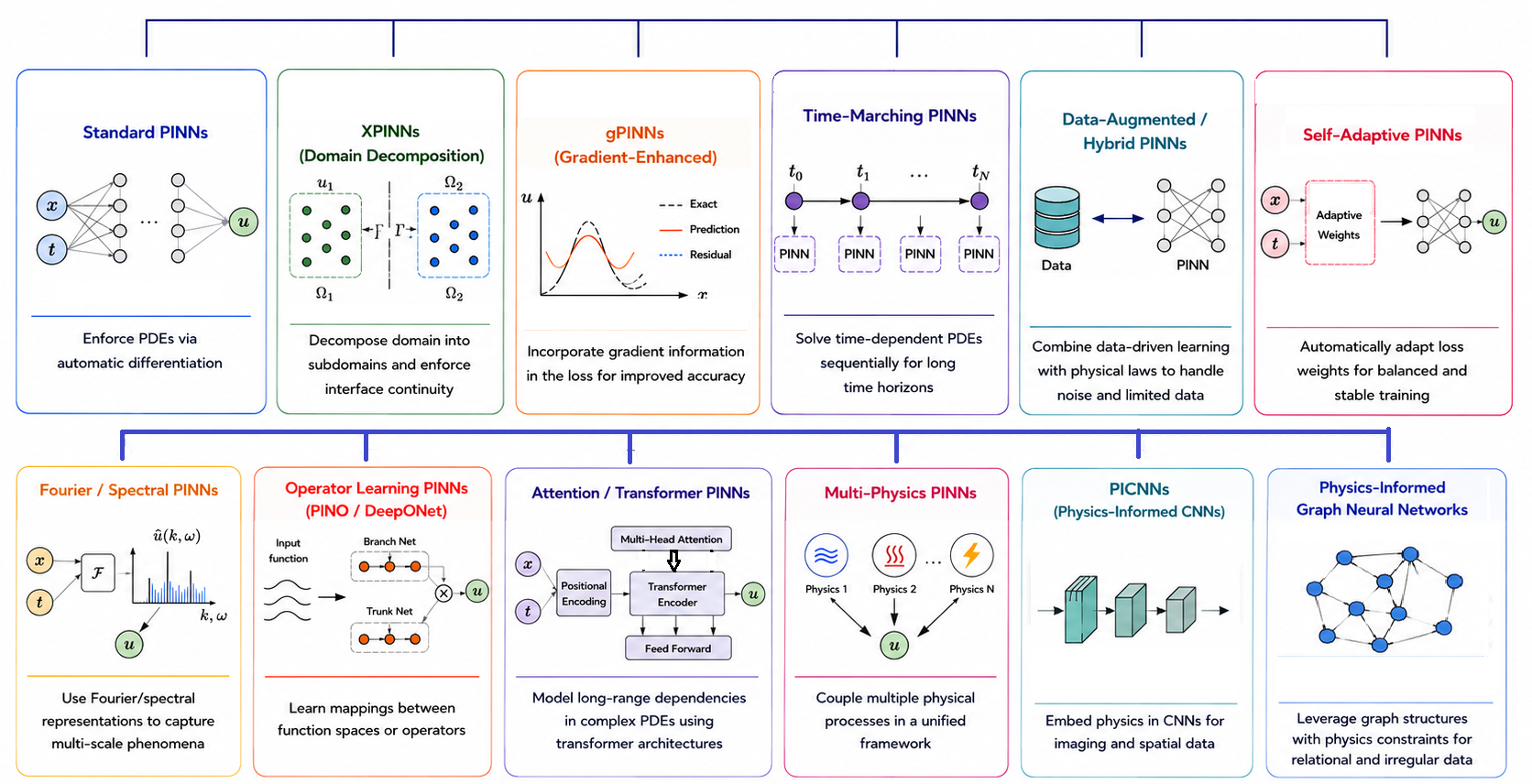}
    \caption{\textbf{Overview of major physics-informed deep learning architectures and their key characteristics.}}
    \label{fig:pidl_variants}
\end{figure}

\vspace{0.2cm}
\noindent \textbf{Evolution and landscape of PIDL}: Tables~\ref{tab:pidl_evolution_1} and \ref{tab:pidl_evolution_2} trace the field from early neural PDE solvers and the original PINN to scalable operator learning and to hybrid, transformer-based, graph-based and uncertainty-aware frameworks. Fig.~\ref{fig:pidl_variants} groups the major architectural families: domain-decomposition and adaptive PINNs, spectral and operator-learning approaches, transformer-based and multi-physics architectures, and physics-informed CNNs and graph neural networks. Taken together, they show a progression from solving one equation at a time towards learning reusable, uncertainty-aware models that integrate data and physics at scale.

\begin{table*}[H]
\centering
\caption{\textbf{Evolution of physics-informed deep learning and related scientific machine learning approaches: foundational developments.}}
\label{tab:pidl_evolution_1}
\small
\renewcommand{\arraystretch}{0.8}
\setlength{\tabcolsep}{4pt}
\begin{tabular}{p{3.6cm} p{4.0cm} p{7.7cm}}
\hline
\textbf{Models} & \textbf{Key Technology} & \textbf{Main Contributions} \\
\hline
Neural PDE Solvers~\citep{lagaris1998artificial} & Neural-network trial solutions & Introduced neural-network-based trial solutions for differential equations, with constraints incorporated into the solution representation. \\ \hline
SINDy~\citep{brunton2016discovering} & Sparse regression & Identified governing equations from data by sparse regression over a library of candidate nonlinear terms. \\ \hline
Deep Ritz Method~\citep{yu2018deep} & Variational neural solver & Reformulated PDE solution as variational energy minimization and optimized the corresponding functional using neural networks. \\ \hline
Deep Galerkin method ~\citep{sirignano2018dgm} & Residual-based neural solver & Minimized the PDE residual and boundary/initial losses on randomly sampled points for mesh-free, high-dimensional PDE solving. \\ \hline
PINNs~\citep{raissi2019physics} & PDE-residual-constrained learning & Established the neural  framework by embedding governing differential equations into network training through physics-based residual losses. \\ \hline
HNNs~\citep{greydanus2019hamiltonian} & Hamiltonian-structured learning & Learned dynamical systems through Hamiltonian representations while preserving the underlying physical structure and conservation properties. \\   \hline
fPINNs~\citep{pang2019fpinns} & Fractional differential constraints & Extended physics-informed learning to fractional differential equations by incorporating fractional operators into the residual formulation. \\ \hline
Adaptive activation PINNs ~\citep{jagtap2020adaptive,jagtap2020locally,jagtap2022deep} & Physics-informed neural network & Introduced trainable scaling parameters in activation functions to accelerate convergence and improve PINN training. \\ \hline
LNNs~\citep{cranmer2020lagrangian} & Lagrangian-structured learning & Learned Lagrangian formulations directly from observational data, providing physically structured modeling of dynamical systems without explicitly specifying canonical momenta. \\ \hline
cPINNs~\citep{cpinn_intro} & Conservative domain decomposition & Introduced domain decomposition with interface conditions to improve the treatment of conservation laws and discontinuous solutions. \\ \hline
XPINNs~\citep{jagtap2020extended} & Space-time domain decomposition & Generalized domain decomposition for PINNs by independently training neural networks over multiple space-time subdomains. \\ \hline
hp-VPINNs~\citep{kharazmi2021hp} & Variational formulation and $hp$-refinement & Combined variational PINNs with domain decomposition and $hp$-refinement to improve numerical approximation of PDE solutions. \\ \hline
B-PINNs~\citep{yang2021b} & Bayesian physics-informed learning & Introduced Bayesian inference into PINNs to quantify uncertainty in PDE solutions and inferred physical parameters. \\ \hline
gPINNs~\citep{yu2022gradient} & Gradient-enhanced residual learning & Augmented the conventional PINN objective with gradients of the PDE residual, providing additional differential constraints during training. \\ 
\hline
Norm-PCGrad~\citep{menon2026tackling} & Gradient-conflict mitigation & Extended PCGrad with gradient normalization to obtain conflict-free update directions for PINN and PIKAN training, including domain-decomposed settings such as XPINN. \\ \hline
\end{tabular}
\end{table*}

\begin{table*}[!t]
\centering
\caption{\textbf{Recent developments in physics-informed learning, neural operators and emerging physics-informed architectures.}}
\label{tab:pidl_evolution_2}
\small
\renewcommand{\arraystretch}{0.8}
\setlength{\tabcolsep}{4pt}
\begin{tabular}{p{3.6cm} p{4.0cm} p{7.7cm}}
\hline
\textbf{Models} & \textbf{Key Technology} & \textbf{Main Contributions} \\
\hline
DeepONet~\citep{lu2021learning} & Branch-trunk operator learning & Learned nonlinear mappings between input and output functions using branch and trunk networks, establishing a general neural-operator framework. \\ \hline
FNO~\citep{li2020fourier} & Fourier-space operator learning & Learned mappings between function spaces through Fourier-domain transformations, enabling efficient operator approximation across discretizations. \\ \hline
Physics-informed DeepONet~\citep{wang2021learning} & Operator learning & Incorporated PDE residual constraints into DeepONet training to learn solution operators with little or no paired data. \\ \hline
PINO~\citep{li2024physics} & Physics-informed neural operators & Combined neural-operator learning with PDE constraints, enabling solution operators to learn from available data while enforcing governing equations at potentially higher resolutions. \\ \hline
Geo-FNO~\citep{li2023geometry} & Geometry-aware Fourier operators & Extended Fourier neural operators to geometrically complex and non-Cartesian domains through geometry-aware transformations. \\ \hline
FourCastNet~\citep{pathak2022fourcastnet} & Large-scale Fourier operator learning & Demonstrated scalable Fourier neural operators for high-resolution global weather forecasting, illustrating the use of neural operators for large-scale scientific modeling. \\ \hline
Neural Operator~\citep{kovachki2023neural} & Function-space operator learning & Established a general framework for learning mappings between infinite-dimensional function spaces and emphasized discretization-invariant operator learning. \\ \hline
PINNsFormer~\citep{zhao2023pinnsformer} & Transformer-based physics-informed learning & Introduced transformer-based physics-informed learning using tokenized space-time representations and self-attention for PDE solution modeling. \\ \hline
PINNacle~\citep{hao2024pinnacle} & Benchmarking and evaluation & Provided a systematic benchmark for evaluating PINNs across PDEs, architectures, optimization strategies and numerical settings. \\ \hline
PIKANs~\citep{wang2025kolmogorov} & KAN-based physics-informed learning & Integrated governing-equation constraints with Kolmogorov-Arnold networks using learnable univariate functions within the network representation. \\ \hline
Multi-physics PINNs~\citep{harandi2024mixed} & Coupled physical constraints & Extended physics-informed learning to coupled physical systems by jointly enforcing multiple interacting governing equations and interface constraints. \\ \hline
Physics-informed diffusion models~\citep{bastek2025physics} & Physics-constrained generative learning & Integrated physical constraints with diffusion-based generative models for inverse problems, reconstruction and scientific simulation. \\ \hline
Uncertainty-aware and multi-fidelity PINNs~\citep{imanov2026multi} & Multi-fidelity and probabilistic learning & Combined heterogeneous-resolution observations, physical constraints and uncertainty-aware formulations for learning under limited or noisy measurements. \\
\hline
\end{tabular}
\end{table*}

\section{Motivation for PIDL in Oncology}
Pure data-driven deep learning has become central to computational oncology, enabling the transformation of raw, noisy measurements into clinically actionable representations. By learning hierarchical features directly from data, deep neural networks have advanced tasks such as tumor segmentation, cancer detection and classification, image-guided interventions, and automated annotation. The following subsections review the evolution of major deep learning architectures and highlight the fundamental challenges that motivate the adoption of physics-informed deep learning.

\subsection{Data-driven deep learning in oncology: An overview}
CNN-based architectures \citep{huang2020unet,ruan2024vm,huang2024segmenting} established the foundation for tumor characterisation and quantitative radiomic analysis by learning clinically relevant patterns from oncologic images. Their ability to capture tumor morphology, intensity, texture, and spatial heterogeneity has supported key clinical tasks, including tumor detection, segmentation, classification, grading, and quantitative assessment of treatment response. In particular, encoder-decoder architectures such as U-Net and its variants have facilitated precise delineation of tumor boundaries and subregions, supporting clinical applications in diagnosis, disease assessment, prognosis, and treatment planning. Subsequent advances in transformer-based and state-space architectures \citep{cao2022swin,hatamizadeh2022unetr} have further enhanced global contextual modeling and feature representation for complex oncologic imaging tasks.

\vspace{0.2cm}
\noindent \textbf{U-Net and its variants  in oncology}: U-Net \citep{ronneberger2015u} and its variants form the foundation of medical image segmentation, implementing encoder-decoder structures with skip connections to preserve spatial details. The baseline U-Net achieves strong performance across modalities such as MRI, CT, X-rays, and microscopy \citep{azad2024medical, siddique2021u}. Subsequent variants improve efficiency and representation, lightweight models such as Half-U-Net \citep{lu2022half} reduce computational cost while maintaining accuracy, whereas architectures like DoubleU-Net \citep{jha2020doubleu} enhance feature learning through stacked networks with pretrained encoders and multi-scale context modules. Multi-scale designs, such as EMED-UNet \citep{shah2023emed}, further improve feature extraction through parallel encoder–decoder pathways. Attention-based extensions, including DSCA-Net \citep{shan2023dsca}, incorporate spatial and contextual attention mechanisms to refine segmentation while reducing the number of parameters. Enhanced skip connections, as in U-Net v2 \citep{peng2025u}, improve semantic feature fusion, while recurrent and residual variants such as R2U-Net \citep{alom2018recurrent} capture complex dependencies for improved performance in lesion segmentation tasks. These developments demonstrate the adaptability of U-Net architectures in achieving accurate, efficient, and robust tumor segmentation.

\vspace{0.2cm}
\noindent \textbf{Transformer-enhanced and hybrid CNN architectures in oncology}: While CNNs are effective for local feature extraction, their limited ability to capture long-range dependencies has driven the development of hybrid and transformer-based architectures \citep{xin2022improved,jiao2023transdose,xiao2023transformers}. Hybrid models such as CoTrFuse \citep{chen2023cotrfuse} combine CNN encoders \citep{tan2019efficientnet} with Swin Transformers \citep{liu2021swin} to integrate local and global features, improving segmentation performance. Similarly, MedFuseNet \citep{chen2025medfusenet} employs parallel convolutional and transformer branches with attention mechanisms, including atrous spatial pyramid pooling and cross-attention fusion, to enhance feature representation. Transformer-based models, such as SegFormer3D \citep{perera2024segformer3d}, extend hierarchical self-attention to volumetric data, achieving efficient and accurate 3D segmentation with reduced computational cost. These architectures integrated global contextual representation learning with long-range dependency modeling, representing a shift from local feature extraction to context-aware tumor analysis in oncologic imaging \cite{peng2021conformer,chen2024transunet}.

\subsection{Data-driven deep learning in oncology: Limitations}
Despite significant progress, deep learning-based methods face fundamental challenges due to their heavy reliance on large, annotated datasets and their inability to intrinsically incorporate governing physical principles.  The main challenges are described as follows:

\vspace{0.2cm}
\noindent \textbf{Data availability and domain generalisation}: Data scarcity remains a fundamental challenge in computational oncology, limiting model performance, generalisation, and clinical reliability. High-quality annotations require expert clinicians to delineate complex anatomical and pathological structures, often in volumetric data \citep{bhalgat2018annotation,qu2023abdomenatlas}. This process is time-consuming, expensive, and subject to inter-observer variability, resulting in limited and inconsistent labelled datasets \citep{wang2021annotation,veiga2022comparative}. In parallel, strict privacy regulations such as health insurance portability and accountability act and general data protection regulation restrict data sharing across institutions, leading to fragmented datasets and reduced diversity. Additionally, variations in scanner hardware, imaging protocols, acquisition settings, and patient demographics introduce significant domain shifts, often leading to poor generalisation of models beyond the training distribution \citep{guo2026impact}. Furthermore, the high cost and long acquisition times associated with imaging modalities such as MRI and PET further constrain the collection of large, representative datasets.

\vspace{0.2cm}
\noindent \textbf{Model interpretability}: Model interpretability remains a fundamental challenge for conventional data-driven deep learning in cancer research, as the learned decision-making process is largely opaque, limiting clinical trust, transparency, and adoption \citep{qamar2023understanding,wani2024deepxplainer}. To improve model explainability, numerous post hoc explainability techniques for artificial intelligence have been proposed. Among these, saliency-based methods improve explainability by translating a model’s prediction into a spatial map of feature importance. For example, gradient-weighted class activation mapping uses the gradients of a target class to weight intermediate feature maps, thereby highlighting image regions that most strongly influence the predicted class. Overlaying these maps on the original scan allows clinicians to assess whether the model relies on clinically plausible structures, such as a tumor or its surrounding tissue, or on irrelevant background features and imaging artifacts \citep{talaat2024grad,mothkur2025grad}. This provides an intuitive, case-specific explanation of the model’s behaviour, although the highlighted regions indicate model sensitivity rather than causal importance.

\begin{figure}[pos=htbp]
    \centering
    \includegraphics[width=1.0\textwidth]{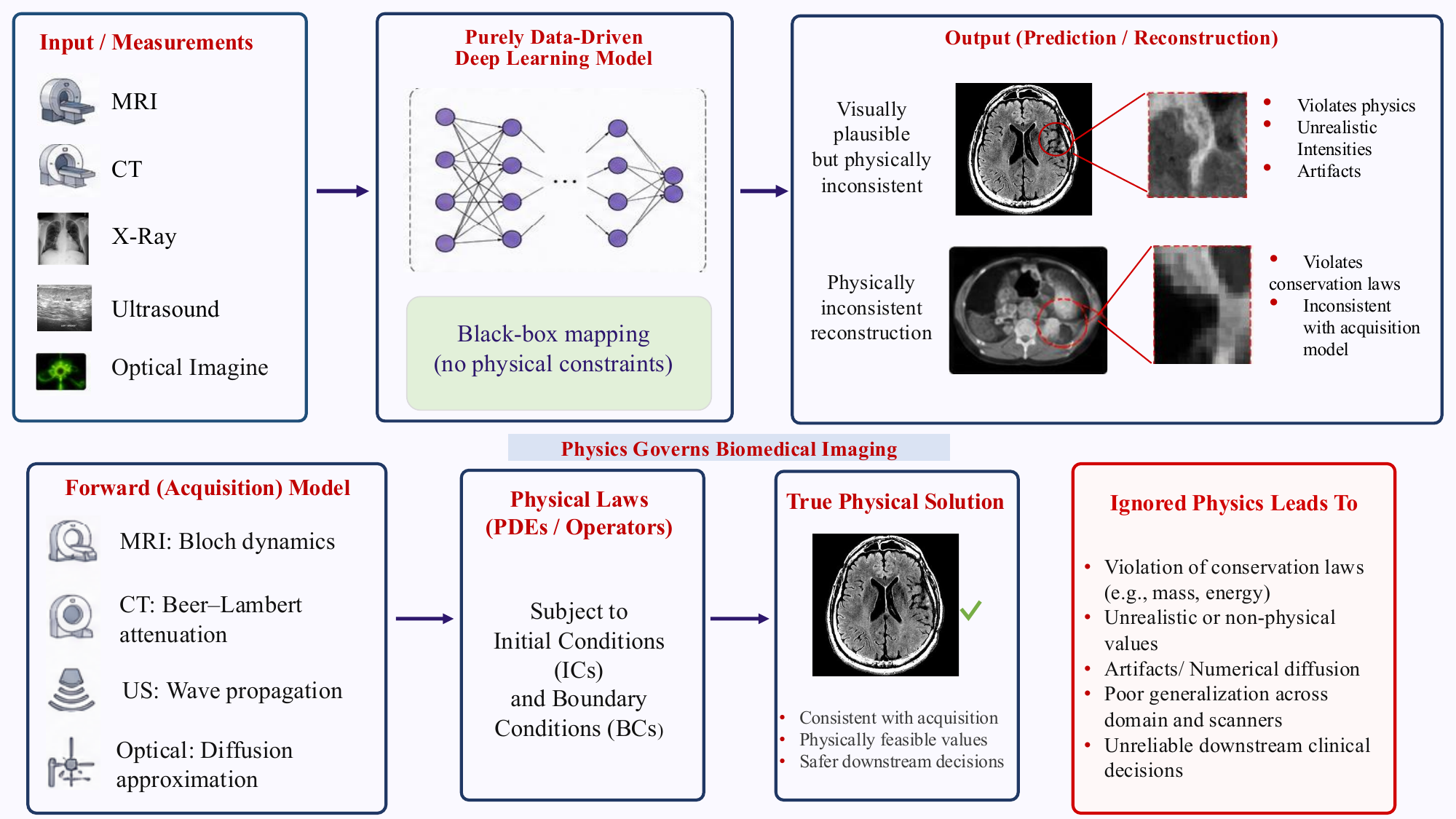}
    \caption{Illustration of the lack of physical consistency in purely data-driven deep learning for oncologic imaging. Although the predicted outputs may appear visually plausible, they can violate image-formation physics, conservation principles, and physiologically feasible ranges, potentially introducing artifacts and reducing clinical reliability. Representative input images are adapted from \citep{nickparvar2021braintumor}.}
    \label{fig:absence_physics}
\end{figure}

\vspace{0.2cm}
\noindent \textbf{Absence of physical consistency}: 
Biomedical imaging modalities, such as MRI, CT, X-ray, ultrasound, and optical imaging, are intrinsically governed by well-established physical principles, commonly expressed as forward models or PDEs. Pure data-driven deep learning approaches, despite their success in image reconstruction, segmentation, and classification, often fail to explicitly enforce governing laws, resulting in outputs that may be visually plausible yet physically inaccurate or inconsistent. Thus, while deep learning has greatly enhanced oncologic imaging pipelines (segmentation, detection, prognosis), its limitations motivate the incorporation of physics priors to improve generalisation and consistency. As illustrated in Fig.~\ref{fig:absence_physics}, purely data-driven learning produces predictions that lack explicit physical consistency in oncological processes. 

\

\section{Task-oriented review of PIDL for oncologic applications}
\label{sec:tasks}
This section reviews the development of PIDL across a broad range of oncologic tasks, spanning image reconstruction, disease characterization, tumor progression modeling, treatment planning, and therapy monitoring. We first outline the oncologic data analysis (ODA) pipeline, which provides the conceptual framework for organizing the task-oriented PIDL applications discussed throughout this section.

\begin{figure}[pos=htbp]
    \centering
    \includegraphics[width=1.0\linewidth]{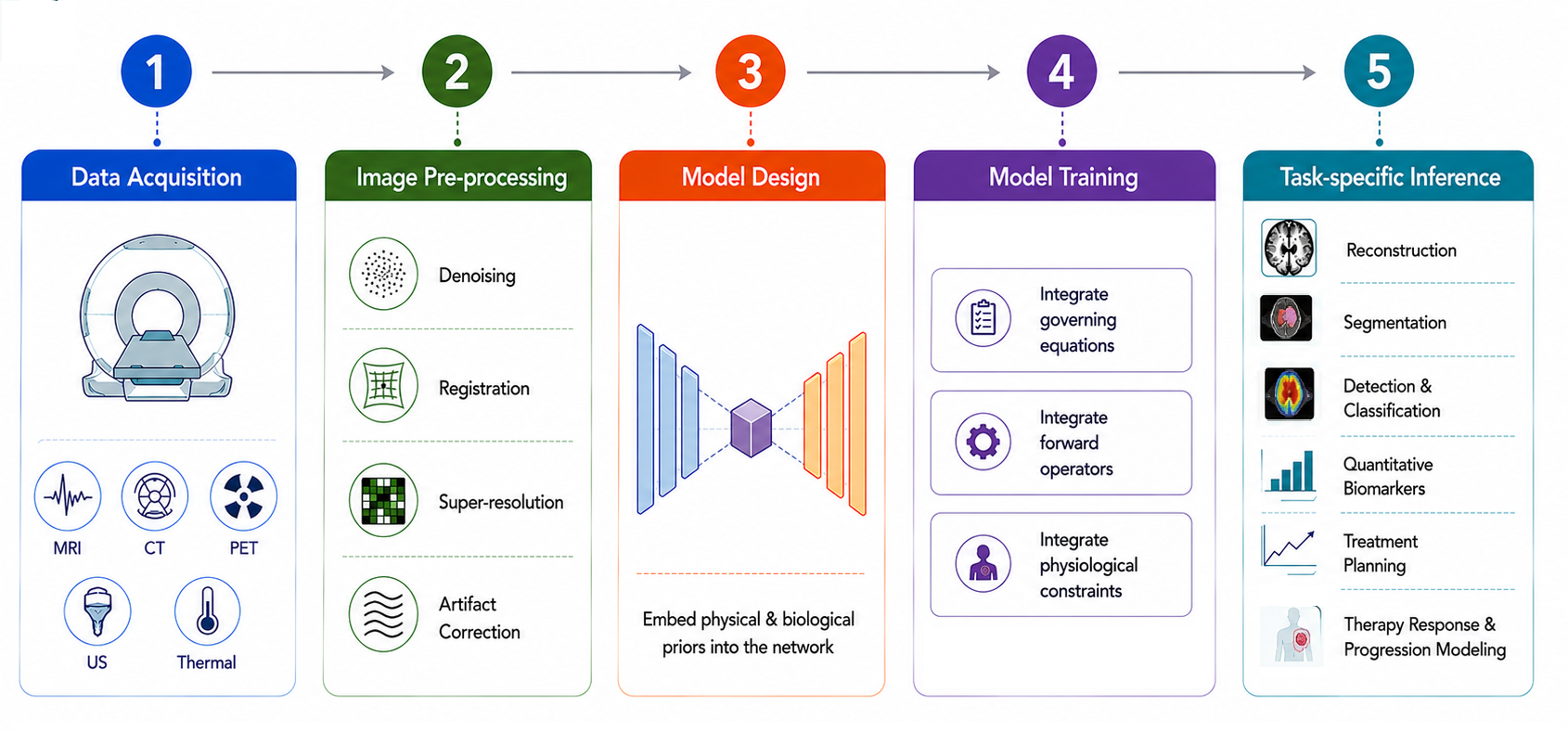}
    \caption{Overview of the physics-informed deep learning  workflow for oncologic imaging.}
    \label{fig:ODA_Pipeline}
\end{figure}

\vspace{0.2cm}
\noindent \textbf{ODA pipeline}: 
The end-to-end ODA workflow comprises five interconnected stages: data acquisition, data pre-processing, model design, model training, and task-specific inference. During data acquisition, physics-informed approaches incorporate imaging acquisition models and measurement physics, including MRI signal encoding, CT attenuation, PET emission, ultrasound propagation, and thermal diffusion, to improve data fidelity and mitigate acquisition-related artifacts. During pre-processing, physics-driven methods support image denoising, registration, super-resolution, and artifact correction while preserving anatomically and physiologically meaningful information. At the model-design stage, physical and biological priors are incorporated into network architectures to promote robust feature representation and capture tumor-specific characteristics. During model training, governing equations, imaging forward operators, conservation laws, and physiological constraints are integrated into the optimization objective through physics-informed loss functions, improving data efficiency and physical consistency. Finally, task-specific inference encompasses downstream oncologic tasks, including image reconstruction, segmentation, tumor detection and classification, quantitative biomarker estimation, treatment planning, therapy-response assessment, and disease-progression modeling. Fig.~\ref{fig:ODA_Pipeline} illustrates this five-stage ODA pipeline.

A solid tumor is not merely a biological mass of proliferating cells but a complex dynamical system governed by interacting physical and biological processes. Its microenvironment and physiological behavior can be described using principles from classical mechanics, fluid dynamics, and thermodynamics \cite{jain1987transport}. For example, the spatial distribution of vital nutrients, such as oxygen and glucose, is governed by reaction-diffusion processes, while limited diffusion in tissue can produce concentration gradients and hypoxic regions. At the macroscopic scale, tumor expansion is commonly described using phenomenological growth laws, such as logistic and Gompertzian kinetics, as well as spatial and continuum-mechanical models \cite{roose2007mathematical}. Tumor angiogenesis further alters vascular transport and tissue perfusion, while the tumor microenvironment behaves as a porous medium in which interstitial fluid flow can be described using Darcy's law. Such transport mechanisms contribute to elevated interstitial fluid pressure and influence the distribution of therapeutic agents within tumors \cite{baxter1989transport}. These coupled physical and biological processes provide the mechanistic basis for incorporating physics into oncologic computational modeling.\par
Tumor-growth modeling has consequently evolved from classical phenomenological descriptions toward spatially resolved, mechanistic, and patient-specific formulations. Classical exponential, logistic, and Gompertz models describe population-level growth, whereas subsequent models incorporate vascular regulation, spatial invasion, chemotaxis, interstitial transport, tumor-tissue interfaces, and vascular-network dynamics. More recent approaches further integrate longitudinal patient-specific observations and AI-based parameter estimation. Representative developments across these modeling levels are summarized in Table~\ref{tab:tumor_growth_models}.
\begin{table}[!t]
\centering
\caption{Evolution of mathematical models for tumor-growth modeling, from classical phenomenological formulations to recent patient-specific mechanistic-AI frameworks.}
\label{tab:tumor_growth_models}
\resizebox{\textwidth}{!}{%
\begin{tabular}{p{2.2cm} p{3.0cm} p{4.3cm} p{4.2cm} p{4.2cm} p{2.0cm}}
\hline
\textbf{Modeling level} &
\textbf{Tumor-growth model} &
\textbf{Mathematical formulation} &
\textbf{Coefficients and variables} &
\textbf{Biological mechanism} &
\textbf{Reference} \\
\hline

Population-level &
Exponential growth &
$\displaystyle \frac{dN}{dt}=rN$ &
$N(t)$: tumor population/volume; 
$r$: intrinsic proliferation rate &
Unconstrained proliferation under effectively unlimited resources &
\cite{collins1956hoeffler}
\\ \hline

Population-level &
Logistic growth &
$\displaystyle
\frac{dN}{dt}
=
rN\left(1-\frac{N}{K}\right)$ &
$N$: tumor population/volume;
$r$: proliferation rate;
$K$: carrying capacity &
Resource limitation and finite carrying capacity &
\cite{bacaer2011verhulst}
\\ \hline

Population-level &
Gompertz growth &
$\displaystyle
\frac{dN}{dt}
=
aN-bN\ln\left(\frac{N}{N_0}\right)$ &
$N$: tumor population;
$N_0$: reference population;
$a$: growth coefficient;
$b$: growth-deceleration coefficient &
Progressive growth-rate deceleration &
\cite{gompertz1815nature}
\\ \hline

Population-vascular &
Dynamic carrying-capacity model &
$\displaystyle
\frac{dV}{dt}
=
\lambda V\ln\left(\frac{K}{V}\right),
~
\frac{dK}{dt}
=
bV-dV^{2/3}K-\mu K-\eta g(t)K$ &
$V$: tumor volume;
$K(t)$: vascular carrying capacity;
$\lambda$: tumor growth parameter;
$b$: vascular stimulation rate;
$d$: endogenous vascular inhibition parameter;
$\mu$: spontaneous vascular loss rate;
$g(t)$: antiangiogenic drug concentration;
$\eta$: treatment coefficient &
Angiogenic regulation of tumor carrying capacity &
\cite{hahnfeldt1999tumor} 
\\ \hline

Spatial &
Fisher-Kolmogorov reaction-diffusion &
$\displaystyle
\frac{\partial u}{\partial t}
=
\nabla\!\cdot(D(\mathbf{x})\nabla u)
+
\rho u\left(1-\frac{u}{K}\right)$ &
$u$: tumor-cell density;
$D(\mathbf{x})$: diffusion tensor;
$\rho$: proliferation rate;
$K$: carrying capacity &
Tumor proliferation and spatial invasion &
\cite{fisher1937wave,kolmogorov2019studies,swanson2000quantitative}
\\ \hline

Spatial-invasive &
Keller-Segel chemotaxis/haptotaxis &
$\displaystyle
\frac{\partial u}{\partial t}
=
D_u\nabla^2u
-\nabla\!\cdot(\chi u\nabla c)
-\nabla\!\cdot(\gamma u\nabla v)
+f(u)$ &
$u$: tumor-cell density;
$D_u$: diffusivity;
$c$: chemical concentration;
$v$: ECM concentration;
$\chi$: chemotactic sensitivity;
$\gamma$: haptotactic coefficient &
Directed migration along chemical and ECM gradients &
\cite{anderson1998continuous}
\\ \hline

Biomechanical &
Interstitial fluid transport / Darcy model &
$\displaystyle
\mathbf{v}=-K_i\nabla p_i,
~
\nabla\!\cdot\mathbf{v}=\Gamma_V$ &
$\mathbf{v}$: interstitial fluid velocity;
$K_i$: hydraulic conductivity;
$p_i$: interstitial fluid pressure;
$\Gamma_V$: fluid exchange term &
Interstitial fluid pressure and fluid--solid transport&
\cite{jain1988mechanisms}
\\ \hline
Vascular-multiscale &
3D-1D vascular tumor-growth model &
$\displaystyle
\frac{\partial u}{\partial t}
=
\nabla\!\cdot(D\nabla u)
+
R(u,c,\ldots),
~
\mathcal{F}_{1D}(\mathbf{q},p)=0$ &
$u$: tissue/tumor state;
$D$: tissue diffusivity;
$c$: nutrient concentration;
$\mathbf q$: vascular flow;
$p$: vascular pressure &
Coupled tumor growth, vascular-network evolution, nutrient transport, and interstitial flow &
\cite{fritz2021modeling}
\\ \hline

Multiphase &
Phase-field tumor-growth model &
$\displaystyle
\frac{\partial\phi}{\partial t}
=\nabla\cdot(M\nabla\mu)
+
R(\phi,c,\ldots),
~
\mu=
\frac{\partial\Psi}{\partial\phi}
-\epsilon^2\nabla^2\phi$ &
$\phi$: tumor phase/order parameter;
$M$: mobility;
$\mu$: chemical potential;
$\Psi$: free-energy density;
$\epsilon$: interface-width parameter &
Tumor-tissue interface evolution, adhesion, and multispecies tumor dynamics &
\cite{wise2008three}
\\ \hline

Mechanistic-patient-specific &
Imaging-informed biomechanistic model &
$\displaystyle
\frac{\partial N}{\partial t}
=
\nabla\cdot(D\nabla N)
+
\rho N\left(1-\frac{N}{\theta}\right),
~
(D,\rho)
=
\operatorname{Calibrate}(y_{1:T})$ &
$N$: tumor-cell density;
$D$: tumor-cell diffusivity;
$\rho$: net proliferation rate;
$\theta$: tissue carrying capacity;
$y_{1:T}$: longitudinal imaging observations &
Patient-specific calibration and forecasting of tumor growth &
\cite{lorenzo2024pilot} 
\\ \hline

Mechanistic-AI &
Patient-specific mechanistic--AI framework &
$\displaystyle
\mathcal{F}(u,\boldsymbol{\mu};\mathbf{x},t)=0,
~
\boldsymbol{\mu}
=
\mathcal{N}_{\theta}(\mathbf{z}_{\mathrm{patient}})$ &
$u$: tumor state;
$\boldsymbol{\mu}$: patient-specific model parameters;
$\mathbf{z}_{\mathrm{patient}}$: patient-specific data;
$\mathcal{N}_{\theta}$: neural parameter estimator &
AI-assisted parameterization of mechanistic tumor models using patient-specific data &
\cite{lorenzo2024patient}
\\
\hline
\end{tabular}}
\end{table}

Traditional data-driven deep learning models learn these relationships primarily from observed data, often requiring substantial training datasets to capture complex biological and physical dependencies. PIDL instead incorporates known physical laws and mechanistic constraints directly into the learning process. Governing equations, imaging forward operators, conservation laws, and physiological relationships can be embedded into the training objective, penalizing predictions that violate established physical or biological constraints. By coupling data-driven representation learning with mechanistic knowledge, PIDL provides a framework for data-efficient and physically consistent modeling of oncologic processes.

For each task, we discuss the relevant physical or biological constraints, representative PIDL strategies, and their clinical relevance. Evaluation across tasks combines conventional quantitative metrics with clinically meaningful endpoints, including image fidelity, lesion detectability, biomarker preservation, diagnostic performance, and therapeutic accuracy. Task-specific datasets, evaluation criteria, and clinical considerations are discussed within the corresponding subsections.

\subsection{Image reconstruction}
\label{subsec:reconstruction}
Image reconstruction is a foundational computational step in the oncologic imaging workflow, providing the anatomical and functional images required for downstream tasks. Because clinical modalities measure indirect physical signals arising from interactions between imaging systems and tissue, recovering diagnostically meaningful images requires solving an inverse problem that is typically ill-posed owing to undersampling, measurement noise, detector imperfections, and dose or time-limited acquisition. Image reconstruction was among the earliest medical-imaging tasks to benefit from physics-informed deep learning, with imaging physics incorporated into network optimisation to improve reconstruction fidelity and quantitative accuracy. The acquisition is modeled through the imaging forward operator~\citep{arridge2019solving}:
\begin{equation}
y_1=\mathcal{A}_1(x_1)+\epsilon_1,
\label{eq:forward}
\end{equation}
where $x_1$ is the unknown image, $y_1$ the measurements, $\mathcal{A}_1$ the modality-specific forward operator, and $\epsilon_1$ the noise; the goal is to recover $x_1$ consistently with both the data and the imaging physics.\par
In MRI, the forward model comprises Fourier encoding, sampling trajectories, and coil sensitivity profiles, with measurements corresponding to undersampled $k$-space; physics-informed reconstruction imposes Fourier-domain data-consistency constraints to suppress aliasing while preserving fine structure~\citep{hammernik2023physics, aggarwal2019modl}. In CT, image formation follows the Beer-Lambert law and the Radon transform. Specifically, the measured X-ray intensity $I$ along a path $l$ is related to the incident intensity $I_0$ and the linear attenuation coefficient $\mu(\mathbf{r})$ through the line-integral model
\begin{equation}
-\log\left(\frac{I}{I_0}\right)
=
\int_l \mu(\mathbf{r})\,dl,
\label{eq:ct_beer_lambert}
\end{equation}
where the logarithmic projection represents the accumulated attenuation along the X-ray path~\citep{xia2023physics}. In discrete form, this forward model can be expressed as $\mathbf{y}_1=\mathbf{A}_1\mathbf{x}_1$, where $\mathbf{x}_1$ denotes the attenuation coefficient distribution, $\mathbf{y}_1$ represents the projection data, and $\mathbf{A}_1$ is the system matrix determined by the scanning geometry~\citep{xia2023physics}. PIDL therefore integrates projection operators and attenuation physics to recover high-quality images from sparse-view or low-dose data~\citep{xia2023physics, han2025physics}. PET reconstruction depends on radioactive decay, detector response, attenuation correction, and Poisson statistics, thereby requiring the preservation of quantitative biomarkers such as standardised uptake values, whereas ultrasound and photoacoustic imaging exploit acoustic wave propagation, beamforming, and optical point-spread functions as physical constraints.\\~\\

In MRI, Fourier encoding and data-consistency layers enable highly accelerated reconstruction while preserving anatomical detail~\citep{hammernik2023physics}, and self-supervised frameworks reduce dependence on fully sampled references, with clinical studies reporting shorter acquisition without loss of lesion detectability~\citep{xie2024deep,almansour2023deep}. In CT, embedding Radon-transform physics improves sparse-view and low-dose reconstruction and, in dual-energy CT, yields more accurate material decomposition~\citep{xia2023physics}. In PET, modeling photon emission, attenuation, and detector response improves low-count reconstruction while preserving metabolic quantification~\citep{rajagopal2022physics}; and in ultrasound tomography the acoustic wave equation has been embedded in deep-unfolded full-waveform inversion to recover quantitative sound-speed maps~\citep{jeong2024investigating,zhang2026physics}.\\~\\
\textbf{\textit{Datasets and evaluation}}: PIDL reconstruction is benchmarked on fastMRI~\citep{zbontar2018fast}, the American association of physicists in medicine  Low-Dose CT Challenge~\citep{mccollough2016tu}, and TCIA collections~\citep{clark2013cancer}, with numerical phantoms (BrainWeb, Shepp-Logan)~\citep{cocosco1997brainweb,shepp1974fourier} for controlled evaluation; image-quality metrics~\citep{wang2009mean} are increasingly complemented by lesion conspicuity, tumor-volume agreement, and biomarker preservation. Overall, PIDL improves fidelity under accelerated, sparse-view, or low-dose acquisitions, though performance depends on forward-model accuracy and adds computational cost.

\subsection{Image denoising}
\label{subsec:denoising}
Image denoising suppresses acquisition noise while preserving tumor morphology, tissue heterogeneity, and quantitative biomarkers needed for lesion detection, delineation, radiomics, and longitudinal assessment. It is especially important in oncology, where dose- and time-reduced protocols (low-dose CT, low-count PET, accelerated and diffusion-weighted MRI) elevate noise and degrade quantitative reliability. Given an observed image
\[
y_2=\mathcal{A}_2(x_2)+\epsilon_2,
\]
with $x_2$ the noise-free image, $\mathcal{A}_2$ the forward operator, and $\epsilon_2$ modality-dependent noise, PIDL addresses this by jointly suppressing noise and enforcing consistency with the acquisition physics.\par
The origin of noise is modality-specific. In low-dose CT, reduced exposure lowers detected photon counts: mean transmitted intensity follows the Beer-Lambert law~\citep{kak2001principles,beer1852bestimmung}, while photon-count fluctuations are Poisson-distributed~\citep{chesler1977noise}, impairing detection of small or low-contrast lesions. In PET, lower activity or shorter scans yield low-count, Poisson-dominated images that bias standardized uptake values; MRI is affected by Rician noise from complex Gaussian $k$-space noise, with further artifacts from parallel imaging, field inhomogeneity, and acceleration that influence quantitative MRI (DWI, DCE-MRI, DSC-MRI, relaxation mapping).\par
PICNNs~\citep{ren2022phycrnet} encode Poisson or Rician statistics through likelihood-based losses, and model-based unrolled frameworks~\citep{arridge2019solving} alternate physics-constrained data consistency with learned denoising. PINNs cast denoising as a PDE-constrained problem via nonlinear diffusion (heat, Perona-Malik)~\citep{quero_crespo_denoising}. A representative nonlinear diffusion model is given by
\begin{equation}
\frac{\partial u(\mathbf{r},t)}{\partial t}
=
\nabla\!\cdot
\left[
g\!\left(\left|\nabla u(\mathbf{r},t)\right|\right)
\nabla u(\mathbf{r},t)
\right],
\qquad \mathbf{r}\in\Omega,\; t>0,
\label{eq:nonlinear_diffusion_denoising}
\end{equation}
where $u(\mathbf{r},t)$ denotes the evolving image intensity, $\nabla u$ represents the image gradient, and $g(\cdot)$ is an edge-stopping diffusion coefficient that reduces diffusion across strong intensity gradients. In the physics-informed formulation, the neural network approximates $u(\mathbf{r},t)$ while automatic differentiation is used to enforce the PDE residual and the associated initial and boundary conditions. This formulation enables noise suppression while retaining important image edges and structures~\citep{quero_crespo_denoising}. Diffusion-probabilistic and transformer models embed acquisition physics into score estimation and attention for robustness at very low dose. Representative measurement-formation approaches include count-domain processing for low-dose CT~\citep{yuan2019countdomain}, Poisson-likelihood model unrolling for PET~\citep{mehranian2021modelbased}, and joint signal/acquisition constraints for quantitative MRI~\citep{zimmermann2024pinqi}.\\~\\
\textbf{\textit{Datasets and evaluation}}: Common resources include the American association of physicists in medicine-Mayo Low-Dose CT Challenge~\citep{mccollough2017lowdose}, TCIA PET/CT collections~\citep{clark2013cancer,clark2013tcia}, and BrainWeb phantoms~\citep{cocosco1997brainweb}. Beyond PSNR/SSIM/RMSE, evaluation increasingly emphasizes modality-specific endpoints (lesion detectability and contrast-to-noise in CT, SUV recovery in PET, ADC and parameter-map fidelity in MRI) and radiomic reproducibility~\citep{zwanenburg2020ibsi}, together with downstream effects on segmentation, planning, and response prediction. A recurring limitation is the use of simplified, stationary noise models that poorly represent scanner- and acquisition-dependent, spatially varying clinical noise.

\subsection{Image registration}
\label{subsec:registration}
Registration establishes spatial correspondence between images acquired at different times, modalities, or physiological states, supporting PET-CT and PET-MRI fusion, longitudinal monitoring, image-guided intervention, adaptive radiotherapy, and contour propagation~\citep{fu2020deep,zou2022review}. It is challenging because respiration, repositioning, growth, edema, surgery, and treatment remodeling produce complex non-rigid deformations that generic smoothness regularization may render mechanically implausible~\citep{sotiras2013deformable,fu2020deep}.

PIDL embeds biomechanical principles into deformation estimation. With fixed and moving images $I_f(\mathbf{x})$ and $I_m(\mathbf{x})$, respectively, and transformation $\boldsymbol{\phi}(\mathbf{x})=\mathbf{x}+\mathbf{u}(\mathbf{x})$, the objective function is
\[
\mathcal{L}
=
\mathcal{L}_{\mathrm{sim}}
+
\lambda_{\mathrm{phys}}\mathcal{L}_{\mathrm{physics}}
+
\lambda_{\mathrm{reg}}\mathcal{L}_{\mathrm{reg}},
\]
where $\mathcal{L}_{\mathrm{sim}}$ measures similarity between $I_f$ and $I_m\circ\boldsymbol{\phi}$, $\mathcal{L}_{\mathrm{physics}}$ penalises biomechanical violations such as lack of elastic equilibrium, inconsistent stress-strain relations, nonphysical volume changes, or excessive strain energy, and $\mathcal{L}_{\mathrm{reg}}$ controls deformation irregularity and topology violations~\citep{bharatha2001evaluation,min2023nonrigid,lopez2023warppinn,amirihezaveh2025physics}.
Several strategies exist. PINN-based methods represent $\mathbf{u}$ with a network and evaluate governing-equation residuals by automatic differentiation; Min et al.\ \cite{min2023nonrigid} used a three-dimensional linear-elastic model for prostate MR-to-ultrasound registration. The biomechanical constraint is governed by the linear-elasticity system comprising the strain-displacement relation, static equilibrium, and constitutive stress-strain relation:
\begin{equation}
\begin{aligned}
\boldsymbol{\varepsilon}
&=
\frac{1}{2}
\left(
\nabla\mathbf{u}
+
\nabla\mathbf{u}^{T}
\right),~~\nabla\!\cdot\boldsymbol{\sigma}
&=
\mathbf{0},~~
\boldsymbol{\sigma}
&=
\lambda\,\mathrm{tr}(\boldsymbol{\varepsilon})\mathbf{I}
+
2\mu\boldsymbol{\varepsilon},
\end{aligned}
\label{eq:linear_elasticity_registration}
\end{equation}
where $\mathbf{u}$ is the displacement field, $\boldsymbol{\varepsilon}$ is the infinitesimal strain tensor, $\boldsymbol{\sigma}$ is the Cauchy stress tensor, and $\lambda$ and $\mu$ are the Lamé parameters determined by the tissue Young's modulus and Poisson's ratio. In the PINN formulation, automatic differentiation evaluates the spatial derivatives in these equations, and the resulting equilibrium and constitutive residuals are incorporated into the training loss to constrain the predicted deformation to be biomechanically plausible~\citep{min2023nonrigid}. Physics  act as regularization: WarpPINN~\citep{lopez2023warppinn} enforced myocardial near-incompressibility in cine-MR registration, a principle transferable to respiratory and organ deformation in oncology. Hybrid approaches couple networks with finite-element or continuum-mechanics solvers that refine a learned deformation~\citep{bharatha2001evaluation}, while Neural-ODE formulations model registration as continuous coordinate evolution~\citep{wu2022nodeo}; Amiri-Hezaveh et al.~\citep{amirihezaveh2025physics} proposed a predictor-corrector neural-ODE enforcing large-deformation elasticity and growth mechanics. Such biomechanical constraints are most valuable when intensity correspondence is weak (PET-CT, PET-MRI, MR-ultrasound) and for modeling mass effect, edema, and postoperative change, although tissue appearing or disappearing through resection or necrosis violates the smooth one-to-one assumption.

Registration is assessed with target-registration error, Dice, Hausdorff distance, inverse-consistency error, and the Jacobian determinant $J_{\boldsymbol{\phi}}(\mathbf{x})=\det(\nabla\boldsymbol{\phi}(\mathbf{x}))$ (non-positive values indicate folding), using landmark datasets such as DIR-Lab 4D-CT~\citep{castillo2009framework,castillo2010fourdimensional}; physics-informed methods additionally report volume-change plausibility, strain consistency, and organ-at-risk propagation~\citep{fu2020deep,zou2022review}. 

\subsection{Biomechanical motion correction}
\label{subsec:motion}
Respiratory, cardiac, and involuntary motion introduces blurring, misregistration, and artifacts in oncologic CT, MRI, and PET, degrading tumor delineation, quantitative imaging, and radiotherapy planning. Unlike similarity-driven registration networks, PIDL regularizes deformation fields with tissue mechanics, mass preservation, and organ-motion dynamics. Typical physical constraints include respiratory-motion periodicity, tissue incompressibility, biomechanical equilibrium, temporal smoothness, and conservation of organ volume. Building on unsupervised dense-deformation estimation~\citep{de2019deep}, Jian et al. in \citep{jian2022weakly} introduced a weakly supervised, biomechanically constrained CT-MRI framework that preserves anatomical integrity, and Boysen et al. \citep{boysen2026biophysics} proposed a biophysics-enhanced neural representation for patient-specific respiratory motion that generates temporally coherent, realistic trajectories.\\
A suitable governing formulation~\citep{boysen2026biophysics} for respiratory motion represents the deformation as the flow generated by a time-dependent velocity field. Specifically, the particle trajectory $\boldsymbol{\phi}_{r,t}(\mathbf{x})$ satisfies the transport system
\begin{equation}
\begin{aligned}
\frac{\partial \boldsymbol{\phi}_{r,t}(\mathbf{x})}{\partial t}
&=
\mathbf{v}_{\theta}
\left(
\boldsymbol{\phi}_{r,t}(\mathbf{x}),s(t)
\right),
\\
\boldsymbol{\phi}_{r,r}(\mathbf{x})
&=
\mathbf{x},
\end{aligned}
\label{eq:respiratory_transport}
\end{equation}
where $\boldsymbol{\phi}_{r,t}(\mathbf{x})$ denotes the position at time $t$ of a tissue point initially located at $\mathbf{x}$ at time $r$, $\mathbf{v}_{\theta}$ is a neural-network parameterized time-dependent velocity field, and $s(t)$ denotes the respiratory surrogate signal. The deformation between two respiratory states can consequently be obtained by integrating the velocity field,
\begin{equation}
\boldsymbol{\phi}_{t_j,t_k}^{\theta}(\mathbf{x})
=
\mathbf{x}
+
\int_{t_j}^{t_k}
\mathbf{v}_{\theta}
\left(
\boldsymbol{\phi}_{t_j,\tau}(\mathbf{x}),s(\tau)
\right)
\,d\tau .
\label{eq:respiratory_deformation}
\end{equation}
In a physics-informed or physics-regularized neural framework, the network is trained to satisfy the transport equation while simultaneously matching the observed image pairs and physiological constraints. This formulation provides a continuous and diffeomorphic representation of respiratory motion and allows motion to be estimated between arbitrary respiratory phases~\citep{boysen2026biophysics}. Complementary physics-constrained registration and motion-aware radiotherapy studies further demonstrate the value of coupling biomechanics with deep learning for motion compensation and adaptive planning~\citep{zou2022review,salari2024artificial}.

\subsection{Super-resolution}
\label{subsec:sr}
Super-resolution (SR) reconstructs a high-resolution image from low-resolution acquisitions, recovering detail lost to hardware, protocol, or image-formation limits. In oncology, resolution affects small-lesion detection, boundary delineation, visualisation of infiltrative disease, characterisation of heterogeneity, and radiomic reproducibility, with the potential to improve micrometastasis detection and response assessment. Unlike interpolation, medical SR is an ill-posed inverse problem in which recovered high-frequency content must remain physically consistent with acquisition. The degradation is commonly written
\[
y=\mathcal{D}\bigl(\mathcal{H}(x)\bigr)+\epsilon,
\]
where $\mathcal{H}$ is the system operator (e.g., point-spread function), $\mathcal{D}$ downsampling, and $\epsilon$ noise; PIDL embeds these operators via data-consistency constraints, point-spread function (PSF) or sampling models, and governing equations, discouraging hallucinated structure.\par
\begin{figure}[pos=htbp]
    \centering
    \includegraphics[width=0.99\linewidth]{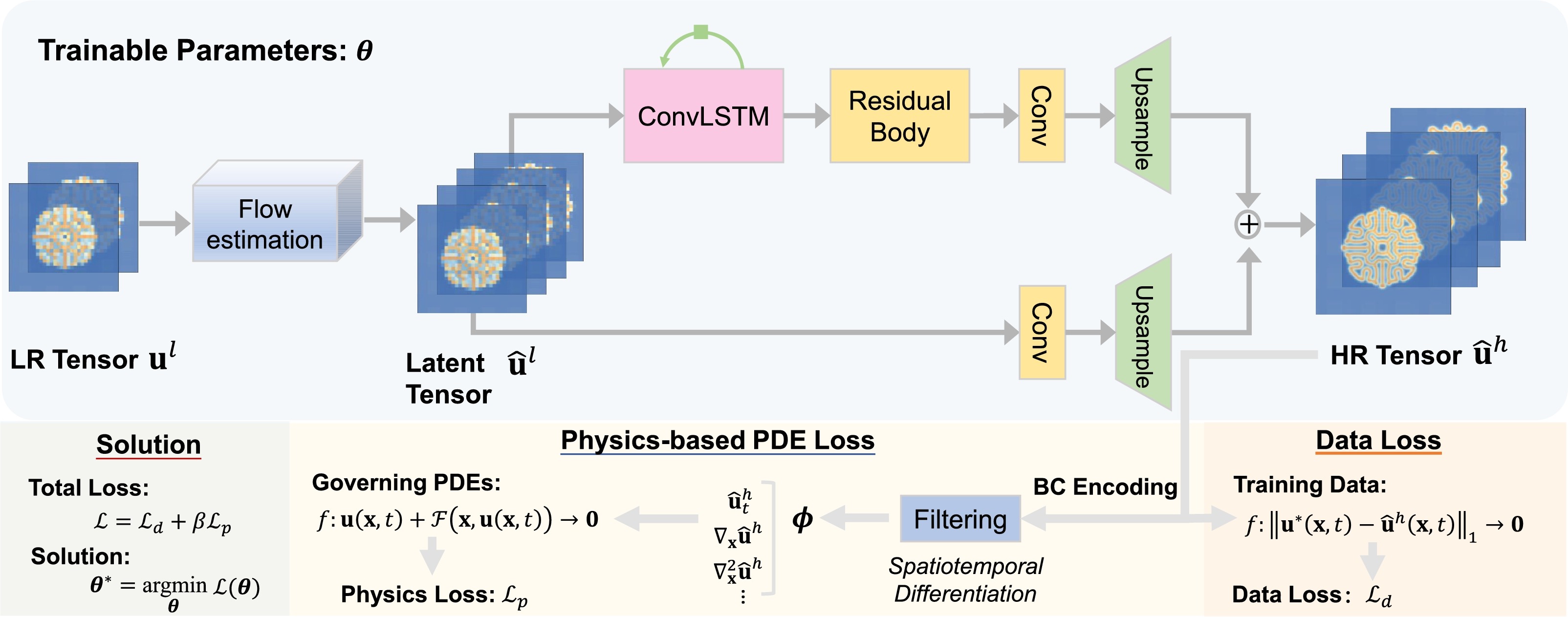}
    \caption{Physics-informed super-resolution architecture integrating ConvLSTM-based latent refinement with data and governing-PDE constraints for high-resolution tensor reconstruction \citep{ren2023physr}.}
    \label{fig:physr}
\end{figure}

Paradigms include model-based networks embedding forward and adjoint operators, physics-informed CNNs encoding blur or PSF models, PINNs posing SR as a PDE-constrained problem, and diffusion, GAN, and transformer models augmented with physics-guided losses; multi-input frameworks use CT or MRI priors to guide SR of functional PET or perfusion imaging. Physics-informed super-resolution can be used to recover high-resolution spatial and temporal information while constraining the reconstruction using modality- or tissue-specific physical models. The PhySR \citep{ren2023physr} architecture provides a representative framework by combining ConvLSTM-based temporal refinement, residual spatial reconstruction, and PDE-based physics constraints, as shown in Fig.~\ref{fig:physr}.
\\~\\
\textbf{\textit{Physics-informed modeling of blood flow}}: A representative hemodynamic  application of physics-informed super-resolution was presented by Fathi et al.~\cite{fathi2020super}  for four-dimensional flow MRI. In this setting, a neural network represents the velocity field $\mathbf{u}(\mathbf{x},t)$ and pressure $p(\mathbf{x},t)$ while enforcing the incompressible Navier-Stokes  equations,
\begin{equation}
\rho\left(
\frac{\partial \mathbf{u}}{\partial t}
+(\mathbf{u}\cdot\nabla)\mathbf{u}
\right)
=-\nabla p+\mu\nabla^{2}\mathbf{u},
\qquad
\nabla\cdot\mathbf{u}=0,
\end{equation}
\noindent where $\rho$ and $\mu$ denote blood density and dynamic viscosity, respectively. Data-fidelity  constraints couple the predicted fields
to the measured complex-valued MRI data, whereas the momentum and
continuity residuals regularize the reconstruction. The trained
network can therefore generate denoised, high-spatial- and
high-temporal-resolution velocity and pressure fields without
requiring precisely segmented vascular geometry or explicitly
prescribed boundary conditions. Although the method was developed
and validated for vascular 4D-flow MRI rather than oncologic imaging,
the same formulation could be adapted to tumor-associated vessels,
where sparse measurements and limited spatial resolution make the
recovery of local perfusion and hemodynamic quantities challenging.\par
A more directly oncologic example was reported by
Sainz-DeMena et al.~\cite{sainz2024exploring}, who used a PINN to
analyze DCE-MRI while incorporating contrast-agent diffusion and
mass conservation into a pharmacokinetic model. Their approach
improved the recovery of vascularization parameters from noisy or
incomplete data, which is particularly relevant to spatially
heterogeneous tumors. Together, these studies illustrate two
complementary roles of physics-informed learning, reconstructing
blood-velocity fields from flow MRI and estimating tumor
microvascular properties from contrast-agent transport.\par
SR remains comparatively underexplored in oncology: most methods originate in cardiovascular or general imaging and require adaptation to heterogeneous tumor morphology, multimodal protocols, and scarce high-resolution references, and balancing enhancement against physical consistency is difficult at high upsampling factors, where clinically misleading structures may appear. Beyond PSNR/SSIM/NRMSE, evaluation should weigh lesion detectability, boundary accuracy, radiomic preservation, and downstream diagnostic and therapeutic impact.

\subsection{Tumor segmentation and boundary delineation}
\label{subsec:segmentation}
Accurate segmentation underpins diagnosis, radiotherapy planning, surgical navigation, radiomics, and response assessment, delineating primary tumors, metastases, subregions, and organs-at-risk (OARs). It remains difficult because tumors show heterogeneous morphology, infiltrative growth, noise, weak boundaries, and inter-patient variability, especially in glioma, liver, pancreatic, and head-and-neck disease~\cite{savjani2022automated}.
Unlike generic semantic segmentation, tumor delineation is tied to biological growth and tissue mechanics-proliferation, migration, angiogenesis, edema, and displacement-commonly modeled by reaction-diffusion, biomechanical, conservation, and transport equations~\cite{hormuth2019mechanism,wong2015tumor,hormuth2017mechanically}. Embedding this knowledge encourages biologically plausible boundaries through reaction-diffusion regularization, biomechanical and mass-conservation constraints, physics-guided losses, hybrid CNN-variational or CNN-level-set formulations, PDE-based refinement, and mechanistic simulators for data augmentation~\cite{el2026tumor,guzel2026dynunet,zhang2025personalized}. A representative governing system  ~\cite{zhang2025personalized} is the Fisher-Kolmogorov reaction-diffusion model for tumor cell-density evolution:
\begin{equation}
\begin{cases}
\displaystyle
\frac{\partial C(\mathbf{x},t)}{\partial t}
=
\nabla\!\cdot\!\left(D(\mathbf{x})\nabla C(\mathbf{x},t)\right)
+
\rho C(\mathbf{x},t)(1-C(\mathbf{x},t)),
& \mathbf{x}\in\Omega,\quad t>0,\\[8pt]
\displaystyle
\nabla C(\mathbf{x},t)\cdot\mathbf{n}=0,
& \mathbf{x}\in\partial\Omega,\quad t>0.
\end{cases}
\label{eq:fisher_kpp_tumor}
\end{equation}
where $C(\mathbf{x},t)$ denotes the normalized tumor cell density, $D(\mathbf{x})$ is the spatially varying diffusion coefficient describing tumor-cell invasion, $\rho$ is the proliferation rate, $\Omega$ is the tissue domain, and $\mathbf{n}$ is the outward normal to the tissue boundary. The diffusion term models spatial invasion, whereas the logistic reaction term $\rho C(1-C)$ represents tumor proliferation subject to a carrying-capacity constraint. In a physics-informed formulation, the neural network approximates $C(\mathbf{x},t)$ and automatic differentiation evaluates the PDE residual, allowing the predicted tumor distribution and its boundary to remain consistent with the assumed growth dynamics~\cite{zhang2025personalized}. Fisher-Kolmogorov reaction-diffusion terms let inferred boundaries reflect disease dynamics rather than intensity alone~\cite{zhang2025personalized,guzel2026dynunet}, which is especially useful for infiltrative tumors.\par
\begin{figure}[pos=htbp]
    \centering
    \includegraphics[width=1.0\linewidth]{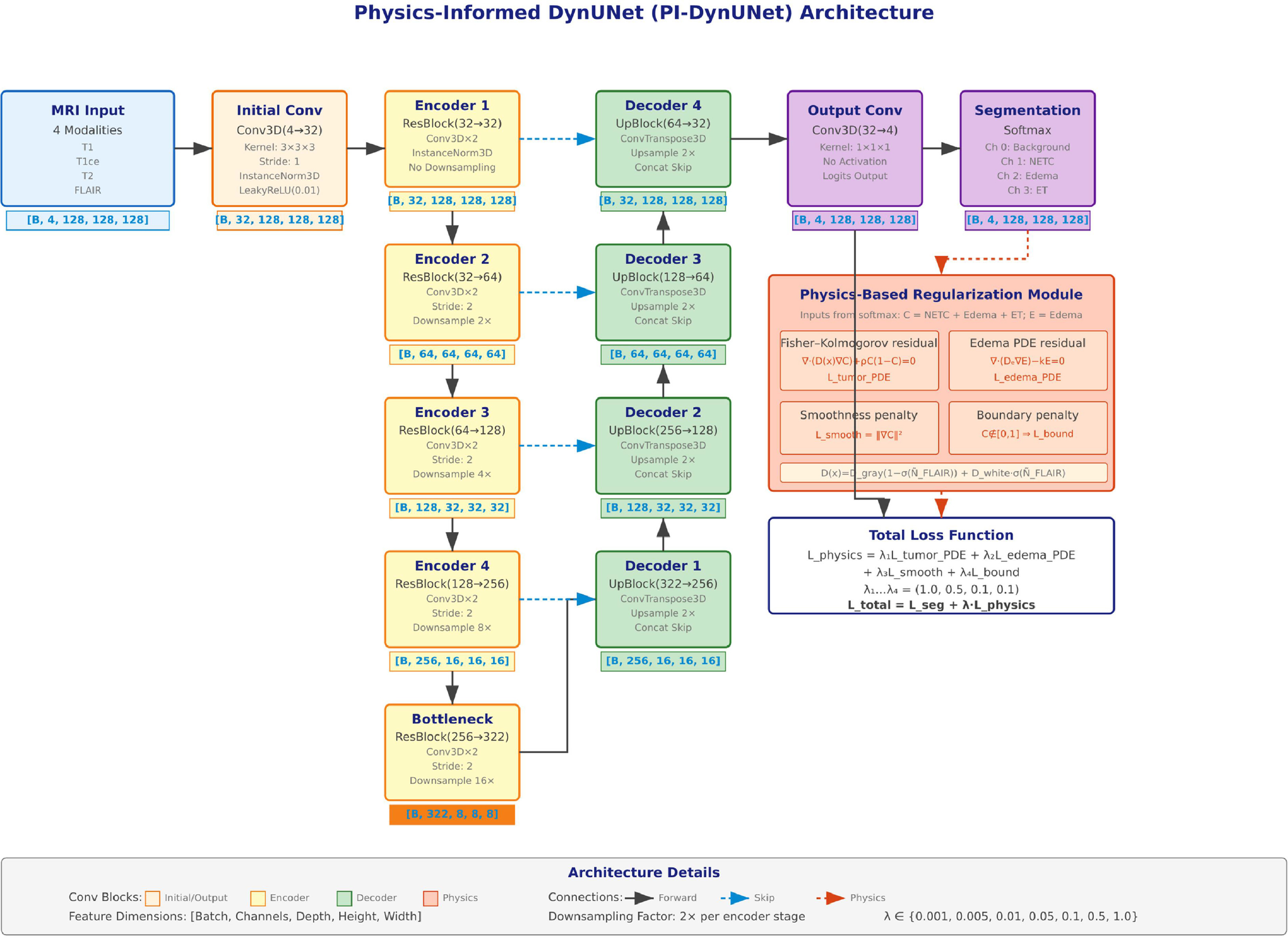}
    \caption{Overview of the PI-DynUNet architecture  for brain metastasis segmentation \cite{guzel2026dynunet}.}
    \label{fig:pi_dynunet}
\end{figure}

El Afari et al.~\cite{el2026tumor} combined U-Net initialization, variational segmentation, and reaction-diffusion modeling in a physics-informed, digital-twin framework that jointly delineates and predicts tumor evolution. Recent work has extended DynUNet with physics-based regularization for brain-metastasis segmentation by incorporating Fisher-Kolmogorov tumor-growth dynamics and edema-transport constraints into the training objective, as illustrated in Fig.~\ref{fig:pi_dynunet} \citep{guzel2026dynunet}.
Beyond overlap metrics (Dice, Jaccard, Hausdorff and surface distance), physics-informed models may report governing-equation residuals, estimated growth or transport parameters, and prediction of future evolution~\cite{el2026tumor,zhang2025personalized}, evaluated on BraTS, LIDC-IDRI, LiTS, KiTS, and TCIA collections. Open challenges include oversimplified reaction-diffusion models that omit necrosis, angiogenesis, hypoxia, immune interaction, mass effect, and treatment change~\cite{hormuth2019mechanism}; computationally demanding and potentially non-identifiable patient-specific parameters~\cite{zhang2025personalized,wong2015tumor}; rarely quantified uncertainty; and a focus on binary or single-lesion segmentation rather than multiclass, multifocal delineation~\cite{savjani2022automated,zhang2025personalized}.

\subsection{Tumor detection and classification}
\label{subsec:detection}
Detection localizes suspicious lesions and classification assigns clinically relevant labels (benign versus malignant, grade, histology, molecular subtype), supporting diagnosis, staging, treatment selection, and prognosis. Despite strong data-driven computer-aided detection (CADe) and diagnosis (CADx) systems, physics-informed detection and classification are the least developed areas in this review, because these are high-level semantic tasks rather than inverse problems, and governing equations cannot be embedded in the prediction objective as naturally as in reconstruction, quantification, or biomechanical modeling~\cite{karniadakis2021physics}.

Physics can nonetheless inform these tasks by supplying quantitative parameter maps as input channels, deriving model-based features, adding auxiliary physiological-estimation objectives, or penalizing inconsistency between predictions and estimated biology, so that decisions depend on proliferation, invasion, transport, vascularity, or deformation rather than intensity alone. Mechanistic models are a promising route: reaction-diffusion and biomechanical models estimate proliferating and infiltrating cell distributions and disease extension beyond visible abnormalities~\cite{wong2015tumor,hormuth2017mechanically}, and patient-specific inverse modeling recovers growth, transport, mechanical, or vascular parameters from longitudinal data~\cite{hormuth2019mechanism,zhang2025personalized}. Personalized reaction-diffusion models coupled with networks and multimodal imaging have estimated infiltration and growth~\cite{zhang2025personalized}, and physics-informed segmentation and evolution models yield mechanistic information potentially useful for detection, grading, or distinguishing recurrence from treatment change~\cite{el2026tumor,guzel2026dynunet}; however, these studies target segmentation, growth prediction, or parameter estimation rather than lesion-level detection and classification directly.

Evaluation uses sensitivity, specificity, accuracy, $F1$ score, AUC, and free-response ROC analysis, and should further test consistency with estimated physiological variables, biomarker preservation, and cross-cohort generalization~\cite{hormuth2019mechanism,zhang2025personalized}. Because most systems remain data-driven and adding physical variables does not automatically improve accuracy or generalization, rigorous comparison with data-driven baselines, external validation, and uncertainty quantification are needed, along with task-specific physics-informed objectives that directly target lesion detection and diagnostic discrimination.

\subsection{Physiological and metabolic quantification}
\label{subsec:quantification}
PIDL has become a powerful framework for quantitatively estimating physiological and metabolic biomarkers from oncologic imaging. Early PINNs estimated hemodynamic and perfusion parameters by embedding tracer kinetics and conservation laws, enabling robust recovery of physiologically meaningful biomarkers from dynamic data~\citep{sarabian2022physics,van2022physics}. Building on this, Sainz et al.\ \cite{sainz2024exploring} incorporated pharmacokinetic transport equations into PINNs for DCE-MRI, improving estimation of vascular permeability, extracellular volume fraction, and perfusion associated with angiogenesis. \cite{rotkopf2024dsc} developed a physics-informed framework for DSC-MRI recovering vascular response functions and perfusion biomarkers, and Fayolle et al.\ \cite{fayolle2025physics} introduced a physics-informed autoencoder for DSC-MRI that leverages analytical perfusion models to derive biomarkers for glioma grading.\par
Beyond physiological quantification, PIDL supports metabolic characterization through thermal imaging: tumor metabolism and angiogenesis increase local heat generation and alter perfusion, producing thermal signatures captured by infrared thermography \cite{perez2023thermal}. Physics-informed frameworks leverage the Pennes' bioheat equation \cite{pennes1948analysis} to relate surface temperature to underlying metabolic and vascular activity. A representative governing equation for tissue temperature is
\begin{equation}
\rho c
\frac{\partial T(\mathbf{x},t)}{\partial t}
=
\nabla\!\cdot
\left(
k\nabla T(\mathbf{x},t)
\right)
+
\rho_b c_b \omega_b
\left(
T_a-T(\mathbf{x},t)
\right)
+
Q_m(\mathbf{x})
+
Q_t(\mathbf{x}),
\label{eq:pennes_bioheat}
\end{equation}
where $T(\mathbf{x},t)$ is the tissue temperature, $\rho$ and $c$ are the tissue density and specific heat capacity, respectively, $k$ is the thermal conductivity, $\rho_b$, $c_b$, and $\omega_b$ denote blood density, blood specific heat, and blood perfusion rate, respectively, $T_a$ is the arterial blood temperature, and $Q_m$ and $Q_t$ represent metabolic and tumor-associated heat-generation terms, respectively. The first term on the right-hand side describes thermal diffusion, the second represents heat exchange between blood and tissue through perfusion, and the final terms account for local metabolic and tumor heat sources. In a physics-informed formulation, the neural network approximates $T(\mathbf{x},t)$ and automatic differentiation evaluates the bioheat-equation residual, while measured surface temperatures can be incorporated to infer unknown thermal or physiological parameters~\citep{perez2023thermal,perez2024transformative}.
Early studies modeled heat transport and temperature distributions in breast tissue~\citep{mukhmetov2023physics,zhao2024application}; Perez-Raya et al. \cite{perez2023thermal} demonstrated PINNs for breast-cancer thermal modeling, and an inverse PINN framework incorporating the Pennes' bioheat equation \cite{pennes1948analysis} and infrared thermography inferred tumor-associated heat sources, enabling tumor-size and localization estimation validated against biopsy-confirmed cases~\cite{perez2024transformative}. Recently, Das and Scindhiya \cite{das2026adaptive} introduced an adaptive residual-guided domain refinement in PINN approach for precise breast-tumor thermal profiling through bio-thermal modeling. Collectively, these quantify thermophysiological biomarkers of metabolism, vascularity, and perfusion, providing a non-invasive avenue for metabolic assessment and early detection.
\subsection{Treatment planning, radiotherapy, and therapy monitoring}
\label{subsec:planning}
Treatment planning and therapy monitoring translate quantitative imaging into personalized therapeutic decisions, including three-dimensional dose estimation, beam optimization, adaptive planning, and longitudinal response assessment. Radiotherapy dose deposition depends on beam energy, attenuation, scattering, tissue composition, patient anatomy, and machine geometry. Conventional planning relies on Monte Carlo or deterministic transport methods, which provide high accuracy but are computationally expensive. In contrast, purely data-driven dose prediction is computationally efficient but may violate dose-deposition physics and generalize poorly across beam configurations, treatment protocols, and patient anatomies. Physics-informed deep learning (PIDL) addresses this gap by incorporating beam geometry, attenuation, transport approximations, dose-deposition characteristics, and dosimetric constraints into the learning process.\par
The underlying radiation-transport physics can be described by the steady-state linear Boltzmann transport equation (LBTE) \citep{zhang2026pinnradiotherapy}, which governs the spatial and angular distribution of photon fluence:
\begin{equation}
\hat{\boldsymbol{\Omega}}\cdot\nabla
\psi(\mathbf{r},E,\hat{\boldsymbol{\Omega}})
+
\mu_t(\mathbf{r},E)
\psi(\mathbf{r},E,\hat{\boldsymbol{\Omega}})
=
\int_{4\pi}\int_{0}^{\infty}
\mu_s
\left(
\mathbf{r},E'\!\rightarrow E,
\hat{\boldsymbol{\Omega}}'\!\rightarrow\hat{\boldsymbol{\Omega}}
\right)
\psi(\mathbf{r},E',\hat{\boldsymbol{\Omega}}')
\,dE'\,d\hat{\boldsymbol{\Omega}}'
+
S(\mathbf{r},E,\hat{\boldsymbol{\Omega}}),
\label{eq:lbte_radiotherapy}
\end{equation}
where $\psi$ denotes the angular photon fluence, $\mu_t$ is the total interaction coefficient, $\mu_s$ is the differential scattering coefficient, $\mathbf{r}$ is the spatial position, $E$ is photon energy, $\hat{\boldsymbol{\Omega}}$ is the propagation direction, and $S$ represents the radiation source. The resulting photon fluence determines energy deposition and absorbed dose, which can be expressed as
\begin{equation}
D(\mathbf{r})
=
\int_{0}^{\infty}
\frac{\mu_{\mathrm{en}}(\mathbf{r},E)}
{\rho(\mathbf{r})}
\Phi(\mathbf{r},E)\,dE,
\label{eq:dose_energy_deposition}
\end{equation}
where $\Phi(\mathbf{r},E)$ is the scalar photon fluence, $\mu_{\mathrm{en}}/\rho$ is the mass energy-absorption coefficient, and $\rho$ is the local tissue density. In PIDL, these radiation-transport and energy-deposition relationships can be incorporated as physics residuals or soft constraints, thereby encouraging predicted dose distributions to remain consistent with the underlying radiation physics~\citep{zhang2026pinnradiotherapy}.\par
A representative application is the physics-informed framework of Sun et al.~\cite{sun2025physics}, which predicts individual beam-dose distributions for intensity-modulated radiotherapy by incorporating the attenuation characteristics of 6-MV photon beams together with beam-dose, dose-gradient, and anatomical-masking losses. Beam-by-beam prediction and crosshair-sampling analysis further improve the accuracy and interpretability of the predicted dose distributions. Such approaches demonstrate how radiation-transport knowledge can be incorporated into deep neural networks without requiring the network to explicitly replace a full Monte Carlo or deterministic dose calculation.\par
Beyond dose prediction, PIDL can support treatment-response modeling through physics-informed digital twins that combine imaging-derived biomarkers with mechanistic descriptions of tumor evolution and therapy response. Mechanistic finite-element models of tumor growth and drug interaction provide physically interpretable predictions of disease progression and treatment response, providing a foundation for patient-specific treatment planning and digital-twin development~\citep{yadav2025avascular,yadav2026adaptive,chhetri2026cbinn}. Camacho-Gomez et al.~\cite{camacho2025physics} developed a patient-specific prostate-cancer digital twin using multiparametric MRI and longitudinal prostate-specific antigen (PSA) measurements. T2-weighted MRI defines anatomy and tumor location, diffusion-weighted imaging provides information related to cellularity, and DCE-MRI-derived $K^{\mathrm{trans}}$ characterizes vascularization. The mechanistic component models tumor-cell concentration together with PSA production, transport, and decay, while a neural network estimates the spatiotemporally varying proliferating-cell fraction, thereby complementing rather than replacing the mechanistic model.\par
PIDL can further support adaptive radiotherapy by integrating repeated imaging, deformable registration, biomechanical models, accumulated-dose mapping, and radiation-transport calculations as anatomy and tumor biology evolve during treatment. Such integration enables the estimated dose distribution and treatment plan to be updated according to changes in tumor geometry, organ position, tissue properties, and treatment response. Consequently, physics-informed digital twins provide a potential framework for linking longitudinal imaging, biological evolution, dose accumulation, and personalized treatment adaptation.\par
Evaluation should therefore extend beyond image-level prediction accuracy to clinically meaningful dosimetric and biological endpoints. For dose prediction, relevant measures include dose MAE, dose-volume-histogram agreement, gamma passing rate, target coverage, conformity and homogeneity indices, and organ-at-risk doses. For digital twins and therapy monitoring, tumor-volume prediction error, longitudinal spatial agreement, biomarker trajectories such as PSA kinetics, and treatment-response accuracy are also important. Key challenges include limited generalization beyond photon therapy, reduced transport accuracy near tissue interfaces, biological and parameter uncertainty, and the scarcity of longitudinal patient data. These limitations motivate uncertainty-aware dose calculation, deliverable treatment optimization, accumulated-dose estimation, and external multicenter validation.
\\~\\
Table~\ref{tab:PIDL_tasks} summarizes the role of physics-informed learning across major oncologic imaging tasks, highlighting the incorporated physiological information, principal anatomical regions, computational architectures, and representative clinical applications. Further, Table~\ref{tab:pidl_effective} summarizes the major oncologic imaging tasks and corresponding advantages of PIDL, highlighting how physics-based constraints address task-specific challenges and improve the physical consistency and reliability of model predictions.

\begin{table*}[!t]
\centering
\caption{Representative PIDL Approaches for Major Oncologic Tasks.}
\label{tab:PIDL_tasks}

\renewcommand{\arraystretch}{1.25}
\setlength{\tabcolsep}{4pt}
\scriptsize

\begin{tabularx}{\textwidth}{
>{\raggedright\arraybackslash}p{2.1cm}
>{\raggedright\arraybackslash}p{3.2cm}
>{\raggedright\arraybackslash}p{2.5cm}
>{\raggedright\arraybackslash}p{3.0cm}
>{\raggedright\arraybackslash}X
}
\toprule

\textbf{Task}
&
\textbf{Physics incorporated}
&
\textbf{ROI / cancer context}
&
\textbf{Architecture / model}
&
\textbf{Representative application}
\\
\midrule

\textbf{Detection}
&
Attenuation, diffusion, perfusion, tracer kinetics, tissue acoustics,
bioheat transfer, and mechanistic tumor-growth information
&
Breast, brain, prostate, lung, and other lesion-bearing organs
&
Physics-informed CNN, hybrid PINN-CNN, inverse PINN, and
physics-guided feature-learning models
&
Lesion localization using physiologically meaningful parameter maps,
including thermal, perfusion, diffusion, metabolic, and acoustic
biomarkers. In thermal imaging, inverse bioheat models have been used
to estimate tumor-associated heat sources and locations
\cite{perez2023thermal,perez2024transformative,camacho2025physics}.
\\
\midrule

\textbf{Segmentation}
&
Reaction-diffusion equations, tumor-growth dynamics, biomechanical
constraints, conservation laws, anatomical priors, and geometric
regularization
&
Glioma, brain metastases, edema, infiltrative tumor margins, and
organs-at-risk
&
U-Net combined with PINNs, DynUNet, variational networks,
level-set models, and hybrid mechanistic-learning frameworks
&
Biologically informed delineation of tumor core, edema, and infiltrative
regions. Physics-informed segmentation frameworks combine image-derived
boundaries with mechanistic tumor-growth models to improve biological
plausibility and support longitudinal prediction \citep{guzel2026dynunet}.
\\
\midrule

\textbf{Classification}
&
Diffusion, perfusion, relaxation, tracer kinetics, vascular response,
metabolic activity, and other quantitative imaging biomarkers
&
Glioma grading, benign-malignant discrimination, and cancer subtype
classification
&
Physics-informed autoencoder, multimodal CNN, hybrid classifier, and
physiology-guided feature-learning network
&
Tumor grading and diagnostic classification using quantitative
physiological biomarkers rather than image intensity alone, including
DSC-MRI-derived vascular and perfusion characteristics
\cite{rotkopf2024dsc,fayolle2025physics}.
\\
\midrule

\textbf{Quantification}
&
Signal-formation equations, tracer-kinetic and pharmacokinetic models,
perfusion and diffusion equations, conservation laws, bioheat transfer,
and transport equations
&
Brain tumors, prostate cancer, breast cancer, and tumor vasculature
&
PINN, inverse PINN, physics-informed autoencoder, neural operator, and
hybrid mechanistic model
&
Estimation of qMRI, DCE-MRI, DSC-MRI, PET-kinetic, hemodynamic, thermal,
and metabolic biomarkers. Governing equations constrain the recovered
parameters to remain physiologically meaningful \citep{sainz2024exploring, fayolle2025physics}.
\\
\midrule

\textbf{Reconstruction}
&
MRI forward operators, Fourier and \(k\)-space consistency, CT projection
and attenuation models, PET system response, point-spread functions,
noise statistics, and acoustic wave propagation
&
Brain MRI, oncologic CT, PET, ultrasound, and multimodal imaging
&
Unrolled reconstruction network, physics-informed CNN, PINN,
PhysicsGAN, transformer, and model-based deep network
&
Accelerated, sparse-view, low-dose, or undersampled image reconstruction
in which predicted images are constrained to remain consistent with
the acquired measurements and scanner physics \citep{yaman2020self}.
\\
\midrule

\textbf{Motion correction and registration}
&
Elasticity, tissue biomechanics, incompressibility, mass and volume
preservation, respiratory dynamics, temporal smoothness, and
deformation regularity
&
Thoracic, abdominal, brain, prostate, and radiotherapy imaging
&
Registration PINN, biomechanically constrained CNN, implicit neural
representation, transformer, and hybrid deformation model
&
Respiratory, cardiac, and treatment-related motion correction using
anatomically plausible deformation fields. Physics-informed constraints
reduce folding, unrealistic tissue compression, and non-physical image
warping \citep{chan2026uncertainty, min2023nonrigid}.
\\
\midrule

\textbf{Treatment planning and dosimetry}
&
Beam geometry, photon attenuation, radiation transport, scattering,
dose deposition, bioheat transfer, tumor-growth dynamics, drug
interaction, and dosimetric constraints
&
Prostate, brain, head-and-neck, breast, and other radiotherapy targets
&
Physics-informed dose-prediction network, differentiable dose model,
PINN, finite-element digital twin, and hybrid mechanistic-neural model
&
Beam-dose prediction, inverse planning, adaptive radiotherapy,
hyperthermia optimization, and patient-specific treatment-response
modeling. Mechanistic digital twins combine imaging biomarkers with
tumor-growth and treatment-response models
\cite{yadav2025avascular,yadav2026adaptive,camacho2025physics}.
\\
\bottomrule
\end{tabularx}
\end{table*}

\begin{table*}[!t]
\centering
\caption{Task-oriented summary of the principal benefits of PIDL in oncological task.}
\label{tab:pidl_effective}
\renewcommand{\arraystretch}{1.35}
\small
\begin{tabularx}{\textwidth}{p{4.0cm}X}
\toprule
\textbf{Oncological application}
&
\textbf{Why PIDL is effective?}
\\
\midrule

Image reconstruction
&
PIDL integrates image-acquisition and signal-formation physics into the
learning process, thereby reducing reconstruction artifacts, improving
measurement consistency, and enabling reliable reconstruction from sparse,
noisy, or undersampled measurements
\cite{peng2022learning,zimmermann2024pinqi,schote2024joint,fujita2025physics}.
\\
\midrule

Image enhancement / denoising
&
Physics-informed methods incorporate modality-specific noise statistics,
image-formation operators, and degradation models, enabling effective noise
suppression while preserving anatomical structures, tissue boundaries, and
clinically relevant image details
\cite{kazemi2024pi, fathi2020super}.
\\
\midrule

Tumor segmentation
&
PIDL embeds anatomical, physiological, and tumor-growth constraints into the
segmentation process, producing more coherent and physically plausible tumor
boundaries, particularly in low-contrast images and limited-annotation
settings
\cite{borges2019physics,guzel2026dynunet,kang2026pmrnet}.
\\
\midrule

Image registration
&
Physics-informed registration incorporates biomechanical tissue models,
elasticity constraints, and deformation regularity, thereby producing
anatomically plausible displacement fields while reducing folding and
unrealistic image warping
\cite{min2023nonrigid,lopez2023warppinn,jian2022weakly,
amirihezaveh2025physics}.
\\
\midrule

Tumor detection / diagnosis
&
PIDL combines imaging features with disease-specific physiological and
biophysical knowledge, which can reduce false-positive detections, improve
robustness to imaging variability, and enhance the interpretability of
diagnostic predictions
\cite{mukhmetov2023physics,zhao2024application}.
\\
\midrule

Disease classification
&
Physics-guided feature learning incorporates quantitative imaging biomarkers,
perfusion dynamics, diffusion behaviour, or other physiologically meaningful
information, thereby improving feature robustness and potentially enhancing
generalization across institutions, scanners, and patient populations
\cite{rotkopf2024dsc,fayolle2025physics}.
\\
\midrule

Quantitative parameter estimation
&
PIDL formulates quantitative imaging as an inverse problem constrained by
governing equations, enabling physically consistent estimation of perfusion,
diffusion, thermal properties, electrical conductivity, elasticity, and
metabolic or hemodynamic parameters
\cite{sarabian2022physics,inda2022physics}.
\\
\midrule

Tumor localization
&
Physics-informed localization incorporates reaction-diffusion tumor-growth
models, multimodal image assimilation, and tissue-biomechanical constraints
to infer physiologically meaningful tumor locations, including infiltrative
regions that may not be clearly visible in conventional imaging
\cite{balcerak2024physics,zhang2025personalized,wong2015tumor}.
\\
\midrule

Treatment planning
(hyperthermia/radiotherapy)
&
PIDL integrates bioheat transfer, radiation-dose deposition, tumor-growth
dynamics, and patient-specific tissue properties to improve thermal-field
prediction, dose estimation, treatment optimization, and individualized
treatment planning
\cite{sun2025physics,balcerak2025individualizing}.
\\
\midrule

Treatment response prediction
&
Physics-informed models represent disease progression and
treatment-induced physiological changes through mechanistic tumor-growth,
drug-transport, and treatment-response equations, leading to more
interpretable longitudinal predictions and supporting patient-specific
treatment assessment
\cite{sun2025physics, achlatis2025physics, tran2021deep}.
\\
\bottomrule
\end{tabularx}
\end{table*}

\section{Modality-oriented review of PIDL for oncologic applications}
\label{sec:modality-specific}
This section provides a modality-specific synthesis that
links the governing physical models to their implementation
in representative PIDL studies. By distinguishing
physics-derived inputs, model-based architectures, and
governing-equation constraints, it clarifies how physical
Knowledge enters learning across imaging modalities.
This synthesis complements the task-oriented review and
supports comparison of the assumptions and roles of physics
in oncologic applications. Tables~\ref{tab:pidl_MRI}
and~\ref{tab:task_modality} summarize representative connections.

\subsection{CT imaging} 
\label{subsec:ct}
CT image formation is governed by energy-dependent X-ray attenuation. Under the monochromatic, scatter-free approximation, the Beer-Lambert law relates the transmitted intensity $I_\ell$ along a ray path $\ell$ to the attenuation distribution $\mu(\mathbf{r})$:
\begin{equation}
I_\ell = I_{0,\ell}
\exp\left[-\int_\ell \mu(\mathbf{r}(s))\,\mathrm{d}s\right],~
p_\ell = -\ln\left(\frac{I_\ell}{I_{0,\ell}}\right)
= \mathcal{R}\mu(\ell),
\label{eq:ct_beer_lambert1}
\end{equation}
here, $s$ denotes distance along the ray, $I_{0,\ell}$ is the
incident intensity, and $\mathcal{R}$ is the ray-integral
projection operator. Polychromatic models additionally account
for the source spectrum and detector response
\citep{hellwege2026unsupervised}.
In DECT, energy-dependent attenuation supports the material
decomposition. Poirot et al. \citep{poirot2019physics} combined a
physics-based material-decomposition mapping with a CNN
correction to estimate non-contrast images. Physics, therefore, 
enters through the network input rather than an explicit
physical residual.  
Moreover, authors in \citet{hellwege2026unsupervised} explicitly embedded a polychromatic forward model in unsupervised DECT material decomposition:
\begin{equation}
\widehat{p}_i^{\,s}
= -\ln\left[
\sum_k S_{s,k}
\exp\left(
-\sum_{m=1}^{2}
\mu_m(E_k)(A\widehat{f}_m)_i
\right)
\right],
~
s \in \{\mathrm{LE},\mathrm{HE}\},
\label{eq:ct_polychromatic}
\end{equation}
where $S_{s,k}$ is the normalized spectrum including detector response, $\mu_m(E_k)$ is the attenuation coefficient of basis material $m$, $A$ is the projection matrix, and $\widehat{f}_m$ is its predicted spatial coefficient map. Training minimizes the projection-domain discrepancy, expressed for one acquisition as
\begin{equation}
\mathcal{L}_{\mathrm{proj}}(\theta)
= \frac{1}{M}
\sum_{s \in \{\mathrm{LE},\mathrm{HE}\}}
\sum_{i=1}^{M}
\left(
\widehat{p}_i^{\,s}(\theta)-p_i^{\,s}
\right)^2,
\label{eq:ct_projection_loss}
\end{equation}
where $M$ is the number of rays per spectrum and $p_i^{\,s}$ denotes the measured log-transformed projection. The differentiable forward model supplies the training
constraint, enabling material decomposition without
ground-truth material images. CT harmonization incorporates a complementary form of scanner physics through the modulation transfer function (MTF). Under linear, shift-invariant and isotropic imaging assumptions, images reconstructed using different kernels can be related approximately by
\begin{equation}
x_b = \mathcal{F}^{-1}\left[
\mathcal{F}(x_a)
\frac{\mathrm{MTF}_b}{\mathrm{MTF}_a}
\right],
\label{eq:ct_mtf_relation}
\end{equation}
here, $\mathcal{F}$ denotes the Fourier transform and the ratio
requires care where the denominator is small. Zarei et al.~\citep{zarei2024physics} incorporated measured MTF information into
a GAN generator to learn CT harmonization. This approach
incorporates scanner physics through network conditioning,
complementing the projection-domain constraints used for DECT.

\subsection{MRI}
\label{subsec:mri}
MRI-based PIDL incorporates acquisition and physiological physics through three complementary mechanisms: data-consistency constraints or unrolled reconstruction, differentiable signal and tracer-kinetic models, and physiological PDE residuals, as in Eq.~\eqref{eq:total_loss}, in the learning objective.

\paragraph{\bf Acquisition physics as data consistency}:
For a multi-coil acquisition, the measured $k$-space data $\mathbf{y}_c$ are related to the image $\mathbf{x}$ by
\begin{equation}
\mathbf{y}_c
=
\mathbf{P}\,\mathcal{F}\,\mathbf{S}_c\,\mathbf{x}
+
\boldsymbol{\epsilon}_c,
~c=1,\dots,N_c,
\label{eq:mri_forward_model}
\end{equation}
where $\mathbf{S}_c$ is the coil-sensitivity map, $\mathcal{F}$ is the Fourier transform, and $\mathbf{P}$ is the sampling operator \citep{pruessmann1999sense}. Stacking the coil-specific operators gives
$\mathbf{A}\mathbf{x}
=(\mathbf{P}\mathcal{F}\mathbf{S}_c\mathbf{x})_{c=1}^{N_c}$.
PIDL incorporates this forward model through a
data-consistency loss or explicit reconstruction layers:
\begin{equation}
\mathcal{L}_{\mathrm{DC}}
=
\left\|\mathbf{A}\hat{\mathbf{x}}-\mathbf{y}\right\|_2^2,
~
\mathbf{x}^{(k+1)}
=
\mathcal{D}_{\boldsymbol{\theta}_k}\!\left(
\mathbf{x}^{(k)}
-
\eta_k\,\mathbf{A}^{H}\!\left(\mathbf{A}\mathbf{x}^{(k)}-\mathbf{y}\right)
\right),
\label{eq:mri_data_consistency}
\end{equation}
where $\hat{\mathbf{x}}$ is the network output and $\mathcal{D}_{\boldsymbol{\theta}_k}$ is a learned denoiser. Specific studies differ mainly in how they extend $\mathbf{A}$ and where they impose the constraint:
\begin{itemize}
\item \textit{\bf Motion correction}: PI-MoCoNet \citep{safari2025physics} applies $\mathcal{L}_{\mathrm{DC}}$ to a U-Net/Swin-Transformer \citep{liu2021swin} output, so that motion artifacts are removed without departing from the acquired data.
\item \textit{\bf Unrolled and diffusion-prior reconstruction}: Fujita et al.~\cite{fujita2025physics} unroll the iteration in Eq.~\eqref{eq:mri_data_consistency} for accelerated clinical 3D-FLAIR. Ahmed et al.~\cite{ahmed2025pinn} replace $\mathcal{D}_{\boldsymbol{\theta}}$ with a score-based diffusion prior and enforce $\mathcal{L}_{\mathrm{DC}}$ at each sampling step.
\item \textit{\bf Main-field inhomogeneity}: Schote et al.~\cite{schote2024joint} augment $\mathbf{A}$ with an off-resonance phase $e^{-i\gamma\Delta B_0(\mathbf{r})t}$, parameterize $\Delta B_0$ in a spherical-harmonic basis, and estimate it jointly with $\mathbf{x}$ for low-field MRI.
\item \textit{\bf Learned sampling}: PILOT \citep{weiss2019pilot} and Peng et al.~\cite{peng2022learning} learn $\mathbf{P}$ itself. The trajectory $\mathbf{k}(t)=\tfrac{\gamma}{2\pi}\int_0^t\mathbf{G}(\tau)\,d\tau$ is optimized jointly with the reconstruction, subject to gradient-amplitude and slew-rate limits $\|\mathbf{G}\|_\infty\le G_{\max}$ and $\|\dot{\mathbf{G}}\|_\infty\le S_{\max}$. PILOT imposes these limits as penalties, whereas Peng et al. \cite{peng2022learning} parameterize $\mathbf{k}(t)$ by a neural ODE.
\end{itemize}

\paragraph{\bf Signal and tracer-kinetic models as differentiable decoders}:
In quantitative MRI, a network $f_{\boldsymbol{\theta}}$ predicts tissue parameters $\mathbf{q}=f_{\boldsymbol{\theta}}(\mathbf{s})$ from measured signals $\mathbf{s}$. A known signal model $g$ then maps these parameters back to the signals expected under the acquisition settings $\boldsymbol{\xi}$, which permits self-supervised training without ground-truth parameter maps:
\begin{equation}
\mathcal{L}_{\mathrm{model}}
=
\left\|
\mathbf{s}
-
g\!\left(f_{\boldsymbol{\theta}}(\mathbf{s});\boldsymbol{\xi}\right)
\right\|_2^2.
\label{eq:mri_model_decoder}
\end{equation}
The choice of $g$ defines the application.\\~\\
\textbf{Relaxometry}: For relaxometry, $g$ is a closed-form steady-state solution of the Bloch equations. For spin echo,
\begin{equation}
S_{\mathrm{SE}}
=
\mathrm{PD}\left(1-e^{-T_R/T_1}\right)e^{-T_E/T_2},
\label{eq:signal_se}
\end{equation}
where $\mathrm{PD}$ is the proton density and $T_R$ and $T_E$ are the repetition and echo times. Jacobs et al.~\cite{jacobs2024generalizable} use such models to synthesize arbitrary contrasts from predicted $(T_1,T_2,\mathrm{PD})$ maps. PINQI \citep{zimmermann2024pinqi} couples an inversion-recovery $T_1$ model with Eq.~\eqref{eq:mri_forward_model}, so that $T_1$ is estimated directly from undersampled $k$-space.\\~\\
\textbf{DCE-MRI}: For DCE-MRI, $g$ is the extended Tofts model \citep{tofts1999estimating},
\begin{equation}
C_t(t)
=
v_p\,C_p(t)
+
K^{\mathrm{trans}}\!\int_{0}^{t}C_p(\tau)\,e^{-K^{\mathrm{trans}}(t-\tau)/v_e}\,d\tau,
\label{eq:tofts}
\end{equation}
where $C_t$ and $C_p$ are the tissue and plasma contrast-agent concentrations, $K^{\mathrm{trans}}$ is the volume transfer constant, and $v_p$ and $v_e$ are the plasma and extravascular extracellular volume fractions. Sainz et al.~\cite{sainz2024exploring} showed that this constraint improves the robustness of $(K^{\mathrm{trans}},v_e,v_p)$ estimation under noise.\\~\\
\textbf{DSC-MRI}: For DSC-MRI, $g$ is the indicator-dilution convolution \citep{ostergaard1996high},
\begin{equation}
C_t(t)
=
\mathrm{CBF}\int_{0}^{t}C_a(\tau)\,R(t-\tau)\,d\tau,
\label{eq:dsc_convolution}
\end{equation}
where $C_a$ is the arterial input function, $R$ is the residue function, and CBF is the cerebral blood flow. Rotkopf et al.~\cite{rotkopf2024dsc} and SPINNED \citep{asaduddin2024spinned} represent $C_t$ and $R$ with networks constrained by Eq.~\eqref{eq:dsc_convolution}, which regularizes this ill-posed deconvolution. Van Herten et al.~\cite{van2022physics} apply the same principle to myocardial perfusion, using the residual of a two-compartment exchange model.

\paragraph{\bf Physiological PDEs as residual constraints}:
When the quantity of interest is a spatiotemporal field that MRI observes only sparsely or indirectly, the governing PDE enters the loss as $\mathcal{L}_{\mathrm{phys}}$ in Eq.~\eqref{eq:total_loss}. The residuals used in the MRI literature are
\begin{subequations}
\label{eq:mri_pde_residuals}
\begin{align}
\mathcal{R}_{\mathrm{NS}}
&=
\rho\left(\partial_t\mathbf{u}+(\mathbf{u}\cdot\nabla)\mathbf{u}\right)
+
\nabla p
-
\mu\nabla^2\mathbf{u},
\qquad
\nabla\cdot\mathbf{u}=0,
\label{eq:navier_stokes}
\\
\mathcal{R}_{\mathrm{EPT}}
&=
\nabla^2 B_1^{+}
+
\omega^2\mu_0\left(\varepsilon-\tfrac{i\sigma}{\omega}\right)B_1^{+},
\label{eq:helmholtz_ept}
\\
\mathcal{R}_{\mathrm{diff}}
&=
\partial_t c
-
\nabla\cdot\left(D\nabla c\right),
\label{eq:diffusion}
\\
\mathcal{R}_{\mathrm{tumor}}
&=
\partial_t u
-
\nabla\cdot\left(D(\mathbf{r})\nabla u\right)
-
\rho\,u\,(1-u).
\label{eq:fisher_kpp}
\end{align}
\end{subequations}
Each residual serves a different inverse problem.\\~\\
\textbf{Flow}: Eq.~\eqref{eq:navier_stokes} super-resolves and denoises 4D-flow velocities $\mathbf{u}$ and recovers the unmeasured pressure $p$. This approach was used with fully connected PINNs \citep{fathi2020super}, on graph edges for irregular vascular domains \citep{kazemi2024pi}, and for refining high-velocity regions \citep{kang2025flow}. Sarabian et al.~\cite{sarabian2022physics} use a reduced-order one-dimensional counterpart of this model together with sparse Doppler and MRI data.\\~\\
\textbf{Electrical-property tomography}: Eq.~\eqref{eq:helmholtz_ept} is the Helmholtz form of Maxwell's equations for the measured transmit field $B_1^{+}$. Treating the conductivity $\sigma$ and permittivity $\varepsilon$ as trainable fields avoids the noise-amplifying second derivatives of conventional EPT \citep{katscher2009determination,inda2022physics,ruan2024magnetic}.\\~\\
\textbf{Molecular transport}: Eq.~\eqref{eq:diffusion} identifies the effective diffusivity $D$ of an MRI tracer from sparse time points \citep{zapf2022investigating}.\\~\\
\textbf{Tumor growth}: Eq.~\eqref{eq:fisher_kpp} is the Fisher-Kolmogorov model \citep{swanson2000quantitative}, and it is the most directly oncologic of the four residuals. MRI supplies the domain, the tissue-dependent diffusivity $D(\mathbf{r})$, and threshold observations of the tumor-cell density $u$. The PINN then infers the patient-specific parameters $(D,\rho)$ and the infiltration beyond the visible margin. El et al.~\cite{el2026tumor} couple this model with U-Net segmentation and uncertainty quantification. Zhang et al.~\cite{zhang2025personalized} pre-train on the PDE and fine-tune to individual glioblastoma patients.
Table~\ref{tab:pidl_MRI} lists each study together with its embedding strategy and governing equation.

\begin{table*}[!t]
\centering
\caption{Representative PIDL methods for MRI. Strategy: DC = acquisition-physics data consistency; MD = differentiable signal/kinetic model decoder; PR = PDE-residual constraint.}
\label{tab:pidl_MRI}
\footnotesize
\renewcommand{\arraystretch}{1.12}
\setlength{\tabcolsep}{4pt}
\begin{tabularx}{\textwidth}{@{}p{1.8cm}p{1.0cm}p{3.4cm}p{3.0cm}X@{}}
\toprule
\textbf{Reference} & \textbf{Strategy} & \textbf{Embedded physics (Eq.)} & \textbf{Objective} & \textbf{Principal contribution} \\
\midrule
\cite{weiss2019pilot} & DC & Trajectory with gradient/slew limits & Acquisition design & Joint learning of hardware-feasible trajectory and reconstruction. \\
\cite{peng2022learning} & DC & Neural-ODE trajectory & $k$-space optimization & Continuous trajectories learned with reconstruction. \\
\cite{schote2024joint} & DC & Off-resonance forward model, Eq.~\eqref{eq:mri_forward_model} & Low-field reconstruction & Joint $B_0$-field and image estimation. \\
\cite{fujita2025physics} & DC & Unrolled forward model, Eq.~\eqref{eq:mri_data_consistency} & 3D-FLAIR reconstruction & Accelerated reconstruction with clinical validation. \\
\cite{ahmed2025pinn} & DC & $k$-space consistency with diffusion prior, Eq.~\eqref{eq:mri_data_consistency} & Reconstruction (fastMRI) & Diffusion-model reconstruction with measurement consistency. \\
\cite{safari2025physics} & DC & $k$-space consistency loss, Eq.~\eqref{eq:mri_data_consistency} & Motion correction & Artifact suppression consistent with acquired data. \\
\cite{jacobs2024generalizable} & MD & Bloch signal models, Eq.~\eqref{eq:signal_se} & Contrast synthesis & Arbitrary contrasts from $(T_1,T_2,\mathrm{PD})$. \\
\cite{zimmermann2024pinqi} & DC+MD & Forward model and IR $T_1$ model & Quantitative $T_1$ mapping & End-to-end quantitative reconstruction. \\
\cite{sainz2024exploring} & MD & Tofts model, Eq.~\eqref{eq:tofts} & DCE-MRI parameters & Noise-robust $K^{\mathrm{trans}}$, $v_e$, $v_p$ estimation. \\
\cite{rotkopf2024dsc} & MD & Indicator dilution, Eq.~\eqref{eq:dsc_convolution} & DSC perfusion & Physics-informed CBF/CBV/MTT estimation. \\
\cite{asaduddin2024spinned} & MD & Indicator dilution, Eq.~\eqref{eq:dsc_convolution} & DSC deconvolution & Training-free, constrained deconvolution. \\
\cite{van2022physics} & PR & Two-compartment exchange & Myocardial perfusion & Physiologically constrained perfusion parameters. \\
\cite{fathi2020super} & PR & Navier--Stokes, Eq.~\eqref{eq:navier_stokes} & 4D-flow super-resolution & High-resolution velocity and pressure. \\
\cite{kazemi2024pi} & PR & Navier--Stokes on graph, Eq.~\eqref{eq:navier_stokes} & 4D-flow super-resolution & Graph learning on irregular vascular domains. \\
\cite{kang2025flow} & PR & Navier--Stokes, Eq.~\eqref{eq:navier_stokes} & Flow refinement & Improved high-velocity regions. \\
\cite{sarabian2022physics} & PR & Reduced-order hemodynamics & Cerebral flow & High-resolution flow from sparse TCD/MRI. \\
\cite{inda2022physics} & PR & Helmholtz, Eq.~\eqref{eq:helmholtz_ept} & EPT & Noise-robust conductivity reconstruction. \\
\cite{ruan2024magnetic} & PR & Helmholtz, Eq.~\eqref{eq:helmholtz_ept} & EPT & Stable electrical-property reconstruction. \\
\cite{zapf2022investigating} & PR & Diffusion, Eq.~\eqref{eq:diffusion} & Molecular transport & Inverse estimation of diffusivity. \\
\cite{el2026tumor} & PR & Fisher-Kolmogorov, Eq.~\eqref{eq:fisher_kpp} & Tumor delineation and growth & Segmentation, growth prediction, and UQ. \\
\cite{zhang2025personalized} & PR & Fisher-Kolmogorov, Eq.~\eqref{eq:fisher_kpp} & Glioblastoma infiltration & Patient-specific pre-trained/fine-tuned PINN. \\
\bottomrule
\end{tabularx}
\end{table*}
\subsection{PET}
\label{subsec:pet}

PET-based PIDL incorporates photon-counting statistics,
attenuation and resolution effects into reconstruction, and
tracer kinetics into parametric imaging. Radioactive decay,
$A(t)=A_0e^{-\lambda t}$ with $\lambda=\ln 2/T_{1/2}$, requires
consistent temporal referencing of measured activity and input
functions. The principal integration mechanisms are
likelihood-based reconstruction, tissue-informed correction,
and kinetic-model constraints.

\vspace{0.2cm}
\noindent \textbf{Measurement physics and learned reconstruction}:
For an activity image $\mathbf{x}\geq 0$, uncorrected coincidence
counts are modeled as conditionally independent Poisson
variables, $y_i\sim\operatorname{Poisson}(\bar y_i)$
\citep{shepp1982maximum}, with
\begin{equation}
\bar{\mathbf{y}}(\mathbf{x})=\mathbf{P}\mathbf{x}+\mathbf{r}+\mathbf{s},~\mathbf{P}
=\operatorname{diag}(\mathbf{n})
  \operatorname{diag}(\mathbf{a})\mathbf{G}\mathbf{H},
~
a_i=\exp\left[-\int_{L_i}\mu(\mathbf{r})\,d\ell\right].
\label{eq:pet_forward_model}
\end{equation}
Here, $\mathbf{G}$ projects activity onto lines of response,
$\mathbf{H}$ approximates resolution effects, and $\mathbf{n}$
contains multiplicative detector-sensitivity factors.
The attenuation coefficient $\mu$ is evaluated at $511$~keV;
$\mathbf{r}$ and $\mathbf{s}$ represent expected random and
scattered coincidences. The corresponding negative
log-likelihood, up to constants, is
\begin{equation}
\mathcal{L}_{\mathrm{Poisson}}(\mathbf{x})
=\sum_i\left[
\bar y_i(\mathbf{x})-y_i\ln\bar y_i(\mathbf{x})
\right].
\label{eq:pet_poisson_nll}
\end{equation}
This objective connects learned image representations directly
to the measured counts. Gong et al.~\cite{gong2019iterative}
embedded a pretrained CNN representation in constrained
likelihood-based reconstruction, using ADMM to alternate
between image and network-input updates. FBSEM-Net
\citep{mehranian2021model} instead unrolled
forward--backward splitting EM, combining a learned
regularization step with an EM data-fidelity step and weighted
fusion. Thus, acquisition physics enters the reconstruction
architecture rather than being applied only after image
formation.

\vspace{0.2cm}
\noindent \textbf{Tissue-informed positron-range correction}:
Positron transport before annihilation introduces
radionuclide- and tissue-dependent blurring
\citep{levin1999calculation}. A general image-space
approximation is
\begin{equation}
x_{\mathrm{obs}}(\mathbf{r})
=\int_{\Omega}
h_{\mathrm{PR}}(\mathbf{r},\mathbf{r}';\mathcal{T})
x(\mathbf{r}')\,d\mathbf{r}'
+\epsilon(\mathbf{r}),
\label{eq:pet_positron_range}
\end{equation}
where $\mathcal{T}$ describes the tissue distribution and
$h_{\mathrm{PR}}$ is a spatially varying positron-range kernel.
Encina-Baranda et al.~\cite{encina2025tissue} incorporated
attenuation-map information into 3D residual encoder--decoder
networks and a tissue-dependent loss for preclinical
$^{68}$Ga PET correction. Here, physics informs the training
data, anatomical inputs, and loss weighting, rather than
entering as an explicit transport-equation residual.

\vspace{0.2cm}
\noindent \textbf{Tracer-kinetic constraints}:
Dynamic PET links tissue time--activity curves to the plasma
input function $C_p(t)$. The two-tissue compartment model
underlying FDG kinetic analysis describes transport and
phosphorylation through
\citep{sokoloff1977deoxyglucose}
\begin{equation}
\begin{aligned}
\frac{dC_1}{dt}
&=K_1C_p-(k_2+k_3)C_1+k_4C_2,\\
\frac{dC_2}{dt}
&=k_3C_1-k_4C_2,
\end{aligned}
\label{eq:pet_2tcm}
\end{equation}
where $C_1$ and $C_2$ denote unphosphorylated and
phosphorylated tracer concentrations, $K_1$ is the influx
coefficient, and $k_2$, $k_3$, and $k_4$ are rate constants.
For consistently decay-corrected concentrations, the measured
tissue signal is modeled as
\begin{equation}
C_{\mathrm{PET}}(t)
=(1-v_b)\big[C_1(t)+C_2(t)\big]
+v_bC_{\mathrm{wb}}(t),
\label{eq:pet_measured_tac}
\end{equation}
where $v_b$ is the blood-volume fraction and $C_{\mathrm{wb}}$
is whole-blood activity concentration.

These kinetics can enter PIDL as differential-equation
residuals or as a differentiable decoder mapping predicted
parameters to time--activity curves. Predicted signals should
be averaged over acquisition frames before comparison with
reconstructed dynamic images, for example using weighted
least squares. For direct estimation from count data, the
kinetic predictions must instead pass through the PET forward
model before evaluating Eq.~\eqref{eq:pet_poisson_nll}.
This links parameter estimation to tracer physiology
\citep{zar2026current}.

\vspace{0.2cm}
\noindent \textbf{Tumor-growth constraints and multimodal assimilation}:
Beyond acquisition and tracer kinetics, PET can constrain
mechanistic tumor models. A Fisher--Kolmogorov model
\citep{swanson2000quantitative} describes normalized
tumor-cell density $u$ as
\begin{equation}
\frac{\partial u}{\partial t}=\nabla\cdot(D\nabla u)+\rho u(1-u),~\mathbf{n}\cdot D\nabla u=0
\quad\text{on }\partial\Omega,
\label{eq:pet_fisher_kpp}
\end{equation}
where $D$ represents tissue-dependent infiltration and $\rho$
is the proliferation rate. MRI-derived tumor regions and
FET-PET uptake provides complementary observation constraints.
Their relationship to cell density is a modeling assumption,
rather than a direct measurement of tumor-cell concentration. Moreover, Balcerak et al.~\cite{balcerak2024physics} combined multimodal
observations with reaction--diffusion constraints to infer
patient-specific glioma distributions. Glioma Optimizing the Discrete Loss (GliODIL)
\citep{balcerak2025individualizing} optimizes discretized
fields and parameters using a composite objective, written
schematically as
\begin{equation}
\mathcal{L}_{\mathrm{GliODIL}}
=\lambda_{\mathrm{PDE}}
       \|\mathcal{R}_h(\mathbf{u},D,\rho)\|_2^2
 +\lambda_{\mathrm{MRI}}\mathcal{L}_{\mathrm{MRI}}
  +\lambda_{\mathrm{PET}}\mathcal{L}_{\mathrm{PET}},
\label{eq:pet_odil_loss}
\end{equation}
where $\mathcal{R}_h$ is the discretized growth-model residual.
The soft physics penalty allows a trade-off between model
agreement and imaging observations. GliODIL is a related
physics-informed inverse method rather than a neural-network
representation of the solution; it is included to distinguish
direct field optimization from PIDL.
Table~\ref{tab:pidl_PET} summarizes these mechanisms.

\begin{table*}[!t]
\centering
\caption{Physics incorporation in representative PET learning
and related inverse methods.}
\label{tab:pidl_PET}
\footnotesize
\renewcommand{\arraystretch}{1.1}
\setlength{\tabcolsep}{4pt}
\begin{tabularx}{\textwidth}{@{}lXXX@{}}
\toprule
\textbf{Reference} &
\textbf{Physical model} &
\textbf{Integration mechanism} &
\textbf{Application} \\
\midrule

\cite{gong2019iterative}
& PET forward model and Poisson likelihood,
Eqs.~\eqref{eq:pet_forward_model}
and~\eqref{eq:pet_poisson_nll}
& Pretrained CNN representation within constrained
likelihood optimization
& PET reconstruction \\

\cite{mehranian2021model}
& PET forward model and Poisson likelihood
& Unrolled EM data-fidelity steps with learned regularization
& PET and MR-guided PET reconstruction \\

\cite{encina2025tissue}
& Tissue-dependent positron-range blurring,
Eq.~\eqref{eq:pet_positron_range}
& Simulation-based training, attenuation-map inputs,
and tissue-dependent loss
& Preclinical $^{68}$Ga resolution correction \\

\cite{balcerak2024physics}
& Reaction--diffusion growth,
Eq.~\eqref{eq:pet_fisher_kpp}
& Joint imaging and physics constraints
& MRI/FET-PET glioma inference \\

\cite{balcerak2025individualizing}
& Discretized growth model,
Eq.~\eqref{eq:pet_odil_loss}
& Direct field optimization with soft physics and data penalties
& Individualized glioma infiltration maps for
radiotherapy planning \\

\bottomrule
\end{tabularx}
\end{table*}
\subsection{Ultrasound}
\label{subsec:us}

Ultrasound-based PIDL incorporates acoustic propagation into
sound-speed reconstruction and fluid mechanics into vascular
flow estimation. Physics enters through simulated training
data and reconstructed inputs, differentiable inversion
steps, or governing-equation residuals.

\vspace{0.2cm}
\noindent \textbf{Acoustic physics and sound-speed estimation}:
Under the linear, constant-density, lossless approximation,
acoustic pressure $p(\mathbf{r},t)$ satisfies
\begin{equation}
\frac{1}{c^2(\mathbf{r})}
\frac{\partial^2p}{\partial t^2}
-\nabla^2p=s(\mathbf{r},t),
\label{eq:ultrasound_acoustic_wave}
\end{equation}
where $c(\mathbf{r})$ is sound speed and $s$ is the source.
More complete models account for density variations and
frequency-dependent attenuation, often represented by
$\alpha(f)=\alpha_0f^y$ \citep{szabo1994time}.
Numerical solvers such as k-Wave provide acoustic simulations
for training and forward modeling \citep{treeby2010kwave}.
Spatial variations in sound speed produce an anomaly when
beamforming assumes a homogeneous medium, motivating
patient-specific sound-speed estimation.\par
Jeong et al.~\cite{jeong2024investigating} trained a CNN to
combine a low-resolution sound-speed map from traveltime
tomography with a reflectivity map from reflection tomography.
The former provides quantitative propagation information,
while the latter supplies tissue-boundary information.
Training used numerical breast phantoms, with additional
evaluation on clinical breast data. Physics therefore enters
through simulation and the complementary reconstructed
inputs, rather than an explicit PDE residual in the loss.\\~\\
\noindent \textbf{Physics-guided full-waveform inversion}:
Full-waveform inversion (FWI) estimates sound speed by matching
simulated and measured receiver signals
\citep{tarantola1984inversion,virieux2009overview}:
\begin{equation}
\widehat c=\arg\min_{c>0}
\left[\mathcal{J}_{\mathrm{d}}(c)
+\lambda\mathcal{R}(c)\right],~\mathcal{J}_{\mathrm{d}}(c)
=\frac{1}{2}\sum_{k,j}\int_0^{T}
\left|p_k(\mathbf{r}_j,t;c)-d_{kj}(t)\right|^2\,dt.
\label{eq:us_fwi}
\end{equation}
Here, $p_k$ is the wavefield for source $k$, $d_{kj}$ is the
signal measured at receiver $j$, and $\mathcal{R}$ is a
regularizer. Wave-equation gradients, commonly evaluated
using adjoint methods \citep{plessix2006review}, connect
signal mismatch to updates of $c$. Repeated wave solves and
sensitivity to initialization motivate learned inversion. Zhang et al.~\cite{zhang2026physics} unfolded FWI into a
sequence of learned updates driven by the data-fidelity
gradient. Schematically,
\begin{equation}
\mathbf{g}^{(m)}=\nabla_c\mathcal{J}_{\mathrm{d}}(c^{(m)}),~
c^{(m+1)}
=c^{(m)}
+\mathcal{U}_{\theta_m}\!\left(\mathbf{g}^{(m)}\right)
\label{eq:us_unfolded_fwi}
\end{equation}
where $\mathcal{U}_{\theta_m}$ predicts an update from the
physics-derived gradient. The method reported real-time
sound-speed reconstruction in simulations and phantom
experiments for edema assessment. It illustrates a
transferable inversion strategy, while oncologic performance
requires separate validation. The forward model guides each
update, but does not guarantee exact data consistency.\\~\\
\noindent \textbf{Microvascular flow and pressure inference}: 
Ultrasound localization microscopy estimates microvascular
velocities by tracking individual microbubbles
\citep{errico2015ultrafast}. Sparse velocity observations can
be supplemented by incompressible-flow constraints. Under
a Newtonian approximation with negligible body forces,
\begin{equation}
\rho_b\left(
\frac{\partial\mathbf{v}}{\partial t}
+(\mathbf{v}\cdot\nabla)\mathbf{v}
\right)=-\nabla p_b+\mu_b\nabla^2\mathbf{v},~\nabla\cdot\mathbf{v}=0
\label{eq:ultrasound_navier_stokes}
\end{equation}
where $\rho_b$, $\mu_b$, $\mathbf{v}$, and $p_b$ denote blood
density, dynamic viscosity, velocity, and pressure. Liang et al.~\cite{liang2025ultrasound} combined
super-resolution ultrasound observations with hemodynamic
simulations through data assimilation, then fitted a PINN
constrained by two-dimensional Navier--Stokes equations.
Its principal loss components can be summarized as
\begin{equation}
\mathcal{L}
=\lambda_{\mathrm{d}}\mathcal{L}_{\mathrm{vel}}
+\lambda_{\mathrm{m}}
\left\langle
\|\mathcal{N}_{\mathrm{mom}}
[\mathbf{v}_{\theta},p_{b,\theta}]\|_2^2
\right\rangle_{\Omega,t}+\lambda_{\mathrm{c}}
\left\langle
|\nabla\cdot\mathbf{v}_{\theta}|^2
\right\rangle_{\Omega,t},
\label{eq:us_ns_pinn_loss}
\end{equation}
where $\mathcal{L}_{\mathrm{vel}}$ fits the assimilated
velocity data, $\mathcal{N}_{\mathrm{mom}}$ is the momentum
residual, and the brackets denote averaging over collocation
points. The study included liver-tumor peritumoral vessels,
providing an oncologic application of flow-constrained
learning. Pressure is inferred indirectly through momentum
balance and requires a reference to fix its spatially
uniform component. Interpretation depends on the assumed
rheology, boundary conditions, and adequacy of the
two-dimensional flow approximation.
Table~\ref{tab:pidl_US} summarizes these distinct routes
for incorporating physics.

\begin{table*}[!t]
\centering
\caption{Physics incorporation in representative
ultrasound learning methods.}
\label{tab:pidl_US}
\footnotesize
\renewcommand{\arraystretch}{1.1}
\setlength{\tabcolsep}{4pt}
\begin{tabularx}{\textwidth}{@{}lXXX@{}}
\toprule
\textbf{Reference} &
\textbf{Physical information} &
\textbf{Integration mechanism} &
\textbf{Application} \\
\midrule

\cite{jeong2024investigating}
& Acoustic propagation and reflection
& Simulation-based training with traveltime and
reflection tomography inputs
& Breast sound-speed reconstruction \\

\cite{zhang2026physics}
& Waveform data fidelity,
Eq.~\eqref{eq:us_fwi}
& Learned updates driven by wave-model gradients,
Eq.~\eqref{eq:us_unfolded_fwi}
& Sound-speed reconstruction for edema assessment;
simulation and phantom validation \\

\cite{liang2025ultrasound}
& Incompressible Navier--Stokes equations,
Eq.~\eqref{eq:ultrasound_navier_stokes}
& Data assimilation and PINN residual penalties,
Eq.~\eqref{eq:us_ns_pinn_loss}
& Microvascular velocity and pressure inference,
including liver-tumor peritumoral vessels \\

\bottomrule
\end{tabularx}
\end{table*}
\subsection{Thermal imaging}
\label{subsec:thermal}

Infrared thermography estimates skin-surface temperature
from measured radiance. Thermal PIDL links these observations
to subsurface heat transport, allowing tissue temperature
and selected thermal or tumor parameters to be inferred
under a bioheat model.

\vspace{0.2cm}
\noindent \textbf{Bioheat transport and surface observations}:
The Pennes equation describes heat conduction, perfusion,
and metabolic heat generation \citep{pennes1948analysis}:
\begin{equation}
\rho c\frac{\partial T}{\partial t}
=\nabla\cdot(k\nabla T)
+\rho_b c_b\omega_b(T_a-T)
+Q_m+Q_t.
\label{eq:thermal_pennes}
\end{equation}
Here, $\rho$, $c$, and $k$ are tissue density, specific heat
capacity, and thermal conductivity; $\rho_b$ and $c_b$
are the corresponding blood properties; $\omega_b$ is
volumetric perfusion per unit tissue volume; and $T_a$
is arterial temperature. The terms $Q_m$ and $Q_t$ denote
basal and additional tumor-associated volumetric heat
generation. Tumor effects can be represented through
spatial variations in heat generation and perfusion,
rather than assuming that every lesion produces the
same thermal response. For equilibrated static acquisitions,
the model is commonly approximated as steady state.

A prescribed chest-wall temperature,
$T=T_{\mathrm{core}}$, can be combined with convective
and radiative heat exchange at the skin:
\begin{equation}
-k\nabla T\cdot\mathbf{n}
=h(T-T_{\infty})
+\varepsilon\sigma(T^4-T_{\mathrm{sur}}^4)
\quad\text{on }\Gamma_{\mathrm{s}},
\label{eq:thermal_skin_bc}
\end{equation}
where $\mathbf{n}$ is the outward normal, $h$ is the
convective coefficient, $T_{\infty}$ is air temperature,
and $T_{\mathrm{sur}}$ is effective radiative-surroundings
temperature. The radiative term uses absolute temperatures,
skin emissivity $\varepsilon$, and the Stefan--Boltzmann
constant $\sigma$. Camera-derived surface temperatures
require calibration accounting for emissivity and
reflected environmental radiation.

\vspace{0.2cm}
\noindent \textbf{Forward and inverse physics-informed learning}:
A PINN represents temperature by $T_{\theta}$ and penalizes
the residual of Eq.~\eqref{eq:thermal_pennes} together with
the boundary conditions. With prescribed tissue and tumor
parameters, this gives a forward temperature solver, as
investigated by Mukhmetov et al.~\cite{mukhmetov2023physics}
and Zhao et al.~\cite{zhao2024application}.
Perez-Raya et al.~\cite{perez2023thermal} applied this
framework to patient-specific breast geometries and
thermal-conductivity estimation.
For steady-state inverse modeling, selected unknown
parameters $\boldsymbol{\eta}$ are optimized jointly with
the network using surface observations:
\begin{equation}
\mathcal{L}(\theta,\boldsymbol{\eta})
=\lambda_{\mathrm{f}}
\left\langle
|\mathcal{R}_{\mathrm{bioheat}}
[T_{\theta};\boldsymbol{\eta}]|^2
\right\rangle_{\Omega}+\lambda_{\mathrm{b}}
\mathcal{L}_{\mathrm{BC}}(\theta,\boldsymbol{\eta})+\frac{\lambda_{\mathrm{s}}}{N_s}
\sum_{j=1}^{N_s}
|T_{\theta}(\mathbf{r}_j)-T_j^{\mathrm{IR}}|^2,
~\mathbf{r}_j\in\Gamma_{\mathrm{s}}.
\label{eq:thermal_inverse_loss}
\end{equation}
Here, $\mathcal{R}_{\mathrm{bioheat}}$ is the steady-state
bioheat residual, the brackets denote averaging over
interior collocation points, and $\mathcal{L}_{\mathrm{BC}}$
enforces the thermal boundary conditions.
Depending on the available data and prior assumptions,
$\boldsymbol{\eta}$ may include conductivity, perfusion,
heat-source strength, or tumor geometry.
Perez-Raya et al.~\cite{perez2024transformative} investigated
inverse estimation of tumor characteristics from surface
temperature data using this physics-informed approach.
Transient formulations additionally require initial
conditions and time-resolved observations.
Residual-based adaptive refinement concentrates training
points where the governing-equation residual is large
\citep{lu2021deepxde,wu2023comprehensive}.
Das et al.~\cite{das2026adaptive} used distribution-based
adaptive refinement for breast thermal simulations and
inverse thermal-coefficient identification. This improves
numerical resolution of the bioheat model; it does not
remove the non-uniqueness of the inverse problem.
Heat diffusion smooths subsurface signatures, and different
combinations of tumor depth, size, perfusion, and heat
generation can produce similar surface temperatures.
Consequently, inferred parameters remain conditional on
the assumed geometry, boundary conditions, and prior
constraints.
\subsection{Hybrid imaging and multiphysics models}
\label{subsec:hybrid}

Hybrid PIDL couples complementary imaging information
through acquisition models, anatomical constraints, or
mechanistic models of disease. The coupling mechanism
determines whether physics guides reconstruction,
parameter estimation, or longitudinal inference.

\vspace{0.2cm}
\noindent \textbf{Anatomy-guided reconstruction}:
In PET/MRI, MR images support attenuation-map estimation
and provide anatomical information for PET reconstruction.
MRI does not directly measure attenuation at $511$~keV;
the attenuation map must be inferred using an additional
mapping or tissue model. A schematic learned reconstruction
objective is
\begin{equation}
\widehat{\mathbf{x}}
=\arg\min_{\mathbf{x}\geq 0}
\left[
\mathcal{L}_{\mathrm{Poisson}}
(\mathbf{x};\mu_{\mathrm{MR}})
+\beta\mathcal{R}_{\theta}
(\mathbf{x},\mathbf{z}_{\mathrm{MR}})
\right],
\label{eq:hybrid_petmr}
\end{equation}
where $\mu_{\mathrm{MR}}$ is the estimated attenuation map,
$\mathbf{z}_{\mathrm{MR}}$ is the anatomical image, and
$\mathcal{R}_{\theta}$ represents an MR-conditioned learned
regularizer. The PET forward model and likelihood remain
those of Eqs.~\eqref{eq:pet_forward_model}
and~\eqref{eq:pet_poisson_nll}.
Rajagopal et al.~\cite{rajagopal2022physics} reviewed these
physics-driven reconstruction strategies, while FBSEM-Net
\citep{mehranian2021model} provides an example of incorporating
MR information into learned regularization within unrolled
PET reconstruction. Anatomical guidance must accommodate
PET uptake patterns without corresponding MR boundaries.
Furthermore, multimodal MRI can constrain the reaction--diffusion model
of Eq.~\eqref{eq:pet_fisher_kpp} without introducing another
growth equation. Structural images define the brain
geometry and tissue distribution, while T1-weighted
contrast-enhanced and FLAIR segmentations constrain
cell-density patterns through assumed observation thresholds.
Zhang et al.~\cite{zhang2025personalized} combined these
constraints with a PINN, using characteristic dimensionless
parameters for pre-training and subsequent patient-specific
fine-tuning. Their formulation supports inference from a
single-time MRI snapshot under the adopted scaling and
observation assumptions. Here, imaging constrains the
visible tumor, while the PDE regularizes the inferred
infiltration beyond its segmented boundary.

\vspace{0.2cm}
\noindent \textbf{Imaging-informed biomarker dynamics}:
Camacho-Gomez et al.~\cite{camacho2025physics} coupled
multiparametric MRI, tumor growth, and prostate-specific
antigen (PSA) dynamics in a prostate digital twin.
T2-weighted MRI provides geometry, diffusion-derived ADC
maps inform cellularity, and DCE-derived
$K^{\mathrm{trans}}$ informs vascular exchange under
model-specific assumptions. The learned component regulates
tumor-cell growth:
\begin{equation}
\frac{\partial c_t(\mathbf{r},t)}{\partial t}
=\phi_{\theta}(\mathbf{r},t)\,
\alpha_t c_t(\mathbf{r},t),
\label{eq:hybrid_prostate_growth}
\end{equation}
where $c_t$ is tumor-cell concentration, $\alpha_t$ is a
growth-rate parameter, and $\phi_{\theta}\in[0,1]$ is the
network-estimated proliferating fraction. Mechanistic
equations describe tissue PSA production, tissue--blood
exchange, and serum clearance, linking spatial tumor
dynamics to measured serum PSA.
The network is trained to reproduce PSA dynamics, while
patient-specific model calibration also uses follow-up MRI.
Subsequent PSA observations guide tumor-growth
reconstruction within the calibrated model.
Thus, learning supplies a component of the mechanistic
simulation rather than replacing its governing equations.
The inferred spatial growth remains conditional on the
imaging-to-parameter mappings and biomarker model.
\subsection{Additional modalities and emerging PIDL applications}
\label{subsec:other}

Additional imaging modalities provide opportunities to
incorporate acquisition physics into learning. The examples
below distinguish demonstrated integration mechanisms from
potential extensions to oncologic applications.

\vspace{0.2cm}
\noindent \textbf{Mammography and digital breast tomosynthesis}:
Mammography and digital breast tomosynthesis (DBT) share
the X-ray attenuation physics discussed in
Section~\ref{subsec:ct}. In DBT, limited angular coverage
makes reconstruction underdetermined. Moriakov et
al.~\cite{moriakov2019deep} incorporated the acquisition
operator into a learned primal--dual reconstruction
scheme, combining projection and backprojection with
trainable updates. Teuwen et al.~\cite{teuwen2021deep}
extended this approach with breast-thickness information
for volumetric density and patient-specific imaging-dose
estimation. Physics therefore enters through the
reconstruction operator and acquisition-specific priors,
rather than requiring another attenuation equation.

\vspace{0.2cm}
\noindent \textbf{SPECT}:
SPECT reconstruction combines Poisson counting statistics
with emission-point-dependent attenuation and
collimator--detector response:
\begin{equation}
y_i\sim\operatorname{Poisson}(\bar y_i),~
\bar y_i=\sum_j a_{ij}x_j+s_i,~a_{ij}\approx g_{ij}
\exp\left[
-\int_{\mathbf{r}_j}^{\mathbf{d}_i}
\mu(\mathbf{r})\,d\ell
\right].
\label{eq:other_spect}
\end{equation}
Here, $x_j$ is voxel activity, $g_{ij}$ represents geometric
sensitivity and detector response, and $s_i$ is expected
scatter. The attenuation factor is written in a ray-path
approximation and depends on the emitting voxel. Yu et al.~\cite{yu2021physics} combined physics-based
attenuation estimation from scatter-window data with CNN
segmentation, then used the resulting attenuation map in
iterative activity reconstruction.
Chen et al.~\cite{chen2022direct} compared direct learned
attenuation correction with indirect approaches that
estimate attenuation maps for reconstruction. These
cardiac SPECT studies illustrate transferable correction
strategies; they do not establish oncologic performance.
The distinction is whether the physical model is explicitly
used in reconstruction or its correction is learned from
paired examples.

\vspace{0.2cm}
\noindent \textbf{Cone-beam CT (CBCT) and radiotherapy imaging}:
CBCT shares CT transmission physics but requires particular
attention to scatter, acquisition geometry, and changing
patient anatomy. These effects motivate physics-guided
scatter correction, reconstruction, and anatomical
registration. Their relevance to adaptive radiotherapy
lies in preserving image geometry and calibrated tissue
densities for dose calculation. Clinical CBCT-guided
adaptive workflows \citep{sibolt2021clinical} establish
the application context, but AI-assisted adaptation alone
does not demonstrate physics-informed learning.

\vspace{0.2cm}
\noindent \textbf{Photoacoustic imaging}:
Photoacoustic imaging couples optical absorption to
acoustic propagation. Under thermal and stress confinement,
a short optical pulse produces initial pressure
\citep{xu2006photoacoustic}
\begin{equation}
p_0(\mathbf{r})
=\Gamma(\mathbf{r})\,
\mu_a(\mathbf{r})\,
\Phi(\mathbf{r}),
\label{eq:other_pa_source}
\end{equation}
where $\Gamma$ is the Gr\"uneisen parameter, $\mu_a$ is
the optical absorption coefficient, and $\Phi$ is fluence,
which depends on tissue absorption and scattering.
The pressure propagates according to
Eq.~\eqref{eq:ultrasound_acoustic_wave}, with
$p(\mathbf{r},0)=p_0(\mathbf{r})$ and
$\partial_t p(\mathbf{r},0)=0$.  Paul et al.~\cite{paul2025model} combined model-derived
data-fidelity gradients with learned regularization for
sparse linear-array reconstruction. This constrains the
acoustic inverse problem. Quantitative absorption and
oxygenation estimation additionally require accounting
for spatially and spectrally varying fluence
\citep{cox2012quantitative}; accurate pressure
reconstruction alone does not resolve that optical
inverse problem.

\vspace{0.2cm}
\noindent \textbf{Optical diffraction tomography}:
Under a scalar-wave approximation, optical scattering
is governed by
\begin{equation}
\nabla^2u+k_0^2n^2(\mathbf{r})u=0,
\label{eq:other_helmholtz_ls}
\end{equation}
where $u$ is the optical field, $k_0$ is the free-space
wavenumber, and $n$ is refractive index.
Saba et al.~\cite{saba2022physics} trained a PINN using
the Helmholtz residual as a physical loss to approximate
the scattering forward model for tomography.
This connects learned field prediction to wave physics
and supports refractive-index reconstruction, while
oncologic diagnostic applications require separate
evaluation.

\vspace{0.2cm}
\noindent \textbf{Histopathology}:
For bright-field stained sections, an approximate
Beer--Lambert model relates optical density to stain
contributions:
\begin{equation}
\mathrm{OD}(\mathbf{r},\lambda)
=-\log_{10}
\frac{I(\mathbf{r},\lambda)}{I_0(\lambda)}
\approx
\sum_{s\in\{\mathrm{H},\mathrm{E}\}}
V_s(\lambda)C_s(\mathbf{r}),
\label{eq:other_stain}
\end{equation}
where $V_s$ denotes the stain absorption spectrum and
$C_s$ an effective stain amount along the optical path.
This approximation underlies stain separation and
normalization \citep{macenko2009method}.
It can provide a physical prior for learned color
processing, but conventional stain normalization alone
is not PIDL.

Table~\ref{tab:task_modality} summarizes selected
task--modality links supported by the studies discussed.

\begin{table*}[!t]
\centering
\caption{Selected task-modality links and mechanisms
of physics incorporation. Entries are illustrative,
not an exhaustive coverage map; inclusion does not
imply clinical validation in oncology.}
\label{tab:task_modality}
\footnotesize
\setlength{\tabcolsep}{4pt}
\renewcommand{\arraystretch}{1.1}
\begin{tabularx}{\textwidth}{
@{}
>{\raggedright\arraybackslash}p{2.5cm}
>{\raggedright\arraybackslash}p{2.5cm}
>{\raggedright\arraybackslash}X
>{\raggedright\arraybackslash}p{2.8cm}
@{}
}
\toprule
\textbf{Task} &
\textbf{Modality} &
\textbf{Physics incorporation} &
\textbf{Examples} \\
\midrule

Reconstruction
& CT and DBT
& Polychromatic forward-model losses or acquisition
operators within learned iterations
& \cite{hellwege2026unsupervised,moriakov2019deep} \\

Emission reconstruction
& PET, PET/MRI, and SPECT
& Poisson likelihood, system response, and
physics-based attenuation correction
& \cite{gong2019iterative,mehranian2021model,yu2021physics} \\

Sound-speed and acoustic-source reconstruction
& Ultrasound and photoacoustics
& Wave-model gradients combined with learned updates
or regularization
& \cite{zhang2026physics,paul2025model} \\

Resolution correction
& PET
& Tissue-dependent positron-range modeling,
simulation, and loss weighting
& \cite{encina2025tissue} \\

Physiological parameter inference
& Ultrasound and thermal imaging
& Navier-Stokes or bioheat residuals combined
with observations
& \cite{liang2025ultrasound,perez2023thermal} \\

Tumor infiltration inference
& MRI and MRI/FET-PET
& Reaction-diffusion constraints with imaging
observation models; PINNs or direct field optimization
& \cite{zhang2025personalized,balcerak2025individualizing} \\

Tumor and biomarker dynamics
& Multiparametric MRI with serum PSA
& Learned proliferation within a mechanistic
growth and biomarker model
& \cite{camacho2025physics} \\

Breast-density and imaging-dose estimation
& DBT
& Unrolled reconstruction with acquisition
geometry and breast-thickness information
& \cite{teuwen2021deep} \\

\bottomrule
\end{tabularx}
\end{table*}
\section{Datasets and data sources}
The reviewed PIDL literature employs heterogeneous data sources spanning public oncologic imaging datasets, clinical cohorts, synthetic simulations, physical phantoms, and experimental measurements. Representative datasets included in the reviewed studies are summarized in Table~\ref{tab:datasets}.

\begin{table*}[t]
\centering
\caption{Representative datasets and data sources used in physics-informed oncologic imaging studies.}
\label{tab:datasets}
\begin{tabular}{p{2.8cm} p{3.6cm} p{6.2cm} p{2.0cm}}
\hline
\textbf{Dataset / Data source} & 
\textbf{Data type} & 
\textbf{Primary use in reviewed studies} & 
\textbf{References} \\
\hline

\textit{BraTS} &
Multimodal brain MRI (T1, T1c, T2, FLAIR) with expert annotations &
Glioma segmentation, tumor delineation, and tumor-growth modeling &
\cite{menze2015brats} \\ \hline

\textit{BraTS-METS 2023} &
Brain MRI with metastatic lesion annotations &
Physics-informed brain metastasis segmentation and tumor-boundary delineation &
\cite{moawad2024brain} \\ \hline

\textit{Stanford BrainMetShare} &
Independent brain metastasis MRI cohort &
External validation of physics-informed segmentation models &
\cite{grovik2020deep} \\ \hline

\textit{TCIA collections} &
MRI, CT, and PET/CT cancer imaging &
Tumor segmentation, reconstruction, quantitative analysis, and model development; includes TCGA-GBM, TCGA-LGG, NSCLC Radiogenomics, AAPM-Mayo Low-Dose CT, and Mayo PET/CT Kidney &
\cite{clark2013cancer} \\ \hline

\textit{LIDC-IDRI} &
Annotated lung CT &
Pulmonary nodule detection, reconstruction, and quantitative analysis &
\cite{armato2011lung} \\ \hline

\textit{KiTS} &
Abdominal CT with kidney and kidney-tumor annotations &
Renal tumor segmentation &
\cite{heller2021state} \\ \hline

\textit{PROMISE12} &
Prostate MRI with gland annotations &
Prostate segmentation &
\cite{litjens2014evaluation} \\ \hline

\textit{QIN-BREAST-02} &
Multiparametric breast MRI, including CE-MRI, DWI, and ADC maps &
Patient-specific tumor characterization and mechanistic tumor modeling &
\cite{yankeelov2019qinbreast02} \\ \hline

\textit{ACRIN 6698 / I-SPY2} &
Longitudinal multiparametric breast MRI with treatment-response information &
Neoadjuvant chemotherapy response modeling &
\cite{partridge2018diffusion,li2020predicting} \\ \hline

\textit{fastMRI} &
MRI acquisition data &
Physics-consistent accelerated image reconstruction &
\cite{zbontar2018fast} \\ \hline

\textit{Clinical and private cohorts} &
Institution-specific MRI, CT, PET, and radiotherapy data &
Tumor modeling, reconstruction, segmentation, dose prediction, and clinical validation &
\cite{babier2021openkbp,sun2026physics,eichhorn2026motion} \\ \hline

\textit{Synthetic, simulated, phantom, and in-vitro data} &
Data generated from governing equations, imaging forward models, physical phantoms, or experimental systems &
PINN training, inverse modeling, reconstruction, and validation &
\cite{qian2022physics,wang2025one,fan2026pit} \\ \hline

\hline
\end{tabular}
\end{table*}

\section{Challenges and clinical translation of PIDL in oncologic imaging}

Despite the growing success of PIDL in oncologic imaging, several fundamental challenges continue to constrain its widespread adoption and clinical deployment. In image reconstruction, a principal open problem is the extension of current methods to fully three-dimensional and spatiotemporal imaging, such as 4D MRI, while faithfully representing complex acquisition physics, non-Cartesian sampling and nonlinear effects. Equally pressing is the need for robust performance under distribution shifts arising from differences in scanner hardware, acquisition protocols and patient populations, together with the reliable integration of uncertainty estimation. Recent work has accordingly turned towards self-supervised learning, which reduces reliance on ground-truth images, and towards approaches that remain effective when the assumed physical model is incomplete, uncertain or misspecified. The principal obstacles are discussed below.

\begin{figure}[pos=htbp]
    \centering
    \includegraphics[width=1.0\textwidth]{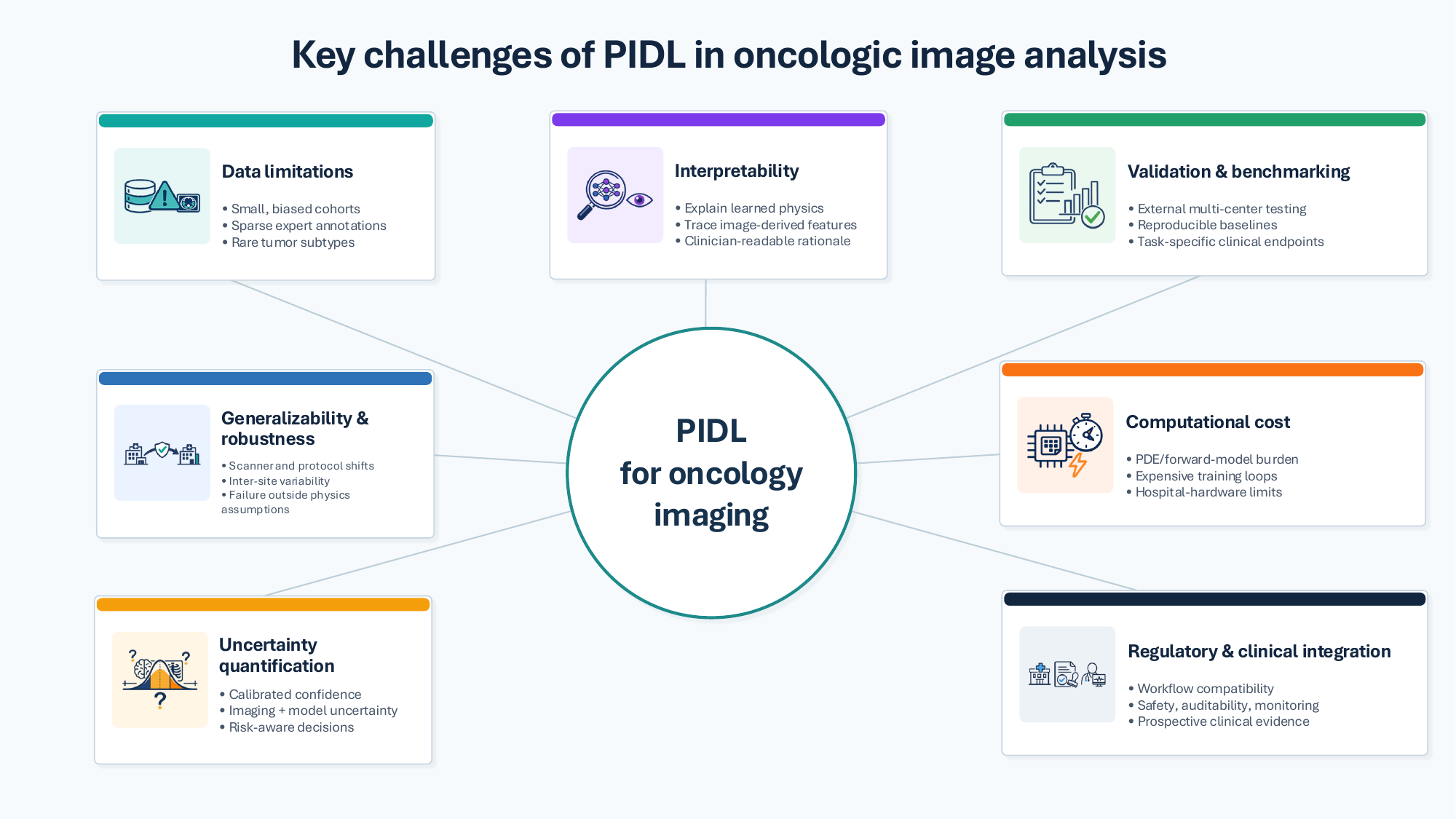}
    \caption{Key challenges limiting the development and clinical translation of PIDL in oncologic image analysis.}
    \label{fig:pidl_challenges}
\end{figure}

\vspace{0.2cm}
\noindent \textbf{Data availability and physics-model fidelity}: Although the incorporation of prior physical knowledge lessens dependence on large annotated datasets, model development nonetheless requires sufficiently diverse imaging data for training, calibration and independent validation \citep{cuomo2022scientific,banerjee2025physics}. In oncology, access to such data is frequently limited by acquisition and annotation costs, patient-privacy requirements, institutional data silos, inter-observer variability and the low prevalence of certain cancer types \citep{bhalgat2018annotation,wang2021annotation,veiga2022comparative}. These limitations are especially consequential when performance must be assessed across institutions, scanner vendors, acquisition protocols and patient populations.

The reliability of PIDL is further contingent on the fidelity of the embedded physical model. Existing frameworks commonly adopt simplified assumptions regarding tissue properties, physiological processes, boundary and initial conditions, imaging-system response or tumor-growth dynamics \citep{cuomo2022scientific,ahmadi2026physics}. Such assumptions may not adequately capture tumor heterogeneity, patient-specific anatomy, evolving treatment response or uncertainty in physiological parameters. Discrepancies between the imposed physics and clinical reality can therefore introduce systematic bias, constrain the learned solution towards an incorrect physical regime and diminish predictive reliability under domain shift \citep{guo2026impact,banerjee2025physics}. Future PIDL frameworks should incorporate uncertainty-aware model calibration, adaptive weighting of physical and data-driven losses, and mechanisms for detecting or correcting physical-model misspecification.

\vspace{0.2cm}
\noindent \textbf{Generalization and robustness}: Achieving robust performance across heterogeneous imaging protocols, scanner vendors, institutions and patient populations remains a central challenge. Physics-based constraints can improve generalization by restricting the learned solution to physically plausible regimes, yet they do not eliminate domain shift or model misspecification \citep{banerjee2025physics,ahmadi2026physics}. Differences in acquisition parameters, reconstruction pipelines, hardware characteristics, patient anatomy and disease phenotype can substantially alter image distributions and degrade performance when models are applied outside the conditions represented during development \citep{guo2026impact}. This vulnerability is most pronounced when the embedded physical assumptions, parameter values or boundary conditions are inaccurate for a new clinical setting. Robust clinical deployment will therefore require external validation across independent institutions, scanner platforms, acquisition protocols and diverse patient cohorts, complemented by sensitivity analyses and strategies for adapting the model when either the data distribution or the assumed physics changes.

\vspace{0.2cm}
\noindent \textbf{Uncertainty quantification and interpretability}: Reliable uncertainty quantification is essential for clinical deployment, as PIDL predictions are affected by measurement noise, uncertain parameters, incomplete physics and model mismatch \citep{linka2022bayesian}. Bayesian physics-informed neural networks (PINNs) and ensemble-based methods offer promising solutions, but their use in oncologic imaging remains limited \citep{casali2025physics}. Interpretability presents a related concern: although physical constraints enhance transparency, PIDL models can still exhibit black-box behavior, limiting clinician trust \citep{khurana2026explainable}.

\vspace{0.2cm}
\noindent \textbf{Computational scalability}: The incorporation of physical constraints inevitably increases computational cost, because PINNs require repeated evaluation of PDE residuals and higher-order derivatives during training \citep{cuomo2022scientific,karniadakis2021physics}. This burden becomes particularly significant for high-resolution three-dimensional imaging, multiphysics models and large clinical datasets. Recent advances, including neural operators, adaptive sampling and hybrid physics-data-driven architectures, have improved training efficiency and scalability; nevertheless, computational cost remains a major impediment to routine clinical deployment \citep{lu2021learning}.

\vspace{0.2cm}
\noindent \textbf{Benchmarking and clinical translation}: The absence of standardized benchmarks and evaluation protocols remains a major barrier to the objective assessment of PIDL methods. Most studies rely on small, single-center or institution-specific datasets, which limits reproducibility and precludes direct comparison across approaches. Progress will require publicly available benchmark datasets, consistent reporting of accuracy, physical consistency, uncertainty and computational cost, and multi-center, prospective clinical validation. Regulatory translation will further demand clear evidence of safety, robustness, interpretability and clinical benefit. Successful integration into routine oncology workflows will thus depend on sustained collaboration among clinicians, medical physicists, imaging scientists, regulatory experts and AI researchers.

In summary, the potential of PIDL in oncologic imaging is tempered by interrelated challenges spanning data availability, generalizability, uncertainty quantification, interpretability, computational cost, validation and clinical integration, as summarized in Fig.~\ref{fig:pidl_challenges}.

\section{Discussion}
\label{sec:optimized_workflows}

Physics-informed deep learning (PIDL) reconciles the representational capacity of deep neural networks with the rigor of physical and physiological modeling. By embedding governing equations, imaging-system physics, biomechanical constraints and domain knowledge directly within the learning process, PIDL confers physical consistency, partial interpretability, robustness and data efficiency that purely data-driven approaches cannot guarantee. This review has examined the theoretical foundations of the field, encompassing physics-informed neural networks (PINNs), physics-informed convolutional neural networks (PICNNs), physics-informed Kolmogorov-Arnold Networks (PIKANs), physics-informed operator-learning (PINO) frameworks, physics-informed generative models (PIGMs) together with their application across the principal oncologic imaging tasks and modalities, namely CT, MRI, PET, ultrasound, thermal imaging, hybrid imaging and emerging modalities. Here, we synthesize this evidence into task-specific design principles, delineate the research directions most likely to define the next generation of PIDL, and identify the conditions required for its clinical translation.

\vspace{0.2cm}
\noindent \textbf{Towards a task-dependent design principle}: A systematic synthesis of existing studies demonstrates that no single physics-informed architecture is optimal across all oncologic imaging tasks. The appropriate strategy is determined by the nature of the inverse problem, the available measurements, the governing physical model, the dimensionality of the data and the clinical objective. A unifying principle emerges from the reviewed literature: physics should be introduced at the stage most directly related to the underlying imaging or biological process. Acquisition physics is most naturally enforced during reconstruction and quantitative imaging; anatomical and biomechanical constraints are better suited to segmentation and motion correction; biological and transport models are better suited to tumor progression and treatment-response modeling; and radiation-transport models should be explicitly coupled to dosimetry and treatment planning. We therefore contend that physics should not be treated as a universal penalty term. Instead, task-specific physics-data coupling is required, in which the physical model, network architecture, loss formulation and validation strategy are selected jointly according to the clinical objective. The workflows below operationalize this principle for the major computational oncology tasks.

\vspace{0.2cm}
\noindent \textbf{Image reconstruction}: The optimal workflow couples the measured acquisition data directly to a differentiable imaging forward model, rather than relying solely on image-domain supervision. The proposed pipeline comprises: (i) preprocessing and normalization of raw measurements; (ii) initialization with a conventional reconstruction or a learned prior; (iii) extraction of image-domain features with a convolutional neural network (CNN), U-Net or transformer; (iv) explicit enforcement of acquisition-domain consistency through the forward model; and (v) joint optimization of data-fidelity, image-regularization and physics-consistency losses. For MRI, k-space consistency and Fourier encoding should be incorporated explicitly, whereas CT reconstruction should account for the X-ray projection and attenuation model. For high-dimensional volumetric reconstruction, operator-learning architectures are preferable to conventional coordinate-based physics-informed architectures on grounds of computational scalability.

\vspace{0.2cm}
\noindent \textbf{Tumor segmentation and boundary delineation}: The reviewed literature indicates that governing physical systems complement image-derived anatomical features. We therefore propose a hybrid workflow in which a U-Net, UNet++, Swin-based encoder-decoder or related architecture first extracts multiscale tumor features, after which tumor-growth, diffusion, edema-transport, biomechanical or modality-specific physical constraints are imposed. The optimization objective should balance segmentation accuracy against physical consistency, for example, by combining cross-entropy losses with physics-based residual or boundary losses. For anatomically complex tumors, boundary-aware physical constraints should take precedence over global PDE constraints. External validation across institutions is indispensable to establish whether the physical prior enhances segmentation robustness, rather than merely improving performance within the training domain.

 \vspace{0.2cm}
\noindent \textbf{Tumor detection and classification}: In detection and classification, physics should constrain the representation, feature extraction and decision function without supplanting the discriminative learning objective. The workflow begins with a physically consistent representation of the available patient-specific information, which may comprise multimodal images, quantitative imaging biomarkers, physiological measurements, molecular profiles, clinical variables or model-derived parameters. Depending on the modality and task, physical information can be introduced through acquisition-aware preprocessing, physics-based feature construction, mechanistic parameters or explicit regularization terms that penalize predictions inconsistent with known anatomical, physiological, metabolic or imaging relationships. The resulting representations can be processed by CNNs, transformers, multilayer architectures or multimodal fusion networks, with the classification objective optimized jointly with a physics-consistency loss that constrains the learned representation or prediction according to the relevant physical model. The choice of constraint should be modality- and disease-specific: PET-based models can incorporate tracer kinetics, attenuation and activity-distribution constraints, whereas MRI-based models can exploit acquisition physics, relaxation characteristics and tissue-dependent signal relationships. For multimodal data, complementary physical and biological information should be fused at the feature or latent-representation level while preserving modality-specific constraints.

\vspace{0.2cm}
\noindent \textbf{Quantitative imaging and parameter estimation}: Quantitative imaging is a particularly appropriate setting for physics-informed learning, because the desired outputs are frequently physical or physiological parameters rather than categorical labels. The problem should accordingly be formulated as a physics-constrained inverse problem: imaging measurements are related to the relevant forward model, a neural network estimates tissue-specific parameters, and the predicted parameters are propagated through the differentiable forward model to reconstruct the observed measurements. The discrepancy between measured and reconstructed observations is then minimized jointly with parameter regularization. Bloch-equation constraints are appropriate for quantitative MRI, pharmacokinetic models for dynamic contrast-enhanced imaging and PET, and tissue- or modality-specific physical models for other quantitative biomarkers. Bayesian (for noisy data) or ensemble (for clean data) formulations should be adopted wherever parameter uncertainty is clinically relevant.

\vspace{0.2cm}
\noindent \textbf{Motion correction}: The optimal strategy combines image-domain motion estimation with acquisition-domain consistency. The recommended workflow comprises motion detection, deformation-field estimation, physics-aware image correction and consistency enforcement against the acquired measurements. For MRI, k-space consistency ensures that corrected images remain compatible with the measured signal, while spatial-smoothness or biomechanical constraints regularize the estimated deformation fields. Transformer- or CNN-based architectures are well suited to motion estimation, and physics-based losses should constrain the corrected output rather than dominate feature learning.

\vspace{0.2cm}
\noindent \textbf{Radiation dosimetry and treatment planning}: The proposed workflow begins with patient-specific image segmentation and tissue-property estimation, followed by the incorporation of radiation-transport or dose-deposition models into the learning framework. A neural surrogate can then predict three-dimensional dose distributions while remaining consistent with the underlying physical model. Physics-based constraints should encompass energy deposition, attenuation, scattering and, where appropriate, tissue-dependent dose response. Evaluation should address both dosimetric accuracy and clinically relevant treatment objectives, rather than relying exclusively on data-driven prediction metrics.

\vspace{0.2cm}
\noindent \textbf{Tumor progression and treatment-response modeling}: These tasks demand the explicit integration of longitudinal observations with the biological mechanisms governing tumor growth, invasion and therapeutic response. We propose a longitudinal physics-informed workflow in which serial imaging and complementary patient-specific measurements first characterize tumor morphology, volume, spatial heterogeneity and quantitative biomarkers. These observations are then coupled with mechanistic models of proliferation, diffusion, invasion or treatment response within a physics-informed temporal framework. Patient-specific parameters are subsequently estimated through inverse learning, enabling the calibrated model to describe tumor evolution and to predict progression or response under specified therapeutic conditions. Uncertainty quantification should account for measurement noise, incomplete longitudinal observations, parameter uncertainty and mismatch between the assumed biological model and the true tumor dynamics.

\vspace{0.2cm}
\noindent \textbf{Multimodal oncologic imaging}: The literature suggests that physics should be incorporated at the modality level before cross-modal fusion. In the proposed workflow, each modality is first processed with its own forward model or physical constraint. The resulting physically consistent representations are then fused by cross-attention, transformer-based fusion or latent-space integration, with shared biological constraints, such as tumor location, tissue properties, perfusion or metabolic activity, providing additional cross-modal regularization. This strategy avoids imposing a single physical model across fundamentally different imaging modalities and affords a more interpretable basis for multimodal integration.

\vspace{0.2cm}
\noindent \textbf{Model selection}: Architectural choice should likewise be guided by the task. Coordinate-based PINNs and PIKANs are most appropriate for solving or identifying low-dimensional governing equations and for estimating unknown physical parameters. PICNNs and physics-guided encoder--decoder architectures are preferable for spatially resolved imaging problems, because convolutional differential operators reduce the computational burden associated with automatic differentiation. Neural operators are particularly attractive when repeated inference is required across varying patient-specific parameters, acquisition conditions or physical states. Transformer-based architectures are suited to long-range spatial or temporal dependencies and to multimodal fusion, whereas hybrid mechanistic--deep learning models are preferable when established biological or physiological models can be coupled with image-derived information.

\subsection{Future Perspectives}
\noindent \textbf{Scientific foundation models, transfer learning and federated learning}: Looking beyond task-specific optimization, future research is expected to progress towards unified, patient-specific imaging frameworks capable of supporting the entire oncologic care pathway, from tumor detection and localization to treatment planning, response assessment and disease monitoring. The emergence of Scientific foundation models (SciFMs) \cite{menon2026scientific} presents a transformative paradigm in this respect. Pretraining on large, multiscale and heterogeneous medical-imaging datasets can improve generalization and reduce dependence on task-specific annotations. Nevertheless, preserving physical consistency, interpretability and robustness when such models extrapolate beyond their training distribution, or encounter extreme regimes, remains a critical challenge. In parallel, transfer learning (TL) \cite{chakraborty2021transfer,prantikos2023physics,kapoor2024transfer,liu2025fast} offers an important means of improving the scalability of PIDL under limited annotated data, by transferring physical representations learned from simulations, synthetic datasets or related anatomical domains to patient-specific oncologic imaging tasks. Future investigations should explore physics-preserving transfer mechanisms and architecture-level adaptation to improve robustness across heterogeneous cancer types, imaging protocols and scanner configurations. Further, federated learning (FL) \cite{yang2023hypernetwork,li2023federated,zhang2025federated,fan2026pifl,chen2026novel} provides a privacy-preserving framework for multi-institutional training of PIDL, enabling collaborative learning across heterogeneous imaging protocols and scanners. The integration of personalization, differential privacy and robust federated optimization is expected to improve the clinical deployment of PIDL in collaborative oncologic imaging.

\vspace{0.2cm}
\noindent \textbf{Digital twins and agentic PIDL}: Digital twins \cite{yang2024data,tang2026physics, shen2026mapping} represent a promising direction for patient-specific PIDL frameworks that continuously integrate multimodal imaging, physiological measurements and clinical data to support tumor characterization, treatment planning and longitudinal therapy monitoring. Recent advances in physics-informed organ modeling, self-updating PINN frameworks and real-time surrogate modeling \cite{kuang2024med,nabila2026physics,karkadakattil2026physics,osman2026cardiovascular} provide the methodological foundation for adaptive oncologic digital twins capable of continuously refining patient-specific predictions. Future research should focus on integrating anatomical constraints, uncertainty quantification and privacy-preserving learning to enable reliable and clinically deployable digital twins for precision oncology \cite{kang2026hard}. Further, agentic artificial intelligence (AI) \cite{sapkota2025ai} offers an emerging route to adaptive PIDL systems that autonomously optimize, validate and refine physics-informed models. Drawing on advances in multi-agent scientific reasoning and physics-audited model verification \cite{jiang2026agenticsciml,toscano2025athena}, future frameworks may enable reliable image reconstruction, tumor characterization, treatment planning and therapy monitoring while maintaining physical consistency.

\vspace{0.2cm}
\noindent \textbf{Uncertainty Quantification}:  Uncertainty-aware \cite{chan2026uncertainty,kong2026uncertainty,venkataramanan2025uncertainty,thakur2026punch,el2026tumor,lee2026evidential} and interpretable PIDL \cite{he2024interpretable,liu2025interpretable} are emerging as pivotal research directions for improving the clinical reliability and trustworthiness of oncologic imaging. Robust quantification of epistemic, aleatoric and model uncertainty is essential for deploying PIDL in high-stakes oncologic workflows, where inaccurate predictions can directly affect tumor targeting, dose calculation and treatment planning \cite{ahmadi2026physics}. Whereas conventional deep learning methods often produce overconfident errors under distribution shift or noisy acquisition protocols, PIDL frameworks offer a distinct structural advantage: governing physical laws and Bayesian formulations act as regularizers that constrain the solution space and flag physically implausible outputs \cite{kong2026uncertainty}. The integration of predictive and epistemic uncertainty into physics-informed models enables confidence-aware predictions and the identification of unreliable outcomes across diverse oncologic imaging tasks, thereby supporting risk-informed diagnostic and therapeutic decision-making. Future research must develop scalable uncertainty-aware techniques tailored to biomedical image simulation, reconstruction and physiological tracking, as well as to the modeling of complex biological dynamics in cancer cell populations, thereby enabling real-time confidence estimation during tumor segmentation, kinetic parameter mapping and longitudinal response assessment.

\vspace{0.2cm}
\noindent \textbf{Interpretability}: The black-box nature of standard neural networks remains a primary barrier to clinical adoption in oncology, as clinicians require transparent justification for diagnostic and therapeutic decisions. Interpretable PIDL addresses this limitation by embedding anatomical, physiological and biomechanical principles directly into network architectures and loss functions, ensuring that model predictions align with known biomedical laws rather than purely statistical correlations \cite{qadir2026physics,menon2026fekan,menon2025anant,zhang2026bubbleokan}. Future developments should prioritize glass-box physics-informed representations, feature-attribution methods mapped to physical quantities (such as mass conservation, diffusion coefficients or wave-propagation parameters) and interactive visualization tools that allow clinicians to interrogate the physical consistency of model outputs. Unified frameworks that combine uncertainty quantification, inherent interpretability and clinician-in-the-loop decision support will be required to enable safe, robust and clinically deployable precision oncology systems.

\vspace{0.2cm}
\noindent \textbf{Clinical translation and deployment}: Clinical translation and real-world deployment constitute the decisive test for PIDL in oncologic imaging. Routine clinical adoption will require rigorous multi-center validation, standardized evaluation protocols and acquisition-agnostic models capable of generalizing across heterogeneous imaging systems and patient cohorts \cite{sun2026physics,stanzione2022oncologic}. Future efforts should also emphasize self-supervised and physics-constrained learning for patient-specific personalization, together with deployment-ready frameworks that support real-time clinical workflows, interoperability and regulatory-compliant validation. The challenges identified in this review, namely data scarcity, model fidelity, domain generalization, computational scalability, benchmarking and regulatory requirements, make clear that successful deployment depends not only on methodological advances but also on standardized evaluation, explainable decision-making and sustained collaboration among clinicians, medical physicists, imaging scientists, regulatory experts and AI researchers.

\subsection{Concluding remarks}
PIDL is evolving from a promising research paradigm into a scientifically grounded framework for precision oncology, and its trajectory points towards a fundamental reconfiguration of how cancer is imaged, modeled and managed. We anticipate that the convergence of scientific foundation models, transfer and federated learning, neural operators, multimodal scientific learning, digital twins, uncertainty-aware Bayesian inference, trustworthy and explainable AI, and agentic scientific intelligence will progressively dissolve the boundaries between image acquisition, reconstruction, interpretation and therapeutic decision-making. In this envisioned paradigm, imaging will cease to be a static diagnostic snapshot and will instead become a continuously updated, physics-governed representation of the individual patient. Foundation models pretrained across scales, modalities and institutions could furnish transferable physical priors, while federated and privacy-preserving optimization allows these priors to be refined across global clinical networks without compromising patient confidentiality. Coupled with mechanistic tumor modeling and multimodal clinical data, patient-specific digital twins may then be calibrated longitudinally to simulate tumor evolution, to test candidate therapeutic strategies in silico before their administration, and to forecast treatment response with quantified confidence. Agentic PIDL systems, operating under physics-audited verification, could further automate model construction, validation and recalibration, shifting the role of the clinician from supervising individual computations to governing adaptive decision-support ecosystems. Realizing this vision will nevertheless depend on resolving the foundational questions identified in this review, namely guaranteeing physical consistency under extrapolation, establishing calibrated and clinically actionable uncertainty, ensuring interpretability that withstands regulatory and ethical scrutiny, and demonstrating clinical benefit through prospective multi-center validation. If these conditions are met, PIDL has the potential to unify imaging physics, tumor biology, mathematical modeling and machine learning within a single computational framework, delivering reliable, interpretable and clinically deployable AI systems that extend beyond correlation-driven learning towards physically consistent and biologically meaningful decision support in cancer care. Such a framework would not merely improve existing oncologic workflows but could establish the computational foundation for truly predictive, preventive and personalized oncology.

\appendix
\section{Abbreviations}

\begin{center}
\renewcommand{\arraystretch}{1.15}
\begin{tabular}{@{}p{0.2\textwidth}p{0.72\textwidth}@{}}
\toprule
\textbf{Abbreviation} & \textbf{Definition} \\
\midrule
AD & Automatic Differentiation \\
BraTS & Brain Tumor Segmentation Challenge \\
CNN & Convolutional Neural Network \\
CT & Computed Tomography \\
DCE-MRI & Dynamic Contrast-Enhanced Magnetic Resonance Imaging \\
DECT & Dual-Energy Computed Tomography \\
DSC-MRI & Dynamic Susceptibility Contrast Magnetic Resonance Imaging \\
FDM & Finite Difference Method \\
FLAIR & Fluid-Attenuated Inversion Recovery \\
GDPR & General Data Protection Regulation \\
HIPAA & Health Insurance Portability and Accountability Act \\
IMRT & Intensity-Modulated Radiation Therapy \\
MedIA & Medical Image Analysis \\
MRI & Magnetic Resonance Imaging \\
NRMSE & Normalized Root Mean Squared Error \\
ODE & Ordinary Differential Equation \\
PDE & Partial Differential Equation \\
PET & Positron Emission Tomography \\
PICNN & Physics-Informed Convolutional Neural Network \\
PIDL & Physics-Informed Deep Learning \\
PIML & Physics-Informed Machine Learning \\
PINN & Physics-Informed Neural Network \\
PINO & Physics-Informed Neural Operators \\
PIGM & Physics-Informed Generative Models \\
PSNR & Peak Signal-to-Noise Ratio \\
qMRI & Quantitative Magnetic Resonance Imaging \\
SSIM & Structural Similarity Index \\
TC & Tumor Core \\
TCIA & The Cancer Imaging Archive \\
U-Net & U-shaped Convolutional Network \\
WT & Whole Tumor \\
\bottomrule
\end{tabular}
\end{center}

\pagebreak


\bibliographystyle{model1-num-names}

\bibliography{cas-refs}



\end{document}